\documentclass[reprint,superscriptaddress,aps,prx,]{revtex4-2}
\usepackage{amsmath}
\usepackage{amssymb}
\usepackage{graphicx}
\usepackage{epstopdf}
\usepackage[colorlinks=true]{hyperref}
\usepackage{physics}
\usepackage{mathrsfs}
\usepackage{comment}

\renewcommand\({\begin{equation}}	
\renewcommand\){\end{equation}}
\renewcommand\[{\begin{eqnarray}}	
\renewcommand\]{\end{eqnarray}}
\newcommand{\al}[1]{\begin{aligned}#1\end{aligned}}
\usepackage[dvipsnames]{xcolor}

\begin{document}

\title{ Quantum computing on magnetic racetracks with flying domain wall qubits}

\author{Ji Zou}
\affiliation{Department of Physics, University of Basel, Klingelbergstrasse 82, 4056 Basel, Switzerland}
\author{Stefano Bosco}
\affiliation{Department of Physics, University of Basel, Klingelbergstrasse 82, 4056 Basel, Switzerland}
\author{Banabir Pal}
\author{Stuart S. P. Parkin}
\affiliation{Max Planck Institute of Microstructure Physics, Weinberg 2, 06120 Halle, Germany}
\author{Jelena Klinovaja}
\affiliation{Department of Physics, University of Basel, Klingelbergstrasse 82, 4056 Basel, Switzerland}
\author{Daniel Loss}
\affiliation{Department of Physics, University of Basel, Klingelbergstrasse 82, 4056 Basel, Switzerland}

\begin{abstract}
 Domain walls (DWs) on magnetic racetracks are at the core of the field of spintronics, providing a basic element for classical information processing.
Here, we show that mobile DWs also provide a blueprint for large-scale quantum computers. 
Remarkably, these DW qubits showcase exceptional versatility, serving not only as stationary qubits, but also performing the role of solid-state flying qubits that can be shuttled in an ultrafast way. We estimate that the DW qubits are long-lived because they can be operated at sweet spots to reduce potential noise sources. Single-qubit gates are implemented by moving the DW, and two-qubit entangling gates exploit naturally emerging interactions between different DWs. These gates, sufficient for universal quantum computing, are fully compatible with current state-of-the-art experiments on racetrack memories. Further, we discuss possible strategies for qubit readout and initialization, paving the way toward future quantum computers based on mobile topological textures on magnetic racetracks.
\end{abstract}

\date{\today}
\maketitle

\section{Introduction}

Magnetic domain walls (DWs) have garnered significant attention in recent years~\cite{Ryu:2013wh,Yang2015uw,Kim2017natmat,Yang:2021wy,Guan:2021vo,Blasing:2018vb,Yoshimura:2016uk,Jiang283,hantao2022prb,sekwon2021prb,donges2022prr,Sinova2014prl,PhysRevLett.117.017202}, owing to their wide-ranging applications in the field of spintronics. They are integral to various logic devices~\cite{Kumar:2022vy,doi:10.1063/5.0042917,Grollier:2020ur,PhysRevLett.121.127701,Daltonenergy}  and are recognized as stable information carriers due to their inherent topological robustness~\cite{TopologyinMagnetism,Zangprl,jiprl2020, jivortex, quantumvortex,Yqwinding}. This has led to their deployment in classical racetrack memories, pushing the boundaries of technology~\cite{Parkin190}. In recent developments, DWs have demonstrated large and tunable mobility on both antiferromagnetic and ferrimagnetic nanotracks~\cite{Ryu:2013wh, Yang2015uw, Kim2017natmat,Yang:2021wy,Guan:2021vo, Blasing:2018vb,Yoshimura:2016uk}, which has significantly enhanced our ability to control the movement of DWs in magnetic nanowires. 

While experimental endeavors have largely focused on the classical regime, the recent technological advancements in spintronics, specifically the ability to stabilize and manipulate nanoscale DWs~\cite{Ryu:2013wh,Yang2015uw,Kim2017natmat,Yang:2021wy,Guan:2021vo,Blasing:2018vb,Yoshimura:2016uk,Jiang283,hantao2022prb,sekwon2021prb,donges2022prr,Sinova2014prl,PhysRevLett.117.017202}, have opened doors to compelling opportunities to breach the quantum frontier and explore  applications of DWs in quantum realms.
Concurrently, there is a mounting quest for quantum computing platforms, fueled by the capability of large-scale universal quantum computers to tackle problems  beyond the reach of their classical counterparts~\cite{shor_1994}. 
Diverse platforms, such as  trapped ions~\cite{Blinov2004nature,volz2006prl,Blatt2008nature}, superconducting circuits~\cite{Koch_pra_2007,Barends_prl_2013}, and quantum dots~\cite{PhysRevA.57.120,Basso2019prl,Qiao:2020ncom,Hendrickx:2021tv,Philips:2022ur,petta2022sa}, are actively pursued. In this context, nanosize spin textures are gaining attention as potential qubits~\cite{christina_prl_2021,xia2023universal,xia2022qubits,PhysRevB.95.144402,PhysRevB.97.064401,ahari2023andreev}, serving as the fundamental building blocks of a quantum computer.

\begin{figure}[t]
	\centering\includegraphics[width=0.95\linewidth]{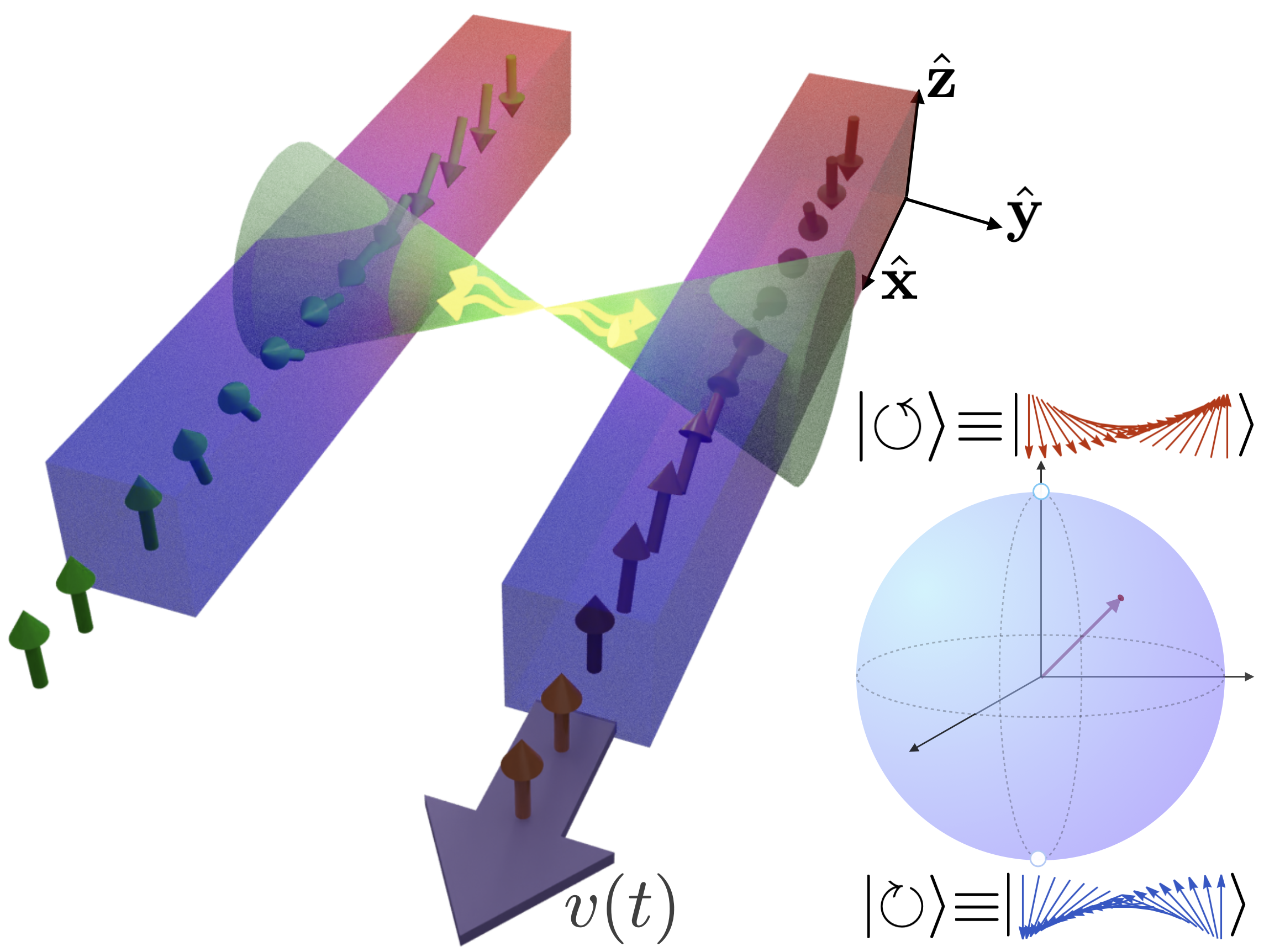}
 \caption{Sketch of two DW qubits on two parallel ferrimagnetic racetracks.  Spins  stand for the uncompensated magnetic moments of the sublattices.  The  spin texture of the right racetrack has a well-definite positive chirality and is in state $\ket{\circlearrowleft}$, whereas the texture within the left one has negative chirality and is in state $\ket{\circlearrowright}$.  Single qubit gates are implemented by shuttling the domain wall and controlling its velocity $v(t)$. Two-qubit entangling gates between different racetracks take advantage of inter-track exchange and dipolar interactions. }
  \label{fig1}
\end{figure}

In this work, we propose a scalable implementation of a universal quantum computer on ferrimagnetic racetracks with mobile DW qubits. The qubit computational space is spanned by two topologically distinct states, each possessing opposite chirality, as depicted in Fig.~\ref{fig1}. Our design takes advantage of the topological nature of the DW and its high mobility, thereby fully harnessing the potential of  DWs within the quantum regime.
The proposed platform brings several crucial advantages. First,  the inherent mobility of DWs uniquely positions them to act as both stationary and solid-state flying qubits. This property expedites entanglement distribution and fosters long-distance coherent quantum communication, eliminating the need for external components such as resonators commonly employed in superconducting qubit platforms. Conservatively estimated, DW qubits can fly coherently for distances of a few micrometers at velocities up to 100 m/s within their coherence time, estimated to be in the microsecond regime.  Second, the strong coupling between the qubit sub-space and  DW motion, originating from the spin Berry phase accumulated by spins within the DW, enables us to achieve fast and high-fidelity single- and two-qubit operations on the scale of 0.1 ns.
Attaining such a substantial effective spin-orbit coupling proves challenging in alternative solid-state qubits~\cite{Nadj-Perge:2010tb,Froning:2021wk,froning2021prr}. 
Third, the racetrack system inherently facilitates scalability, as it allows for the preparation of multiple DWs, thereby providing a natural pathway for scalability in experimental setups. Additionally, racetrack arrays present the potential for a genuinely 3D quantum computing platform~\cite{Parkin190,gu2022three}, enabling higher qubit density and substantial advantages for the development of large-scale quantum computing. Finally, our design is compatible with state-of-the-art experimental spintronic architectures, such as racetrack memory, and can be naturally integrated with advanced DW control technology. This compatibility paves the way for the application of spintronic schemes in future quantum information technology.

\begin{figure*}
	\centering\includegraphics[width=1\linewidth]{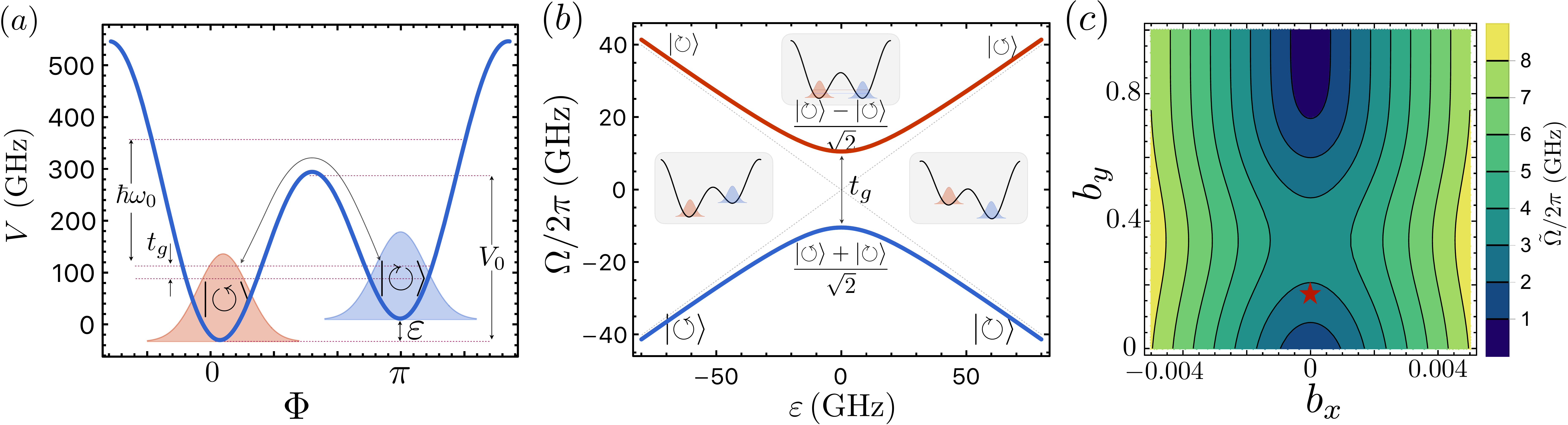}
 \caption{The DW qubit. (a) The effective potential energy $V(\Phi)$. Two states with opposite chirality, which span the Hilbert space of the DW qubit, are localized at two minima of the potential. They are hybridized by the tunnelling $t_g$ through the barrier $V_0$, that is controlled by $B_y$ and $K_y$. The detuning $\varepsilon$ of the two minima is proportional to $B_x$.   (b) Energy dispersion of the  Hamiltonian~\eqref{eq3} against $\varepsilon$.  A finite value of $B_x$ favours states with a well-defined chirality. (c) The physical qubit splitting $\tilde{\Omega}$ as a function of the dimensionless magnetic field $b_x$ and $b_y$ with zero shuttling velocity. The red star  marks the qubit operational point used for estimations. Here, we used the parameters given in Table~\ref{table1}. }
  \label{fig2}
\end{figure*}

\section{Qubit from Domain Wall Chirality}
We consider a quasi-one-dimensional  two-sublattice ferrimagnet described by the  effective low-energy  Lagrangian density:
\( \mathcal{L}= \frac{\hbar^2N}{8J} (\dot{\vb n} -\vb h \times \vb n )^2 + \hbar N S_e \dot{\vb n}\cdot \vb A - \mathcal{U}(\vb n), \label{eq1} \)
where $J>0$ is the antiferromagnetic exchange coupling  and  $\vb h=g\mu_B \vb B/\hbar$ with  external magnetic field $\vb B$  and electronic $g$ factor.  The first term is the kinetic energy density of the N{\'e}el  vector $\vb n$ in the presence of a magnetic field~\cite{ji_2021_neel}, while the second term is the spin Berry phase due to the uncompensated  moments with excess spin $S_e$ per site and vector potential $\vb{A}(\vb n)$ defined by $\nabla_{\vb n}\times \vb A=-\vb n$. We  assume that the  direction of the excess spin is locked along the N{\'e}el  vector~\cite{daniel_1992_prl,daniel_1997_prb}, as only the low-energy spin dynamics is concerned. 

The energy density $\mathcal{U}(\vb n)$ can be separated into two parts. The first contribution with larger energy is $\mathcal{U}_1(\vb n)= NK_z[(\partial_x \vb n)^2- n_z^2 ]/2$ and defines a DW  of width $\lambda=Sa\sqrt{J/K_z}$. Here,  $K_z$ is the easy axis anisotropy along the $z$ direction, $S$ is  the average spin, $a$ is the  lattice spacing, and $N$ is the total number of  spins within a  DW. 
We remark that we use a dimensionless spatial coordinate $x$ throughout this work, measuring distance in the unit of $\lambda$. 
The DW profile  that minimizes $\mathcal{U}_1$ is  $n_x+in_y=e^{i\Phi}\sech(x-X), n_z=\tanh(x-X)$, where for concreteness we used the boundary condition $n_z(\pm\infty)=\pm 1$. The two zero-modes $\Phi$ and $X$, whose variations leave $\mathcal{U}_1$ invariant,  physically represent the position of DW in real space and its azimuthal angle in spin space.  The second  energy contribution to $\mathcal{U}(\textbf{n})$  is   $\mathcal{U}_2=NK_yn_y^2 -\hbar N S_e\vb h \cdot \vb n$,  where  $K_y$ defines the easy $xz$ plane and the second term is the Zeeman energy. The hierarchy of energy scales in the system is given by $J\gg K_z\gg K_y,~\hbar S_e |\vb h|$.

\begin{table*}
\caption{Domain wall qubit parameters. We assume $J=300\, \text{K}, K_z=3\, \text{K}, K_y=0.1\, \text{K}, a=5\, \text{\AA}, \, \lambda=5\, \text{nm}, B_x=B_z=0, B_y=1~\text{T},\, N=100, S=1, S_e=0.01, l_p=2.25\, \text{nm}$, shuttling velocity $v=20\, \text{m/s}$, temperature $T=50\, \text{mK}$, and Gilbert damping parameter $\alpha=10^{-5}$.}
\begin{center}
\begin{tabular}{c c c c c c c} 
 \hline\hline
  Qubit splitting $\tilde{\Omega}$    \;\;\; &  $l_{\text{so}}$   \;\;     & \;\;\;   $T_1$ \;\;\;\; & \;\;\;  $T_2$ \;\;\;&   Maximal shuttling speed    \;\;\;& Coherent shuttling distance  \;\;\;& Quality factor  \\ [0.5ex] 
 \hline
 5~GHz & 1.5~\text{nm}   \;\; & 0.23 $\mu$s \;    & 0.15 $\mu$s & 100 m/s & 3~$\mu$m & $2\times10^3$  \\ 
 \hline\hline
\end{tabular}
\end{center}
\label{table1}
\end{table*}

The two zero-modes are crucial to encode a qubit in the DW. To illustrate this concept, we first focus on the dynamics of $\Phi$ alone, that defines the computational space. The coordinate $X$ plays a critical role for implementing qubit gates and defining the physical qubit frequency, which  will be discussed later.   The Lagrangian for $\Phi$ is $L=M\dot{\Phi}^2/2 -V(\Phi)$, where the potential energy is,
\( V(\Phi)=2NK_y  ( \sin^2\Phi -2b_x\cos\Phi -2b_y \sin \Phi   ) ,   \)
and $b_i=\pi\hbar S_eh_i/4K_y$ is the dimensionless magnetic field, relative to  the easy plane anisotropy.  The detailed derivation is provided in Appendix~\ref{app_a}.
This Lagrangian describes a particle with effective mass  $M=N\hbar^2/2J$ moving in a double-well potential [see Fig.~\ref{fig2}(a)], whose low-energy states can be understood with the path integral quantization~\cite{atland}. The two states with opposite chirality $\ket{\circlearrowleft}\equiv \ket{\Phi=0}$ and $\ket{\circlearrowright}\equiv\ket{\Phi=\pi}$ correspond to macroscopically distinct topological spin textures and span our computational space.
The easy-plane anisotropy $K_y$ breaks the original U(1) symmetry down to $\mathbb{Z}_2$, where the two chirality states are equally preferred. This  $\mathbb{Z}_2$  is further broken by an external field $b_x$, resulting in a detuning energy $\varepsilon=-8NK_yb_x$ favouring one chirality over the other, see Fig.~\ref{fig2}(b). We remark that the Dzyaloshinskii-Moriya interaction, which prefers DWs with certain chirality observed in recent experiments~\cite{Ryu:2013wh, Yang2015uw, Kim2017natmat,Yang:2021wy,Guan:2021vo, Blasing:2018vb,Yoshimura:2016uk}, acts as an effective magnetic field along the racetrack. It can be compensated by turning on $b_x$ in the opposite direction.  For small DW sizes, containing $N\approx 10^2$ spins which is feasible experimentally~\cite{dwsize}, the DW is in the quantum regime, and the two chiralities hybridize with a tunnel splitting $t_g\approx 4\hbar \omega_0\sqrt{S_{\text{inst}}/2\pi \hbar} \exp{-S_{\text{inst}}/\hbar}$. Here, $S_{\text{inst}} \approx 4V_0/\omega_0$ is the instanton action with the  tunneling barrier $V_0=2NK_y(1-b_y)^2$ and the  level spacing $\hbar\omega_0=2 \sqrt{2JK_y(1-b_y^2)}$ (Appendix~\ref{app_b}). We emphasize that the tunneling rate $t_g/\hbar$ is highly tunable by the external magnetic field $b_y$: the barrier $V_0$ is suppressed for larger values of $b_y$, resulting in a larger $t_g$.  
 
Because $\hbar\omega_0/t_g\propto e^{4V_0/\hbar \omega_0}$ is large when $b_y<1$~\cite{anharmo}, the subspace spanned by $\{\ket{\circlearrowleft}, \ket{\circlearrowright}\}$ is well isolated from higher energy levels, and we obtain the effective DW qubit Hamiltonian  (Appendix~\ref{app_b}),
  \( H_{\text{Q}}=  \frac{\varepsilon}{2}\sigma_z -\frac{t_g}{2} \sigma_x,  \label{eq3} \)
  where  $\sigma_i$ are Pauli matrices and  qubit energy is $\hbar\Omega=\sqrt{\varepsilon^2+t_g^2}$  (on the scale of 20~GHz), as shown in Fig.~\ref{fig2}(b). We anticipate that  the physical qubit frequency $\tilde{\Omega}$ [depicted in Fig.~\ref{fig2}(c)] is suppressed by an order of magnitude   because of the strong coupling between $\Phi$ and $X$, as detailed below.

\section{Effective Hamiltonian for Flying domain wall qubits}
 To implement single- and two-qubit gates, we take advantage of the coupled dynamics of the  two soft modes $X$  and $ \Phi$, which  originates from  the finite excess spin $S_e$ in ferrimagnets. The spin Berry connection $\textbf{A}$ and the magnetic field $\textbf{h}$ in Eq.~\eqref{eq1} give rise to an effective spin-orbit interaction between $X$ and $\Phi$, and enable qubit operations by controlling the spatial motion of the DW~\cite{golovach2006prb}. 
 

To obtain the effective Hamiltonian for the DW flying on the magnetic racetrack,  we start from the Hamiltonian for the two soft modes $X, \Phi$: 
\( H= \frac{\hat{P}_\Phi^2}{2M} +V(\hat{\Phi}) + \frac{[\hat{P} +\hat{A}(\hat{\Phi}) ]^2}{2M}+ \frac{M\omega_p^2}{2} [\hat{X} -X_0(t) ]^2, \label{s16}  \)
with $\hat{A}=-\pi M h_x \sin\hat{\Phi}/2 +\pi M h_y\cos\hat{\Phi}/2 -2N\hbar S_e \hat{\Phi}$.  Here we have quantized both $\hat{\Phi}$ and $\hat{X}$, with $[\hat{\Phi}, \hat{P}_\Phi]=[\hat{X}, \hat{P}]=i\hbar$. We assumed the minimum of the DW confining potential is located at $X_0(t)$, which is a classical parameter that can be accurately controlled in experiment by several different means,  enabled by the recent progress in spintronics~\cite{Ryu:2013wh, Yang2015uw, Kim2017natmat,Yang:2021wy,Guan:2021vo, Blasing:2018vb,Yoshimura:2016uk, Jiang283,hantao2022prb,sekwon2021prb,donges2022prr}, such as  by  magnetic or electric fields, or  spin-polarized electric current. We model the DW confinement  with a harmonic potential with frequency $\omega_p$ which we assume to be comparable to the level spacing $\omega_0\sim 300~\text{GHz}$ in our estimation. 

We now project $\hat{\Phi}$ onto the qubit space (Appendix~\ref{app_b}), yielding  the Hamiltonian
\(  H=  \frac{\varepsilon}{2}\sigma_z- \frac{t_g}{2}\sigma_x  + \frac{(\hat{P} + \vb*\beta \cdot \vb*\sigma   )^2}{2M} + \frac{M\omega_p^2}{2} [\hat{X}-X_0(t)]^2, \label{s18} \)
with effective spin-orbit interaction vector 
$ \vb* \beta = ( - \pi Mh_x \gamma_x/2, 0,  \pi M h_y \gamma_z/2  + 2N \hbar S_e \bar{\gamma}_z  ). $
Here we dropped the constant part of the gauge field because it can be gauged away by a spin-independent transformation. $\gamma_x, \gamma_z$ and $\bar{\gamma}_z$ are constants, depending on system parameters. Their explicit expressions are given in Appendix~\ref{app_b}. 

Let us now derive the effective Hamiltonian of the flying DW qubit on the magnetic racetrack. 
It is convenient to introduce the notation: 
$\vb*\beta=\sqrt{\beta_x^2+\beta_z^2} (\cos\beta, 0, \sin\beta)$,  
with $\cos\beta=\beta_x/\sqrt{\beta_x^2+\beta_z^2} $ and $\sin\beta=\beta_z/\sqrt{\beta_x^2+\beta_z^2} $. We switch  to a frame moving with the domain wall by: 
$ H_m=     \hat{T}^\dagger H \hat{T} -\dot{X}_0(t)  i\hbar \hat{T}^\dagger \partial_{X_0} \hat{T},$  with the displacement operator in real space $ \hat{T}= \text{exp}[-i\hat{P}X_0(t)/\hbar], $~\cite{stefano2022prl,stefano2021prb}
which effectively shifts $\hat{X}\rightarrow \hat{X}+X_0(t)$ and introduces the Galilean term $-i\hbar \hat{T}^\dagger \partial_t\hat{T}=-\dot{X}_0\hat{P}$~\cite{anatoli_2017_phyreport}.
We then obtain the following effective Hamiltonian in the moving frame: 
\( H_m\! =\! \frac{\varepsilon}{2}\sigma_z \! - \! \frac{t_g}{2}\sigma_x+ \frac{(\hat{P}+\vb*\beta \cdot \vb*\sigma  )^2}{2m} + \frac{m\omega_p^2\hat{X}^2}{2} \! -\! \dot{X}_0(t)\hat{P}.     \)
To investigate the effect of the large effective spin-orbit interaction on the qubit dynamics, we perform a gauge transformation:
$ \mathcal{H}_m= \hat{ \mathcal{G} }^\dagger H_m \hat{ \mathcal{G} },$ with    $ \hat{ \mathcal{G} }=\text{exp}(-i\hat{X}\vb*\beta\cdot \vb*\sigma/\hbar)$ being a spin-dependent displacement in phase space.
It effectively shifts $\hat{P}\rightarrow \hat{P}-\vb*\beta \cdot \vb*\sigma$ and locally rotates the spin space in a position-dependent way.  We obtain the following Hamiltonian
\(  \al{ \mathcal{H}_m=  \hbar \omega_p \hat{a}^\dagger & \hat{a}  -\dot{X}_0(t)\hat{P}+ \dot{X}_0(t) \vb*\beta \cdot \vb*\sigma \\  & - \frac{t_g}{2} \hat{ \mathcal{G} }^\dagger\sigma_x \hat{ \mathcal{G} } + \frac{\varepsilon}{2} \hat{ \mathcal{G} }^\dagger \sigma_z \hat{ \mathcal{G} }.  } \)
Here, we quantized the spatial degree of freedom $\hat{X}$ with the bosonic opertors $\hat{a}, \hat{a}^\dagger$.   We now project the orbital dynamics onto its ground state $\ket{0}$ defined by $\hat{a}\ket{0}=0$, which leads to the effective Hamiltonian of the flying DW qubit on the magnetic racetrack:
\( \mathcal{H}_m= \frac{\tilde{\varepsilon}(t)}{2} \sigma_z - \frac{\tilde{t}_g}{2} \sigma_x. \label{eq6}  \)
with the renormalized tunneling splitting $\tilde{t}_g$ and the detuning $\tilde{\varepsilon}$, 
\( \al{ \tilde{t}_g=&- \frac{2\hbar \cos\beta \dot{X}_0(t)}{l_{\text{so}}} -\varepsilon \sin\beta \cos\beta (1-e^{-l_p^2/l^2_{\text{so}}}) \\ & +t_g (\cos^2\beta+\sin^2\beta e^{-l_p^2/l_{\text{so}}^2}), \\
          \tilde{\varepsilon}= & \frac{2\hbar\sin\beta \dot{X}_0(t)}{l_{\text{so}}} +\varepsilon (\sin^2\beta+\cos^2\beta  e^{-l_p^2/l_{\text{so}}^2} ) \\ & -t_g (1-e^{-l_p^2/l_{\text{so}}^2}) \sin\beta \cos\beta.    }\)
          Here, $l_p=\sqrt{\hbar/M\omega_p}$ is the  characteristic length  of the harmonic potential which reflects the uncertainty of the DW position, and $l_{\text{so}}=\hbar/|\vb*\beta|$ is the spin-orbit length (the qubit state is flipped by spin-orbit interactions after a distance $\pi l_{\text{so}}/2$).   For realistic system parameters, we estimate that these lengths both are on the scale of a few nanometers, see Table~\ref{table1}. 
  We emphasize that such large spin-orbit interactions are challenging to reach in alternative solid-state qubits, where $l_\text{so}$ is about  $10 \sim100$~nm~\cite{Nadj-Perge:2010tb,Froning:2021wk,froning2021prr,fasth2007prl,Camenzind:2022tr,Li:2015us}.
          With the parameters that we assumed, we have $\beta_x\ll \beta_z$ (Appendix~\ref{app_b}).  Thus we can approximate $\cos\beta=0$, $\sin\beta=1$, in which case parameters above are reduced to 
$ \tilde{t}_g = t_g \text{exp}(-l_p^2/l_{\text{so}}^2)$ and $ \tilde{\varepsilon}=  2\hbar \dot{X}_0(t)/l_{\text{so}} +\varepsilon$.   It is clear that  the effective tunneling $\tilde{t}_g$  in the presence of  the spin-orbit interaction~\cite{stefano2021prx,stefano2021prl,dmytruk2018prb}  is suppressed
and the effective detuning $2\hbar \dot{X}_0(t)/l_{\text{so}}$ is proportional to the  externally tunable velocity $v\equiv \dot{X}_0(t)$, which emerges because of the broken spatial inversion symmetry caused by the motion of the DW.

We stress that   the velocity $v$ of the domain wall should be slow enough such that we can safely project the dynamics of the system onto the qubit space and the orbital ground state. This adiabatic condition explicitly implies: $v\ll\omega_p l_p, \omega_0 l_{\text{so}}$ (we estimate $\omega_p l_p\approx 4725\, \text{m/s}$ and $\omega_0 l_{\text{so}}\approx 3150\, \text{m/s}$). For this reason, we restrict the velocity to $v\lesssim 100\, \text{m/s}$. Besides, the large spin-orbit interaction emerging in our system is a tool to effectively manipulate the spin state. However, if this quantity becomes too large, our effective qubit theory fails because the variables $X, \Phi$  become strongly coupled and one cannot reliably separate their dynamics. This coupling becomes more pronounced in large domain walls, where the system behaves more classically.
 In order to use the low-energy sectors of  $\Phi$ as our qubit space and the motion of the domain wall  given by $X$ as a control knob, we require that the effective spin-orbit interaction is smaller than $\hbar\omega_p, \hbar\omega_0\approx 15K$. With the parameters we used, we estimate that the spin-orbit energy is $\hat{P}\hat{A}/M\sim S_e J\sqrt{\omega_p/\omega_0}\approx 3\,\text{K}$,
and it is therefore safely smaller than the level spacing $\hbar\omega_p, \hbar \omega_0$.

The Hamiltonian~\eqref{eq6} is $\mathcal{H}_m=\hbar \tilde{\Omega}\hat{\sigma}_z/2$ in the diagonal basis where we perform single- and two-qubit gates;  we label this frame by the Pauli matrices $\hat{\sigma}_i$. The  energy of the DW qubit  $\hbar \tilde{\Omega}=\sqrt{\tilde{\varepsilon}^2+\tilde{t}_g^2}$ is in the GHz range, as shown in Fig.~\ref{fig2}(c).
  We note that $\tilde{\Omega}$ first grows  as we increase $b_y$ since  the tunneling barrier is suppressed, whereas it decreases at larger values of $b_y$ because of the renormalization of the tunneling energy due to the large spin-orbit interaction.

\begin{figure}
	\centering\includegraphics[width=\linewidth]{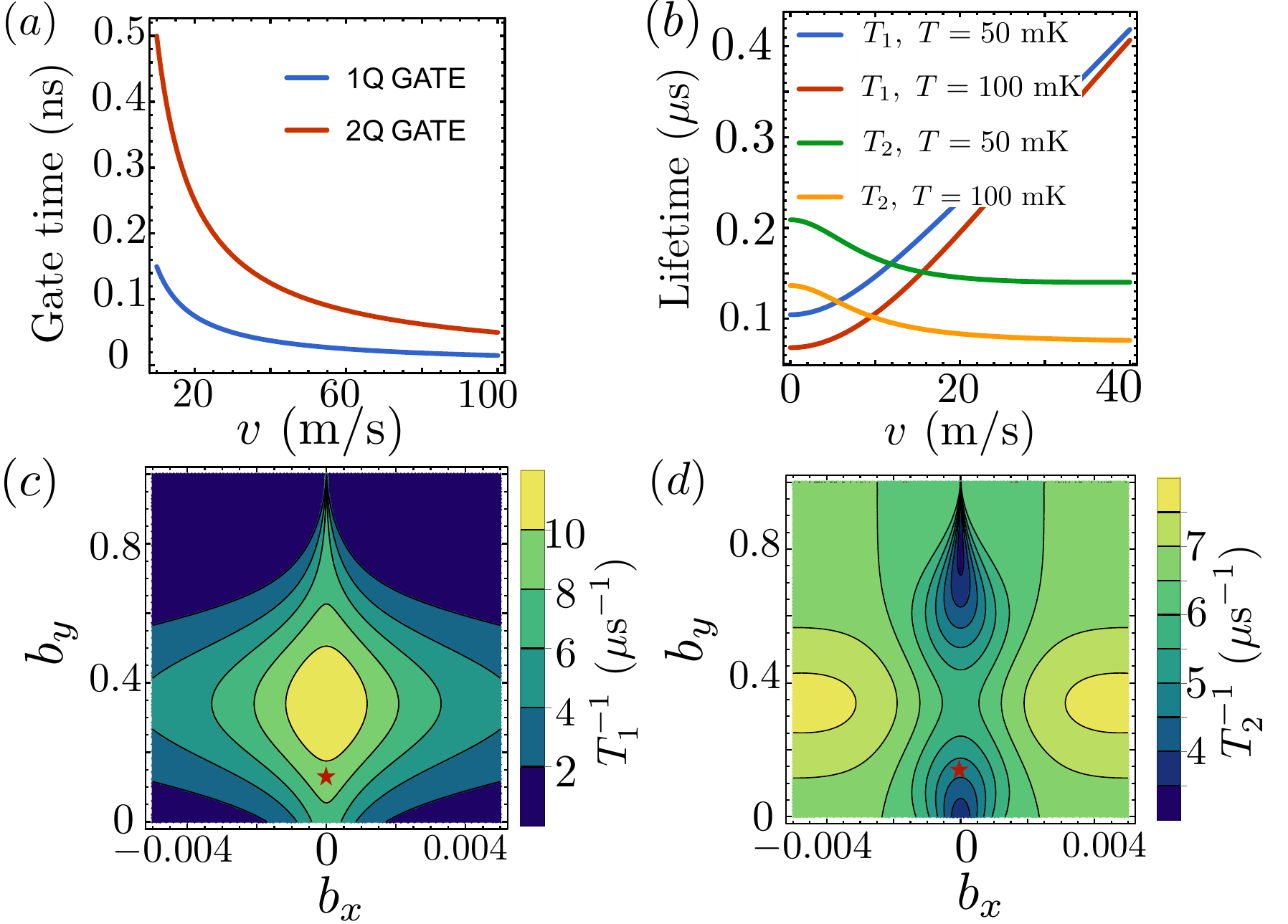}
 \caption{ DW qubit performance.  (a) Single- and two-qubit gate time as a function of the shuttling velocity $v$.  (b) Qubit relaxation $T_1$ and dephasing $T_2$ time as function of $v$ at different temperatures $T$.   (c) The relaxation rate $T_1^{-1}$ of a stationary qubit  as a function of dimensionless magnetic fields $b_x, b_y$.   (d) The dephasing rate $T_2^{-1}$ of a stationary qubit  as a function of dimensionless magnetic fields $b_x, b_y$.   The red star marks the working point of the qubit used in our estimations.}
  \label{fig3}
\end{figure}

\section{Single-qubit gates} 
The  effective spin-orbit interactions in our system enable fast qubit manipulation. Here, we propose to realize single qubit rotations by moving the DW on the racetrack~\cite{golovach2006prb}. When the qubit is shuttled across the racetrack with uniform velocity, $\tilde{\varepsilon}(t)$ is constant and $\mathcal{H}_m$ in Eq.~\eqref{eq6} yields the unitary time-evolution $U(t)=\text{exp}(-i\tilde{\Omega}\, t \, \hat{\vb m}\cdot \vb* \sigma/2)$ that describes a rotation of an angle $\tilde{\Omega}t$ around $\hat{\vb m}=(\sin\theta, 0, \cos\theta)$ [$\tan\theta=-\tilde{t}_g/\tilde{\varepsilon}$]. Therefore, single qubit rotations (around tunable axes in the $xz$ plane) are achieved by modulating the DW velocity or the applied field $B_y$.

Alternatively, one could perform gates by shaking the qubit by a time-dependent velocity profile $\dot{X}_0(t)=v_0 s(t) \cos(\omega_d t +\phi_0)$~\cite{golovach2006prb}, where $v_0$ is the velocity amplitude, $s(t)$ is a dimensionless envelope function, $\omega_d$ is the drive frequency, and $\phi_0$ is the initial phase~\cite{Krantz_2019_apr}. In the rotating frame, the Hamiltonian reads,
\(  {\mathcal{H}}_r=\frac{\hbar \, \delta\omega}{2}\hat{\sigma}_z +\frac{ \hbar \Omega_R s(t)}{2} ( \cos \phi_0 \hat{\sigma}_x+ \sin\phi_0 \hat{\sigma}_y),   \)
with the detuning frequency $\delta\omega=\tilde{\Omega}/\hbar-\omega_d$ and $\Omega_R= v_0/l_{\text{so}}$. At resonance $\hbar\omega_d=\tilde{\Omega}$, the in-phase pulse $\phi_0=0$ yields spin rotations around $x$ axis (and out-of-phase pulse yields rotations around $y$), 
\( \hat{R}_x^{\phi_0=0}=\exp{-i\Omega_R \int^t_0 d\tau \, s(\tau) \frac{\hat{\sigma}_x}{2} }, \)
resulting in typical Rabi oscillations with frequency  $\Omega_R$. The single-gate operational time $\Omega_R^{-1}$ is on the scale of 0.1~ns when $v_0\sim 20~\text{m/s}$, as shown in Fig.~\ref{fig3}(a).

\section{Two-qubit gates}
An entangling two-qubit gate, supplemented with single-qubit gates is sufficient for universal quantum computation~\cite{nielsen}. Here, we propose a way to implement controlled NOT gate with DW  qubits by taking advantage of spin-spin interactions of DWs moving in different racetracks.

We consider two DW qubits on two parallel racetracks or two tracks on top of each other
separated by a thin nonmagnetic spacer layer. Inter-track exchange  between the N{\'e}el  vectors in the two racetracks  yield the energy
\(\al{ H^{(2)}&=N^2 \int dx_1dx_2\; \mathcal{J}_{\alpha\beta} (x_1- x_2)  n_1^\alpha(x_1) n_2^\beta(x_2)\\
 & \approx N^2 \mathcal{J}_{\alpha\beta} \int dx \, n_1^{\alpha}(x) n_2^{\beta}(x) ,  } \)
where $\alpha, \beta$ are spin indices and $\vb n_1(x_1), \vb n_2(x_2)$ describe magnetic textures living on two racetracks with $x_1, x_2$ being the coordinates for these two tracks (we have again scaled $x\rightarrow \lambda x$). In the equation above, we also have assumed the interaction to be local, $\mathcal{J}_{\alpha\beta} (x_1- x_2)  \approx \mathcal{J}_{\alpha\beta} \delta(x_1- x_2) $. 
The interaction strength can be tuned experimentally by varying the distance between the racetracks or by modulating the spacer-layer thickness~\cite{parkin_1990_prl,parkin_1991_prl}. 
Assuming the two DWs are separated by a distance $D(t)$  along the direction ($x$) of the racetracks, we project the interaction onto the qubit subspace resulting in (Appendix~\ref{app_c}),
\(  \mathcal{H}^{(2)}= \frac{2N^2\mathcal{J}_{xx} D}{\sinh D} \hat{\sigma}^x_1\otimes \hat{\sigma}^x_2. \label{eq_inter} \)
 This interaction decreases  exponentially when the two DWs are far way ($D>1$, recall we measure the distance in units of $\lambda$), while when they approach each other, the two qubits become entangled by the two-qubit gate
\( \hat{\mathcal{U}}^{(2)}=\exp{-\frac{i\pi^2 N^2\mathcal{J}_{xx}}{\hbar v} \hat{\sigma}^x_1\otimes \hat{\sigma}^x_2}.  \)
Here we assume that one qubit travels to the other with constant velocity $\dot{D}=v$, and we note that $\hat{\mathcal{U}}^{(2)}$ generates a controlled NOT gate up to single-qubit rotations when $N^2\mathcal{J}_{xx}/\hbar v=1/ 4\pi$.  For DW interactions with $N^2 \mathcal{J}_{xx} \sim 50~\text{MHz}$ which is achievable in experiments, we require $v\sim 20~\text{m/s}$, corresponding to a two-gate operational time $\lambda/v\sim 0.2~\text{ns}$, as shown in Fig.~\ref{fig3}(a). A stronger interaction allows for a larger shuttling velocity thus a shorter gate time.

\section{Lifetime of domain wall qubits}
Decoherence of the DW qubit arises from couplings between spins and various environmental degrees of freedom, such as phonons and electrons.  Here we first derive the dissipative forces experienced by the two soft modes $X$ and $\Phi$, from which we can infer their fluctuations according to the fluctuation-dissipation theorem. We start with  the Rayleigh dissipation function $ R[\vb n] \propto \alpha N\hbar S \int dx\, \dot{\vb n}^2$, 
where, again,  we have scaled $x\rightarrow \lambda x$, $\alpha$ is the Gilbert damping, and $N\hbar S$ is the total angular momentum within the DW. This contribution translates into $ R[X,\Phi] \propto \alpha N \hbar S (\dot{X}^2+\dot{\Phi}^2)$,
leading to  friction forces acting on $X, \Phi$: $ F_\Phi=  -\partial_\Phi R=-\tilde{\alpha} \dot{\Phi}$ and $ F_X=  -\partial_X R=-\tilde{\alpha} X$ 
where $\Tilde{\alpha}=\alpha N\hbar S$. This implies random fluctuating forces acting on $X, \Phi$, that can be modelled by a fluctuating potential,
$\delta V=\xi(t)\Phi+ \xi(t) X. $
The stochastic forces are fully characterized by the classical ensemble average and their correlation function: 
$ \langle \xi(t)\rangle=0,  \langle \xi(t)\xi(t^\prime)\rangle =S(t-t^\prime),  $
where $S(t)$ is related to the dissipation parameter via the fluctuation dissipation theorem: 
$  S_\xi(\omega)=\Tilde{\alpha}\hbar \omega \coth (\hbar \omega/2k_BT),   $
with $S(t)=\int d\omega/2\pi\; S_\xi(\omega)e^{-i\omega t}$.  Let us first focus on the fluctuation in $\Phi$, which leads to two effects|fluctuation in the detuning and fluctuation in the tunneling barrier height $V_0$ between two chirality states: $ \delta H= \xi\sigma_z/2+ \partial_{V_0}t_g \xi \sigma_x/2$. This translates into 
$ \delta \mathcal{H}_\Phi= \xi(t) \sigma_z/2 + \partial_{V_0}\tilde{t}_g \xi(t) \sigma_x/2 , $
in the qubit shuttle Hamiltonian. 

For the spatial degree of freedom, the fluctuating potential gives rise to a fluctuation in the domain wall position: $X_0\rightarrow X_0+ \xi(t)/M\omega_p^2$, which leads to a fluctuating term in the Hamiltonian of the qubit shuttle: $\delta \mathcal{H}_X=\dot{\xi}(t)l_p^2/(\omega_pl_{\text{so}}) \sigma_z$. Combined with $\delta\mathcal{H}_\Phi$,  we obtain  the following noise Hamiltonian for the DW on a racetrack in the DW co-moving frame,
\( \delta \mathcal{H}_m= \frac{\xi(t)}{2}  {\sigma}_z +  \frac{\xi(t)}{2}   \frac{\partial\tilde{t}_g}{\partial V_0}   {\sigma}_x + \frac{l_p^2 \dot{\xi}(t)  }{\omega_p l_{\text{so}}}   {\sigma}_z, \label{eq_noise} \)
resulting in relaxation and  dephasing.  
The relaxation rate $\Gamma_1=T_1^{-1}$ is given by  $\Gamma_1=\sin^2\theta (\omega_p^2l^2_{\text{so}}+4l_p^2\tilde{\Omega}^2 )S_\xi(\tilde{\Omega}) /2(\hbar \omega_pl_{\text{so}})^2$ (Appendix~\ref{app_d}), which is $\sim 1~ \mu\text{s}^{-1}$ as shown in Fig.~\ref{fig3}(c). The dephasing rate is given by $\Gamma_2=T_2^{-1}=\Gamma_1/2+\Gamma_\varphi$ with the pure dephasing $\Gamma_\varphi=\cos^2\theta S_\xi(0) /2\hbar^2$, where $\tan\theta=-\tilde{t}_g/\tilde{\varepsilon}$. We point out that $b_x=0$ is the sweet spot for a stationary qubit as shown in Fig.~\ref{fig3}(d), where the pure dephasing is absent. This sweet spot is slightly shifted when the qubit shuttles with a finite velocity on the racetrack (Appendix~\ref{app_d}).

To estimate the coherence time, we  use   Gilbert damping  $\alpha=10^{-5}$, operational temperature $50~\text{mK}$, and DW velocity $v=20\, \text{m/s}$.  With these realistic parameters, we find that the DW qubit has a moderately long coherence time  with $T_1\approx 0.23~\mu\text{s}$ and $T_2\approx  0.15~\mu\text{s}$, and the quality factor, i.e. the number of coherent Rabi oscillations within the coherence time, is rather large $Q=vT_2/l_{\text{so}}\approx 2\times 10^3$. The dependence of the qubit coherence times on the DW velocity and for different operating temperatures is shown in Fig.~\ref{fig3}(b). The distance that the DW can travel before losing the coherence is $vT_2\approx 3\, \mu \text{m}$. We remark that in our estimation, we used the conservative value of $\alpha=10^{-5}$; in the milli-Kelvin regime where the DW qubit is operated, we expect $\alpha$ to be smaller~\cite{okada2017, marier2017}, resulting in longer coherence times.  As a result, these DWs are not only attractive alternative platforms to implement magnetic-based large-scale quantum computers, but  could  also be used as coherent quantum links  to distribute entanglement over long distances  between different types of qubits such as spin qubits~\cite{Daniel2013prx}  or nitrogen-vacancy (NV) center qubits~\cite{Fukami2021prx,Candido_2020,Burkard2019prb,zou2022prb}.

\section{Initialization and readout}
Reliable state preparation and readout are crucial for a complete proposal for scalable quantum computers. A possible initialization scheme to achieve DW with a well-defined chirality is obtained by cooling down the system sufficiently slowly and applying 
a magnetic field $B_x$  aligned to the racetrack, see Fig~\ref{fig2}(b). Arbitrary product states can be initialized in the system by sequentially applying single-qubit rotations to each qubit.  

We discuss two possible schemes for DW qubit readout. The first approach relies on the recent advances in nanoscale imaging techniques. 
NV centers have been utilized for imaging nanoscale DWs~\cite{Song_science_2021, Finco_Natcom_2021}, and they hold potential as non-invasive quantum sensors for assessing the chirality of DW qubits. Additionally, these projective measurements can be harnessed for preparing states with a definite chirality.
 An alternative strategy  to measure the chirality of the qubit is to adapt well-developed readout techniques for spin qubits. For conducting nanotracks, the chirality readout could be performed by a paramagnetic dot that is comparable or smaller than the DW: electrons near the center of a DW  can tunnel into the dot, whose polarization becomes linked to the DW chirality and can be measured by conventional methods~\cite{PhysRevA.57.120}. A 75\%-reliable measurement of the chirality can be  obtained in this approach (Appendix~\ref{app_e}).  For insulating nanotracks, the DW could be magnetically coupled to a spin qubit, whose state can be readout by various standard means~\cite{Elzerman:2004wt,Lai:2011vv,Pla:2013ta,PhysRevX.8.021046,Morello:2010vm,pnsci2010}.

\section{Conclusion}
We proposed a platform for scalable quantum computers based on mobile DWs on magnetic racetracks. The quantum information stored in the chirality of the DW textures can be efficiently manipulated and  transferred along the racetrack by shuttling the qubits. In state-of-the-art settings, the qubit response is fully tunable by varying an applied global  magnetic field and adjusting the DW velocity. 
Finally,  we remark that the proposed qubit operations by shuttling applies also to other magnetic qubits, e.g. based on skyrmions~\cite{christina_prl_2021}, opening up to new possibilities to integrate different classical spintronic components into the next generation of quantum processors.

\begin{acknowledgements}
This work was supported by the Georg H. Endress Foundation and by the Swiss National Science Foundation, NCCR SPIN (grant number 51NF40-180604).  S.S.P.P. and  B.P. were  supported by the Deutsche Forschungsgemeinschaft (DFG, German Research Foundation) – project no. 403505322, Priority Programme (SPP) 2137.
\end{acknowledgements}

\appendix
\section{Effective Action of Magnetic Domain Walls} \label{app_a}

We consider a quasi-one-dimensional two sublattice ferrimagnetic system with the following  Hamiltonian:
\( H\!=\! J\sum_{\langle i,j\rangle}\vb{S}_i\cdot \vb S_j -\Tilde{K}_z\sum_i (S^z_i)^2 +\Tilde{K}_y \sum_i(S^y_i)^2 -\hbar \sum_i \vb h\cdot \vb S_i, \)
where $J>0$ is the antiferromagnetic exchange coupling, and $\Tilde{K}_z, \Tilde{K}_y>0$  define the $z$ axis to be the easy axis and the $xz$ plane to be the easy plane, respectively. Here, $\vb h \equiv g\mu_B \vb B/\hbar$ is related to the magnetic field  $\textbf{B}$ and  $g$ is the electronic $g$ factor.  We  defined the spin operator without $\hbar$, so all parameters $J, \Tilde{K}_z, \Tilde{K}_y, \hbar |\vb h|$ have the dimension of energy. We denote the average and the excess spin per unit cell as $S$ and $S_e$, respectively. We assume  the magnetic racetrack is aligned along $x$ direction. With the state-of-the-art technology, the racetrack can be made to be atomically thin in the $z$ direction and the width of the track can be made to be around 10 nm in the $y$ direction. So low-energy spin dynamics is frozen in these two direction. We treat the system as a quasi-one-dimensional system and focus on the spin dynamics in the direction of the racetrack.

We find a continuum description of the low-energy dynamics of our system by closely following the procedure used to derive the effective description of a two sublattice antiferromagnet~\cite{qftcmt}.  The key differences are that we now have a weak ferromagnetism and a Berry phase contribution  because of the excess spin $S_e$. As only the low-energy dynamics is concerned, we assume the ferromagnetic order is a slaved degree of freedom, whose direction is locked with the N{\'e}el  vector $\vb n(x, t)$. We then obtain the following Lagrangian: 
\(  L[\vb n(x,t)] \! =\! N\! \!\! \int dx \, \Big[ \frac{\hbar^2}{8J} (\dot{\vb n} -\vb h \times \vb n )^2 + \hbar  S_e \dot{\vb n}\cdot \vb A \Big]  - U(\vb n), \label{s2}  \)
with the potential energy 
\( \al{ U[\vb n(x,t)] = & \frac{NK_z}{2} \int dx\, \big[ (\partial_x\vb n)^2-n_z^2  \big] \\
    & +N \int dx\,(K_y n_y^2-\hbar S_e \vb h\cdot \vb n). } \)
Here, we renormalized the microscopic anisotropies as $K_z=2\tilde{K}_z(S^2+S_e^2)$ and $K_y=\tilde{K}_y (S^2+S_e^2)$. 
We have also rescaled the spatial coordinate $x\rightarrow \lambda x$, such that $x$ above is dimensionless and we measure the spatial distance in the unit of the domain wall size $\lambda\equiv Sa \sqrt{J/K_z}$ ($a$ is the lattice constant). We  use this dimensionless coordinate $x$ throughout this appendix and also the main text.
 We also introduced the parameter $N\equiv \lambda N_A/a$, that corresponds to the total number of spins within a domain wall, with $N_A$ being the number of spins of the cross section ($yz$ plane) of the quasi-one dimensional system.
The first term in $L[\vb n]$ is the typical kinetic energy  of the N{\'e}el  vector, $\propto (\dot{\vb n} -\vb h \times \vb n )^2$, in the presence of a magnetic field, whereas the second term is the  spin Berry phase due to the net spin $S_e$, where the vector potential $\vb{A}(\vb n)$ is  defined by $\nabla_{\vb n}\times \vb A=-\vb n$. We assume $\vb n=(\sin\theta\cos\phi, \sin\theta\sin\phi, \cos\theta)$, and we use the gauge $\dot{\vb n}\cdot \vb A =\dot{\phi}(\cos\theta-1)$, where the Dirac string is aligned to the $-z$ axis. 

We now define the domain wall configuration. To this end, we assume that $J$ is the largest energy in the problem, and the easy $z$ axis anisotropy energy $K_z$ is the second largest, yielding the hierarchy of energies $J \gg K_z\gg K_y, \hbar S_e |\vb h|$. In our estimation, we take $J\sim 300~\text{K}, K_z\sim 3~\text{K}, K_y\sim 0.1~\text{K}$. 
  With this assumption, we can treat the second term of $U[\vb n]$ as a perturbation to the first term.  Let us assume the boundary condition to be $n_z(x=\pm \infty)=\pm 1$ for concreteness. We  minimize the first term in the potential energy, yielding~\cite{kim2022}
\(n_x+in_y=e^{i\Phi}\sech(x-X), \;\;\;   n_z=\tanh(x-X),   \label{s4}\)
where $X$ stands for the position of the domain wall in the racetrack and $\Phi$ stands for the azimuthal angle of the domain wall in  spin space. These coordinates are two zero modes, corresponding to the spontaneous symmetry breaking of the translation in real space and rotation in spin space.

We can now derive the effective action for the two collective coordinates $X, \Phi$~\cite{daniel2016prb,kim2022}. We plug  Eq.~\eqref{s4} into the Lagrangian~\eqref{s2}, and we obtain,
\begin{widetext}
\(\al{ &\int dx\, (\dot{\vb n} -\vb h\times \vb n)^2 = 2 (\dot{X}^2 +\dot{\Phi}^2) +2 \Big[ \pi (h_x\sin\Phi -h_y\cos \Phi)\dot{X} -2h_z \dot{\Phi}  \Big] -2(h_x\cos\Phi+h_y\sin\Phi)^2, 
\\  &\int dx\, \dot{\phi}(\cos\theta-1)=-2X\dot{\Phi}, \;\;\;\int dx\, n_y^2 = 2\sin^2\Phi ,\;\;\;\int dx\, \vb h\cdot \vb n =\pi (h_x \cos\Phi +h_y \sin\Phi) -2h_z X. }\)
Putting all terms together, we find the following effective action for $X, \Phi$,
\(\al{ L(X,\Phi)= & \frac{M}{2}\dot{\Phi}^2 -N\Big[ 2K_y\sin^2\Phi  +  \frac{1}{4J} \big( \hbar h_x \cos\Phi +\hbar h_y\sin\Phi\big)^2  -\pi \hbar S_e (h_x\cos\Phi +h_y\sin\Phi)   \Big]  \longrightarrow  \text{Lagrangian for}\; \Phi
\\ &+ \frac{M}{2}\dot{X}^2 - \frac{1}{2}M\omega_p^2X^2    \;\;\;   \;\;\; \;\;\;  \;\;\;   \;\;\;   \;\;\; \;\;\;  \;\;\;  \;\;\;   \;\;\; \;\;\;  \;\;\;   \;\;\; \;\;\;  \;\;\;  \;\;\;   \;\;\; \;\;\;  \;\;\;  \;\;\; \;\; \;\;\; \;\; \;\;\; \;\; \;\;\; \;\; \;\;\; \;\;\;\;\; \;\; \;\;\; \;\;\;  \longrightarrow   \text{Lagrangian for}\; X
\\ & + \frac{\pi M}{2} (h_x\sin\Phi-h_y\cos\Phi)\dot{X} -2N\hbar S_e X \dot{\Phi} ,     \;\;\;   \;\;\; \;\;\;  \;\;\; \;\;\; \;\; \;\;\; \;\; \;\;\; \;\;\;\;\; \;\; \;\;\; \;\; \;\;\; \;\; \;\;  \longrightarrow    \text{Couplings between}\; \Phi\; \text{and}\; X
     }  \label{s6}   \)
     \end{widetext}
where we set $h_z=0$ because it is not needed to define a domain wall qubit, and we added a confining potential $M\omega_p^2X^2$ for the domain wall along the racetrack. We take $\hbar\omega_p\sim 15K$ in our estimation and thus the associated harmonic characteristic length is $\sqrt{\hbar/M\omega_p}\sim 2.25~\text{nm}$ (with $J\sim 300~\text{K}, N\sim 100, \lambda\sim 5~\text{nm}$), where  
 the effective mass $M\equiv N \hbar^2/2J$  for $\Phi$ and $X$ is proportional to the domain wall size and inversely proportional to the stiffness $J$.  
The magnetic fields $h_x, h_y$ have two effects. First, they introduce effective anisotropies $\propto ( \hbar h_x \cos\Phi +\hbar h_y\sin\Phi )^2 $. Because these contributions are much smaller than other potential terms in the Lagrangian for $\Phi$, we will neglect them in our treatment. 
Second, the magnetic fields $h_x, h_y$ induce a finite magnetization that accumulates spin Berry phase, and yields an effective coupling $\propto (h_x\sin\Phi-h_y\cos\Phi)\dot{X}$  between the two coordinates. 
The excess spin $S_e$ also results in two effects. One is the potential energy $\propto (h_x\cos\Phi +h_y\sin\Phi) $ in the Lagrangian of $\Phi$, through the Zeeman coupling to applied magnetic fields. This contribution is crucial to define a domain wall qubit because it allows to engineer the potential energy of $\Phi$. Secondly, the net spin also accumulates spin Berry phase, leading to the coupling $X\dot{\Phi}$.  Finally, we remark that because the action is proportional to $N$, larger domain walls behave classically, while to observe quantum effect the domain wall needs to be small.

\section{Construction for the Orthonormal Basis  of the Computational Space of the Domain Wall Qubit} \label{app_b}
\begin{figure*}
	\centering\includegraphics[width=0.78\linewidth]{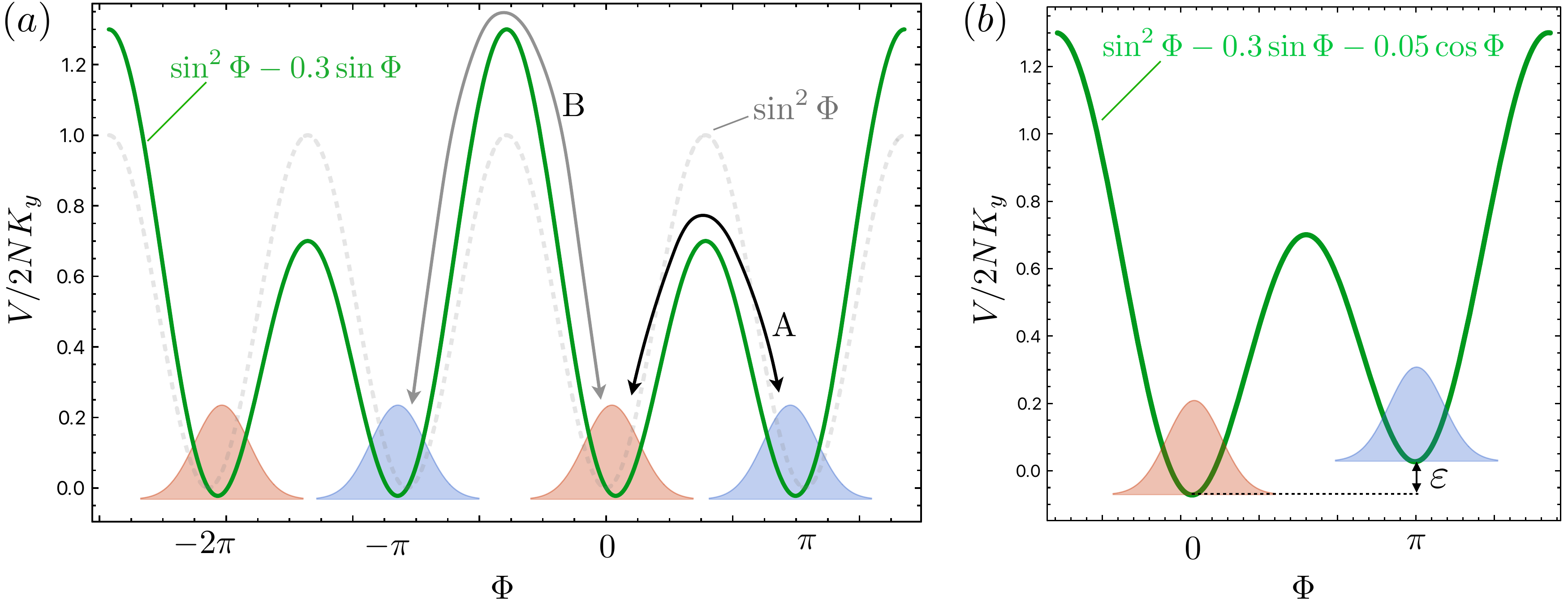}
 \caption{Construction of the computational space of DW qubits.  (a). The potential energy of the collective coordinate $\Phi$.  The dashed gray curve shows the case of zero  magnetic field $b_y=0$, see Eq.~\eqref{s7}.  The green curve corresponds to a finite magnetic field $b_y=0.15$. We take $b_x=0$ for both cases. Because of $b_y$, the tunneling barrier at $\Phi=\pi/2+2\pi \mathbb{Z}$ is suppressed, whereas the barrier at $-\pi/2+2\pi\mathbb{Z}$ is increased.  (b).  A detuning $\varepsilon$ due to a finite $b_x$, where we set $b_x=0.025$.  }
  \label{smf1}
\end{figure*}

To define the domain wall qubit, we focus on the angular degree of freedom with the following potential,
\( V(\Phi)=2NK_y(\sin^2\Phi-2b_x\cos \Phi - 2b_y\sin \Phi),  \label{s7} \)
where we introduced the dimensionless magnetic field $b_i\equiv \pi \hbar S_e h_i/4K_y$, which   is the ratio of the Zeeman energy to the easy plane anisotropy. We first consider the case $b_x=0$, corresponding to the symmetric double well potential $ V(\Phi)=2NK_y(\sin^2\Phi- 2b_y\sin \Phi)$, which has two minima determined by $\sin \Phi_{\pm}=b_y$.
We focus on the regime where the dimensionless magnetic field $b_y$ is finite but small. Thus the two minima of the potential lie at $\Phi_-\approx 2\pi \mathbb{Z}$ (corresponding to domain wall textures with positive chirality) and $\Phi_+\approx\pi+ 2\pi \mathbb{Z}$ (corresponding to domain wall textures with negative chirality), as shown in Fig.~\ref{smf1}(a) where we use $b_y=0.15$ (corresponding to $B_y\approx 1~\text{T}$). We note that $b_y$ suppresses the barrier at $\pi/2+2\pi \mathbb{Z}$ and increases the barrier  at $-\pi/2+2\pi\mathbb{Z}$. As a result, the tunneling process between two minima is dominated by the process A, shown in Fig.~\ref{smf1}(a). Thus we will focus on $\Phi\in[-\pi/2, 3\pi/2]$.  By standard calculation using the instanton technique, we conclude that  two chirality states would  hybridize yielding with a tunnel splitting $t_g\approx 4\hbar \omega_0\sqrt{S_{\text{inst}}/2\pi \hbar} \exp{-S_{\text{inst}}/\hbar}$, where $S_{\text{inst}} \approx 4V_0/ \omega_0$ is the instanton action, with the  tunnel barrier $V_0=2NK_y(1-b_y)^2$ and the level spacing $\hbar\omega_0=2 \sqrt{2JK_y(1-b_y^2)}$.  We remark that the two chirality states are localized at the two minima of the potential energy, which correspond to spin textures lying within the easy-$xz$-plane. On the other hand, higher energy states outside the  subspace have a significant probability of deviating from the easy-$xz$-plane, which is energetically unfavorable. As a result, they have higher energies compared to the two chirality states.

 As we discussed in the main text, we use the subspace spanned by $\{ \ket{\uparrow}\equiv \ket{\Phi_-},   \ket{\downarrow}\equiv \ket{\Phi_+}  \}$ as the computation space of our domain wall qubit. We now explicitly construct the wavefunctions of the basis states in the $\Phi$ representation. These wavefunctions allow us to obtain an effective qubit Hamiltonian by projecting the Lagrangian for $\Phi$  onto the qubit subspace. 
The natural choice for the wavefunctions of the two basis is given by two Gaussian eigenstates localized at the two minima: 
\( \bra{\Phi}\ket{\Tilde{\uparrow}}=\frac{\exp[-\frac{(\Phi-\Phi_-)^2}{2\Phi_0^2}  ]}{\pi^{1/4}\sqrt{\Phi_0}},\;\;\; \bra{\Phi}\ket{\Tilde{\downarrow}}=\frac{  \exp[-\frac{(\Phi-\Phi_+)^2}{2\Phi_0^2}]  }{\pi^{1/4}\sqrt{\Phi_0}},   \)
where $ \Phi_0= \sqrt{\hbar/M\omega_0}$ is the localization length of the wavefunction. However, the basis is not orthogonal to each other, with a small overlap given by,
\(p= \bra{\Tilde{\downarrow}}\ket{\Tilde{\uparrow}}=\exp{-\frac{(\Phi_+-\Phi_-)^2}{4\Phi_0^2}},  \)
which vanishes  when the two minima are  far away from each other compared with $\Phi_0$.  To construct an orthonormal basis for the chirality states, we first introduce the symmetric and antisymmetric states (with zero overlap with each other) and then normalize them: 
\( \ket{S}=\frac{ \ket{\Tilde{\uparrow}} +\ket{\Tilde{\downarrow}} }{\sqrt{2(1+p)}}, \;\;\; \ket{A}=\frac{ \ket{\Tilde{\uparrow}} -\ket{\Tilde{\downarrow}} }{\sqrt{2(1-p)}}. \)
We then transform back to the  chirality basis: 
\( \mqty(\ket{\uparrow} \\ \ket{\downarrow} ) =\frac{1}{\sqrt{2}}\mqty(1 & 1\\ 1 & -1) \mqty(\ket{S} \\ \ket{A})=P  \mqty(\ket{\Tilde{\uparrow}} \\ \ket{\Tilde{\downarrow}} ),\)
with
\( P= \Big( \frac{1}{\sqrt{1+p}} + \frac{1}{\sqrt{1-p}}  \Big) \frac{I}{2} + \Big(\frac{1}{\sqrt{1+p}} - \frac{1}{\sqrt{1-p}} \Big)\frac{\sigma_x}{2}.  \)
 Here the matrix $P$ connecting the old basis and the new basis is  symmetric and hermitian, and any operator $A$ in the orthonormal basis is related by $A=P\Tilde{A}P$ to the corresponding operator  $\Tilde{A}$ in the non-orthogonal basis.

We can now project the potential terms in $V(\Phi)$ onto the qubit space.   The $\cos\Phi$ and $\sin\Phi$ potentials translate into 
\begin{widetext}
\( \cos\Phi\longrightarrow \sqrt{\frac{1-b_y^2}{1-p^2}}e^{-\frac{\Phi_0^2}{4}}\sigma_z\equiv \gamma_z \sigma_z, \;\;\; \sin\Phi\longrightarrow \frac{b_y-p^2}{1-p^2} e^{-\frac{\Phi_0^2}{4}} I+ \frac{(1-b_y)p}{1-p^2} e^{- \frac{\Phi_0^2}{4}} \sigma_x \equiv \gamma_0I+\gamma_x \sigma_x,  \label{s12} \)
\end{widetext}
in the orthonormal basis. As an estimation of $\gamma_z, \gamma_0, \gamma_x$, we take the system parameters used in the main text: $\hbar\omega_0=15\, \text{K},\ J=300\, \text{K},\ N=100$, and $b_y=0.15$. We then obtain $\Phi_+-\Phi_-=2.85, \Phi_0=0.6, p=3.5\times 10^{-3}$, and  $\gamma_z=0.9, \gamma_0=0.14, \gamma_x=2.7\times 10^{-3}$. We note that $\gamma_z$ is close to $1$, whereas $\gamma_x$ is three orders of magnitude smaller than $\gamma_z$, as it is related to the tunneling process. In the main text, we thus use $\gamma_z=1$, for simplicity. For the Lagrangian of $\Phi$, 
\( L(\Phi)=\frac{M}{2}\dot{\Phi}^2 - 2NK_y (\sin^2\Phi -2b_y\sin\Phi)  + \pi \hbar N S_e h_x \cos\Phi,   \)
the first two terms yield in the qubit basis $-t_g\sigma_x/2$ ($t_g$ is the tunneling splitting obtained by instanton calculation) and the last term gives us $\varepsilon \sigma_z/2$ with $\varepsilon=-2 \pi \hbar N S_e h_x\gamma_z=-8NK_yb_x \gamma_z$ (by using what we obtained above). We thus derive the effective  Hamiltonian for a stationary domain wall qubit
\( H_{\text{s}} = \frac{\varepsilon}{2}\sigma_z- \frac{t_g}{2}\sigma_x,  \label{s14} \)
given in Eq.~(3) in the main text. We note that, here we have turned on  a finite $b_x$ which gives rise to a finite detuning energy for the two chiraility states, as shown in Fig.~\ref{smf1} (b) where we set $b_x=0.025$, corresponding to a magnetic field $B_x\approx 160 ~\text{mT}$. We require both $t_g$ and $\varepsilon$ to be much smaller than the level spacing $\hbar\omega_0$ and  tunneling barrier $V_0$ so that two chirality states are well-localized within two wells. In this case, the instanton calculation we performed is justified. We finally work out the projection of $\Phi$ onto the qubit space. It is given by ${\Phi}\rightarrow -\bar{\gamma_z}\sigma_z+\pi/2$ with $\bar{\gamma}_z=(\pi/2-\arcsin b_y)/\sqrt{1-p^2}$ (which can be approximated by $\pi/2$). This is useful in the derivation of the effective Hamiltonian for flying domain wall qubits in  the main text.

\begin{figure}
	\centering\includegraphics[width=\linewidth]{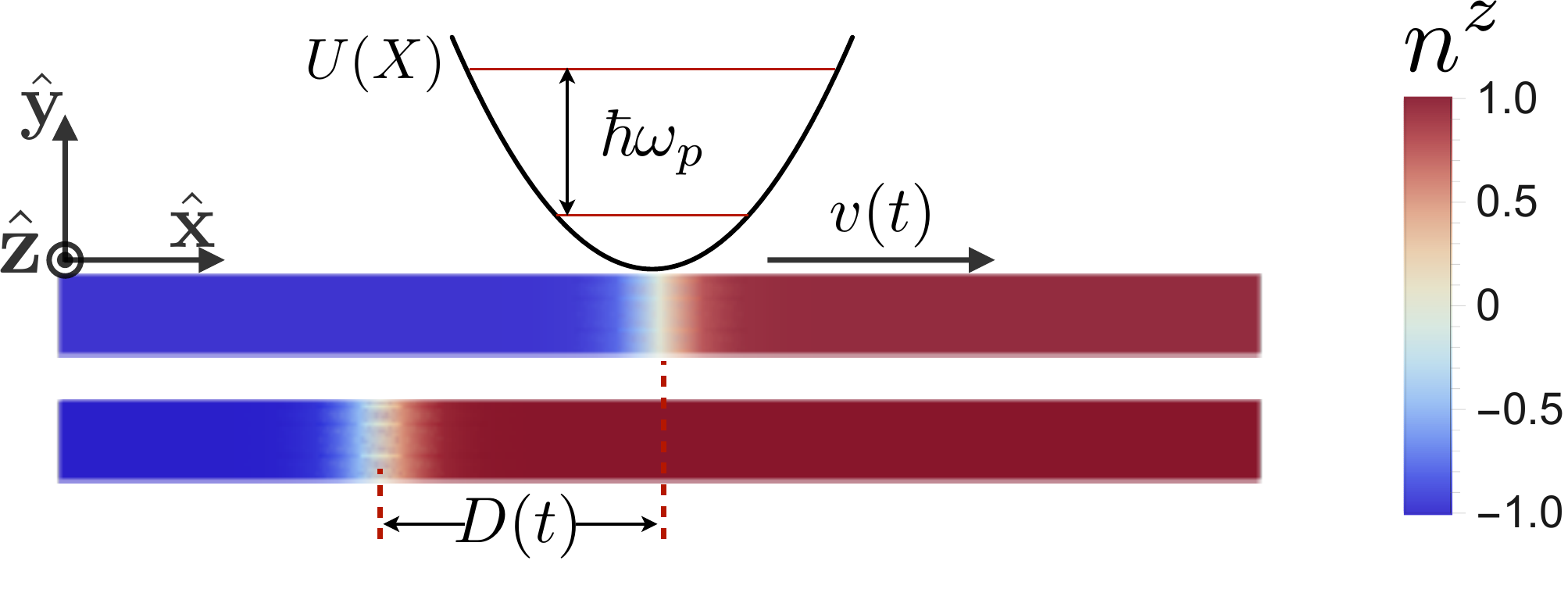}
 \caption{{Two-qubit interaction Hamiltonian.} A top view of two domain walls sitting on two parallel magnetic racetracks with a distance $D(t)$ from each other.  One domain wall is moving with a velocity $v(t)$. A two-qubit gate can be achieved as this domain wall passes by the other one.}
  \label{smf2}
\end{figure}

\section{ Two-qubit Interaction Hamiltonian}\label{app_c}
Here we derive the Eq.~\eqref{eq_inter} in the main text.  The interaction between two domain walls sitting on two tracks can be found by substituting $\vb n_1=\vb n_0(\Phi_1, X_1)$ and $\vb n_2=\vb n_0(\Phi_2, X_2=X_1+D)$, where $\vb n_0$ is the domain wall configuration that we obtained in Eq.~\eqref{s4} and $D$ is the distance between the two domain wall along the direction of the track, as depicted in Fig.~\ref{smf2}. We obtain the following interaction Hamiltonian in terms of collective coordinates:
 \begin{widetext}
\(\al{ H^{(2)}&=\frac{2N^2D}{\sinh D} \big( \mathcal{J}_{xx}\cos\Phi_1 \cos\Phi_2 +  \mathcal{J}_{yy}\sin\Phi_1 \sin\Phi_2  +\mathcal{J}_{xy}\cos\Phi_1 \sin\Phi_2  +\mathcal{J}_{yx}\sin\Phi_1 \cos\Phi_2  \big) \\
                     &+ \pi N^2\tanh\frac{D}{2}  \big( \mathcal{J}_{zy}\sin\Phi_2 -\mathcal{J}_{yz}\sin\Phi_1 +\mathcal{J}_{zx} \cos\Phi_2  - \mathcal{J}_{xz}\cos\Phi_1   \big).     }  \label{s30} \) 
                     \end{widetext}
 Let us first look at terms in the second line.   We project these terms onto qubit space and obtain: 
 \(\al{ H^{(2)}\supset  \pi^2 N^2 \tanh\frac{D}{2} \big(  \mathcal{J}_{zy} \gamma_x \sigma_2^x   - & \mathcal{J}_{yz} \gamma_x \sigma_1^x  +\mathcal{J}_{zx} \gamma_z \sigma_2^z \\ & - \mathcal{J}_{xz}\gamma_z \sigma_1^z     \big)    . }\)
We recall  that $\gamma_z\sim 1$ and $\gamma_x\sim 10^{-3}$ as we estimated before. Therefore, we would have terms $\propto \sigma^z_i$ which renormalize the qubit detuning $\tilde{\varepsilon}$. These terms are finite when two domain walls are far way from each other ($D>1$), which  just reflects that a qubit on one track can ``see" the $n^z$ of the other track as  these terms   originate from interactions $\propto n^z_1 n^x_2, n^z_2 n^x_1$.      We remark that these single-qubit terms  are absent when the interaction is isotropic $\mathcal{J}^{\alpha\beta}\propto \delta^{\alpha\beta}$. We may also make local interaction centers similar to a quantum point contact where qubits are made to interact (by making two tracks closer to each other at these centers). Then we can get rid of these single-qubit renormalization terms since a qubit on one track cannot ``see" the other when it is not at these centers. 
             
    We now project terms in the first line of Hamiltonian~\eqref{s30} onto the qubit space and obtain      
    \begin{widetext}   
\(  \tilde{\mathcal{H}}^{(2)}= \frac{2N^2D}{\sinh D} \big[ \mathcal{J}_{xx} \gamma_z^2 \sigma^z_1\otimes  \sigma^z_2 +  \mathcal{J}_{yy} (\gamma_0+\gamma_x\sigma_1^x)\otimes  (\gamma_0+\gamma_x\sigma_2^x)    + \mathcal{J}_{xy} \gamma_z  \sigma_1^z \otimes  (\gamma_0+\gamma_x\sigma_2^x)      +  \mathcal{J}_{yx} \gamma_z(\gamma_0+\gamma_x\sigma_1^x) \otimes  \sigma_2^z    \big].  \)
\end{widetext}
We note that this interaction decays exponentially as the qubit distance is beyond the domain wall size $D\gg 1$ (recall we measure the distance in unit of domain wall size $\lambda$).  We estimated before that $\gamma_z\sim 1$ and $\gamma_0\sim 0.1, \gamma_x\sim 10^{-3}$. Thus the first term in the interaction is much larger than the others and we will neglect these small corrections (and approximate $\gamma_z\approx 1$). In the main text, we assume $b_x=0$ and  thus the stationary Hamiltonian takes the form of $-t_g\sigma_x/2$. We switch to the diagonal basis (by $\sigma_x\rightarrow -\hat{\sigma}_z, \sigma_z \rightarrow \hat{\sigma}_x$) and the interaction Hamiltonian reads,
\( \mathcal{H}^{(2)}= \frac{2N^2\mathcal{J}_{xx} D}{\sinh D} \hat{\sigma}^x_1 \otimes  \hat{\sigma}^x_2,  \)
corresponding to Eq.~\eqref{eq_inter} in the main text.

\section{Detailed Discussion on Qubit Noise} \label{app_d}

The fluctuating terms in the domain wall qubit Hamiltonian are 
\(\al{ \delta \mathcal{H}_m & = \frac{\xi(t)}{2} \sigma_z+ \partial_{V_0}\tilde{t}_g \frac{\xi(t)}{2} \sigma_x +  \frac{l_p^2 \dot{\xi}(t) }{\omega_p l_{\text{so}}}   \sigma_z \\
& \equiv \zeta_z(t) \sigma_z + \zeta_x(t)\sigma_x, } \)
with $\zeta_z=\xi/2+l_p^2\dot{\xi}/\omega_pl_{\text{so}}  $ and $\zeta_x= \xi\partial_{V_0} \tilde{t}_g/2$. When the domain wall is moving with uniform velocity, we can diagonalize $\mathcal{H}_m =\tilde{\varepsilon}\sigma_z/2-\tilde{t}_g\sigma_x/2$ by a rotation in spin space: $\sigma_z\rightarrow \cos\theta \hat{\sigma}_z -\sin\theta \hat{\sigma}_x, \sigma_x \rightarrow \cos\theta \hat{\sigma}_x +\sin\theta \hat{\sigma}_z$, where $\sin\theta=-\tilde{t}_g/\hbar \tilde{\Omega}, \cos\theta=\tilde{\varepsilon}/\hbar \tilde{\Omega}$ with $\hbar \tilde{\Omega}=\sqrt{\tilde{t}^2_g+\tilde{\varepsilon}^2}$. Then the total Hamiltonian becomes: 
\( \mathcal{H}_m + \delta  \mathcal{H}_m \rightarrow \frac{\hbar \tilde{\Omega}}{2} \hat{\sigma}_z +\zeta_z^\theta (t) \hat{\sigma}_z + \zeta_x^\theta(t) \hat{\sigma}_x, \)
where $\zeta^\theta_z(t)=\zeta_z\cos\theta +\zeta_x\sin\theta$ and $\zeta^\theta_x(t)=-\zeta_z\sin\theta +\zeta_x \cos\theta$. Finally, we have
\( \Gamma_1=\frac{2}{\hbar^2} S_{\zeta^\theta_x}(\tilde{\Omega}),\;\;\Gamma_2=\frac{\Gamma_1}{2}+\Gamma_\varphi,  \)
with  $\Gamma_\varphi=  (2/\hbar^2)  S_{\zeta^\theta_z}(\omega=0)$.
We can write these results in a more explicit form: 
\( \al{ \Gamma_1&= \frac{ S_\xi(\tilde{\Omega})}{\hbar^2}  \Big[ \frac{\sin^2\theta}{2} \Big( 1+ \frac{4l_p^4 \tilde{\Omega}^2}{\omega_p^2l^2_{\text{so}}} \Big)  -\sin\theta\cos\theta \partial_{V_0} \tilde{t}_g \\ &  \;\;\; + \frac{\cos^2\theta}{2} (\partial_{V_0} \tilde{t}_g)^2 \Big],  
\\ \Gamma_\varphi&=\frac{S_\xi(0)}{\hbar^2}  \Big[ \frac{\cos^2\theta}{2} +\sin\theta\cos\theta \partial_{V_0} \tilde{t}_g   + \frac{\sin^2\theta}{2}(\partial_{V_0} \tilde{t}_g)^2 \Big].    } \)
Using the parameters of the main text, we  estimate the dimensionless number $\partial_{V_0} \tilde{t}_g$ in our system  to be $\partial_{V_0} \tilde{t}_g\approx0.02\ll 1$. We can thus set $\partial_{V_0} \tilde{t}_g=0$ in the expressions above resulting in 
  \( \Gamma_1= \frac{ \sin^2\theta S_\xi(\tilde{\Omega})}{2\hbar^2}   \Big( 1+ \frac{4l_p^4 \tilde{\Omega}^2}{\omega_p^2l^2_{\text{so}}} \Big),\,\;\;\;  \Gamma_\varphi= \frac{ \cos^2\theta S_\xi(0)}{2\hbar^2},    \)
which are the expressions we give in the main text [in the discussion below Eq.~\eqref{eq_noise}].

\begin{figure}
	\centering\includegraphics[width=\linewidth]{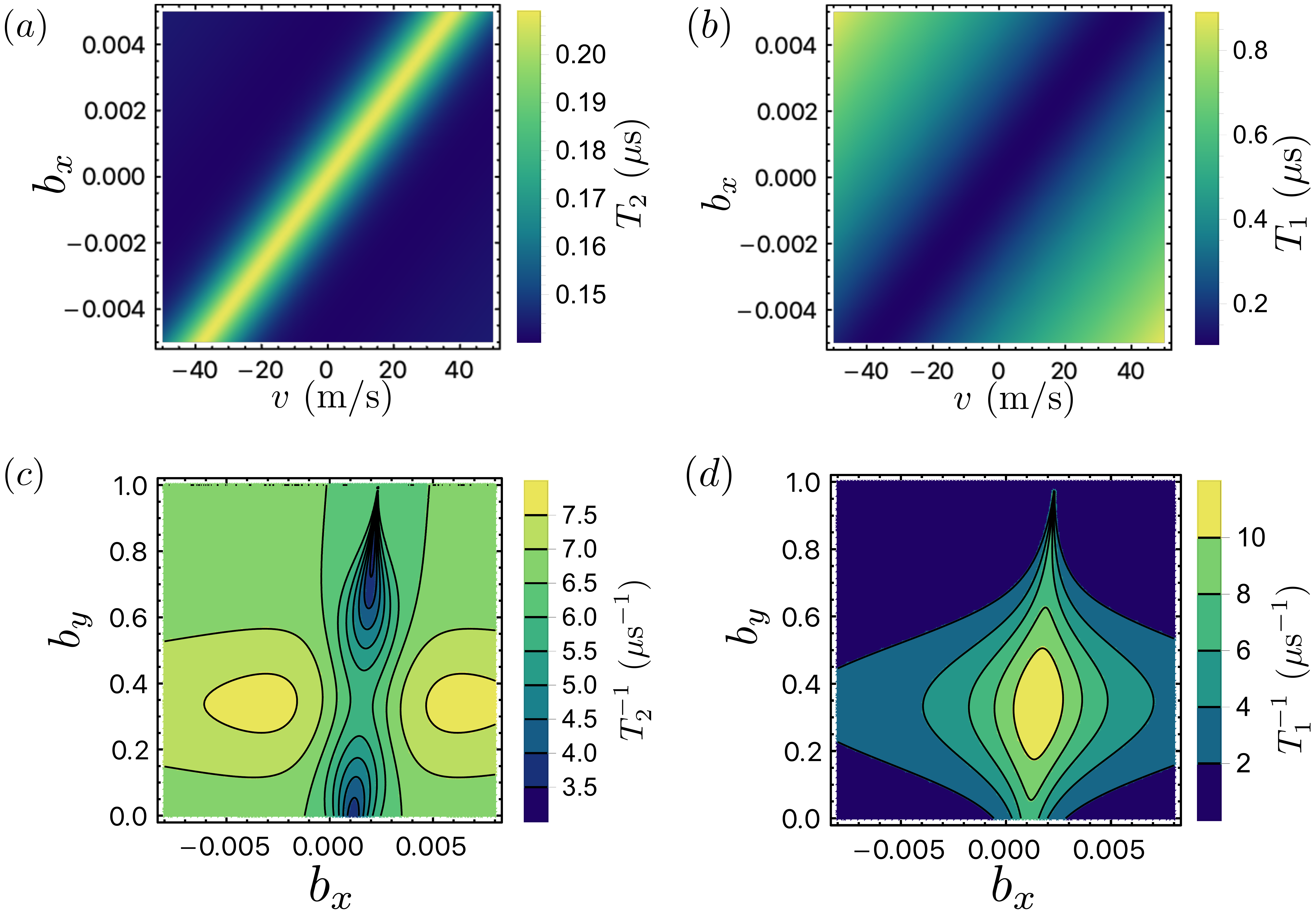}
 \caption{ \text{Coherence time of DW qubits.} (a) $T_2$ as a function of dimensionless magnetic field $b_x$ and shuttling velocity $v$. (b) $T_1$ as a function of $b_x$ and $v$.  (c) The dephasing rate $T_2^{-1}$ as a function of $b_x$ and $b_y$. (d) The relaxation rate $T_1^{-1}$ as a function of $b_x$ and $b_y$.     We fixed $b_y=0.15$ in (a) and (b).  In (c) and (d), we fixed the velocity to be $v=10~\text{m/s}$.  }
  \label{smf3}
\end{figure}

We point out that the sweet spot is at $b_x=0$ for a stationary DW qubit, since $\tilde{\varepsilon}=0$ (so $\cos\theta=0$)  thus the pure dephasing $\Gamma_\varphi$ vanishes in this case, as we discussed in the main text. When we shuttle the qubit with a finite velocity $v$, the sweet spot is  shifted to $b_x=\hbar v/l_{\text{so}} NK_y$, proportional to $v$. In this case, we have $\tilde{\varepsilon}=0$ and have zero pure dephasing.  As shown in Fig.~\ref{smf3}(a), the $T_2$ is maximal on the line given by $b_x=\hbar v/l_{\text{so}} NK_y$.     For example, the sweet spot is $b_x=1.2\times 10^{-3}$ (corresponding to $B_x=8~\text{mT}$) when $v=10~\text{m/s}$. Here we assumed $b_y=0.15$ (corresponding to $B_y=1~\text{T}$). The relaxation time $T_1$ is minimal at the sweet spot since $\sin\theta=1$ thus $\Gamma_1\propto \sin^2\theta$ reaches its maximum in this case, as shown in Fig.~\ref{smf3}(b). It should be clear that $T_2$ is the main limiting factor (with parameters we used for estimation). Therefore, we would like to work at the sweet spot to extend the qubit lifetime. It is also clear from Fig.~\ref{smf3}(c) and (d) that the sweet spot (the symmetry axis) is shifted in the presence of a finite shuttling velocity [compared to the Fig.~\ref{fig3}(c) and (d) in the main text where the velocity is zero].

\section{Qubit Readout with  Paramagnetic Dots} \label{app_e}
\begin{figure}[h]
	\centering\includegraphics[width=\linewidth]{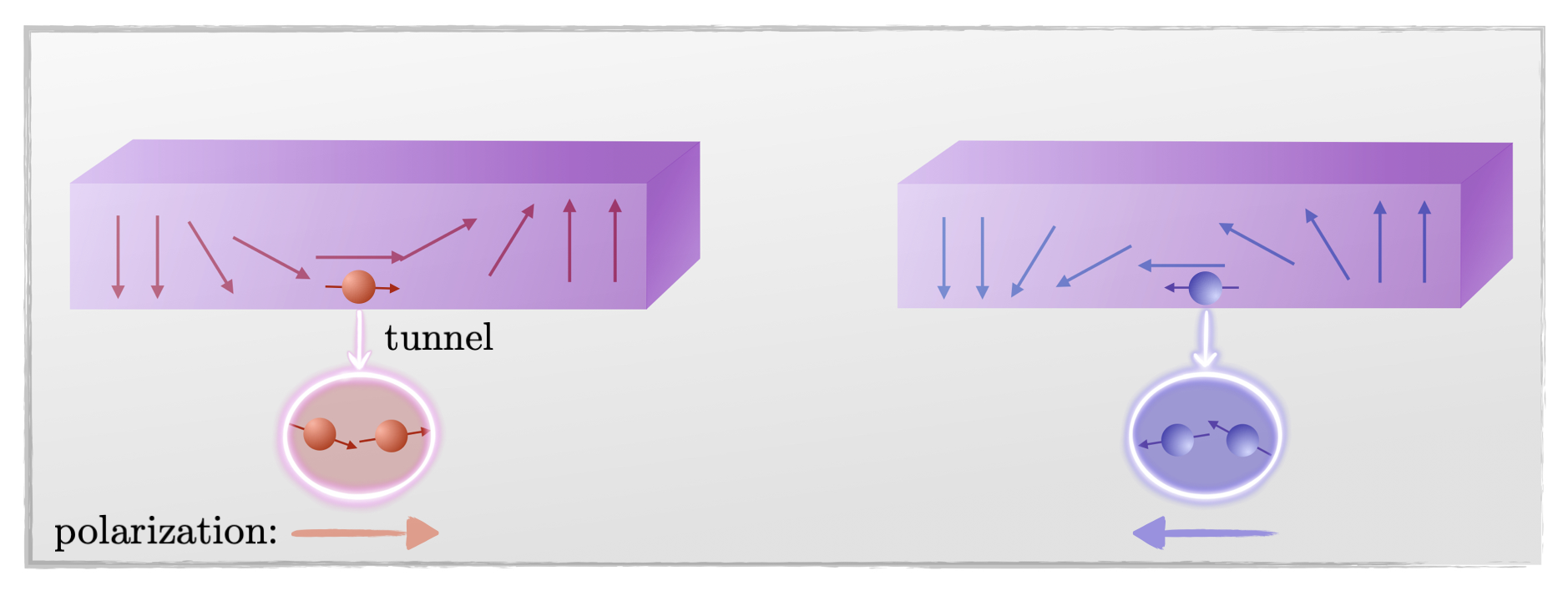}
 \caption{\text{Qubit readout via a paramagnetic dot}.  When the dot is comparable with the domain wall size, electrons with similar directions would tunnel to the dot, leading to the   formation of a ferromagnetic domain. When we interpret the magnetization direction of the domain in the right (left) hemisphere as positive (negative) chirality state, a 75\%-reliable measurement of the chirality is obtained.}
  \label{smf4}
\end{figure}

Here we provide some details about the qubit readout with a paramagnetic dot, as sketched in Fig.~\ref{smf4}. To read out the chirality state, we require that the size (the diameter) of the paramagnetic dot to be comparable or smaller than the domain wall size ($\sim 5\,\text{nm}$) which is feasible in experiments. Therefore, directions of the electrons that tunnel to the paramagnetic dot would be approximately along the same direction, as shown in Fig.~\ref{smf4}. We point out that this readout method does not require these directions to be perfectly parallel to each other.   
The tunneling event results in the formation of a ferromagnetic domain within the paramagnetic dot. The magnetization direction of this domain can be reliably determined using conventional measurement methodologies.
We can parameterize the magnetization direction with $(\theta, \phi)$, and the measurement outcomes would form a continuous set instead of two discrete values. This scenario is explained by the general
formalism of positive-operator-valued measurements. 

 If the magnetization direction  $(\theta, \phi)$  in the right
hemisphere is interpreted as positive chirality  state and in the left hemisphere
as negative chirality state, a 75\%-reliable measurement of the chirality is obtained. The argument is similar to the original proposal of quantum dot qubit. Assuming the paramagnetic dot is isotropic, the positive measurement operators would be projectors into the overcomplete set of spin-1/2 coherent states: $\ket{\theta,\phi}=\cos(\theta/2)\ket{\uparrow}+e^{i\phi}\sin(\theta/2)\ket{\downarrow}$. Then the reliability of the measurement is 
\(  \frac{1}{2\pi}\int_{U}d\Omega |\bra{\uparrow}\ket{\theta,\phi}  |^2 =\frac{3}{4}, \)
where $U$ denotes integration over the right hemisphere.

\begin{thebibliography}{85}%
\makeatletter
\providecommand \@ifxundefined [1]{%
 \@ifx{#1\undefined}
}%
\providecommand \@ifnum [1]{%
 \ifnum #1\expandafter \@firstoftwo
 \else \expandafter \@secondoftwo
 \fi
}%
\providecommand \@ifx [1]{%
 \ifx #1\expandafter \@firstoftwo
 \else \expandafter \@secondoftwo
 \fi
}%
\providecommand \natexlab [1]{#1}%
\providecommand \enquote  [1]{``#1''}%
\providecommand \bibnamefont  [1]{#1}%
\providecommand \bibfnamefont [1]{#1}%
\providecommand \citenamefont [1]{#1}%
\providecommand \href@noop [0]{\@secondoftwo}%
\providecommand \href [0]{\begingroup \@sanitize@url \@href}%
\providecommand \@href[1]{\@@startlink{#1}\@@href}%
\providecommand \@@href[1]{\endgroup#1\@@endlink}%
\providecommand \@sanitize@url [0]{\catcode `\\12\catcode `\$12\catcode
  `\&12\catcode `\#12\catcode `\^12\catcode `\_12\catcode `\%12\relax}%
\providecommand \@@startlink[1]{}%
\providecommand \@@endlink[0]{}%
\providecommand \url  [0]{\begingroup\@sanitize@url \@url }%
\providecommand \@url [1]{\endgroup\@href {#1}{\urlprefix }}%
\providecommand \urlprefix  [0]{URL }%
\providecommand \Eprint [0]{\href }%
\providecommand \doibase [0]{https://doi.org/}%
\providecommand \selectlanguage [0]{\@gobble}%
\providecommand \bibinfo  [0]{\@secondoftwo}%
\providecommand \bibfield  [0]{\@secondoftwo}%
\providecommand \translation [1]{[#1]}%
\providecommand \BibitemOpen [0]{}%
\providecommand \bibitemStop [0]{}%
\providecommand \bibitemNoStop [0]{.\EOS\space}%
\providecommand \EOS [0]{\spacefactor3000\relax}%
\providecommand \BibitemShut  [1]{\csname bibitem#1\endcsname}%
\let\auto@bib@innerbib\@empty
\bibitem [{\citenamefont {Ryu}\ \emph {et~al.}(2013)\citenamefont {Ryu},
  \citenamefont {Thomas}, \citenamefont {Yang},\ and\ \citenamefont
  {Parkin}}]{Ryu:2013wh}%
  \BibitemOpen
  \bibfield  {author} {\bibinfo {author} {\bibfnamefont {K.-S.}\ \bibnamefont
  {Ryu}}, \bibinfo {author} {\bibfnamefont {L.}~\bibnamefont {Thomas}},
  \bibinfo {author} {\bibfnamefont {S.-H.}\ \bibnamefont {Yang}},\ and\
  \bibinfo {author} {\bibfnamefont {S.}~\bibnamefont {Parkin}},\ }\bibfield
  {title} {\bibinfo {title} {Chiral spin torque at magnetic domain walls},\
  }\href@noop {} {\bibfield  {journal} {\bibinfo  {journal} {Nature
  Nanotechnology}\ }\textbf {\bibinfo {volume} {8}},\ \bibinfo {pages} {527}
  (\bibinfo {year} {2013})}\BibitemShut {NoStop}%
\bibitem [{\citenamefont {Yang}\ \emph {et~al.}(2015)\citenamefont {Yang},
  \citenamefont {Ryu},\ and\ \citenamefont {Parkin}}]{Yang2015uw}%
  \BibitemOpen
  \bibfield  {author} {\bibinfo {author} {\bibfnamefont {S.-H.}\ \bibnamefont
  {Yang}}, \bibinfo {author} {\bibfnamefont {K.-S.}\ \bibnamefont {Ryu}},\ and\
  \bibinfo {author} {\bibfnamefont {S.}~\bibnamefont {Parkin}},\ }\bibfield
  {title} {\bibinfo {title} {Domain-wall velocities of up to 750 m/s driven by
  exchange-coupling torque in synthetic antiferromagnets},\ }\href@noop {}
  {\bibfield  {journal} {\bibinfo  {journal} {Nature Nanotechnology}\ }\textbf
  {\bibinfo {volume} {10}},\ \bibinfo {pages} {221} (\bibinfo {year}
  {2015})}\BibitemShut {NoStop}%
\bibitem [{\citenamefont {Kim}\ \emph {et~al.}(2017)\citenamefont {Kim},
  \citenamefont {Kim}, \citenamefont {Hirata}, \citenamefont {Oh},
  \citenamefont {Tono}, \citenamefont {Kim}, \citenamefont {Okuno},
  \citenamefont {Ham}, \citenamefont {Kim}, \citenamefont {Go}, \citenamefont
  {Tserkovnyak}, \citenamefont {Tsukamoto}, \citenamefont {Moriyama},
  \citenamefont {Lee},\ and\ \citenamefont {Ono}}]{Kim2017natmat}%
  \BibitemOpen
  \bibfield  {author} {\bibinfo {author} {\bibfnamefont {K.-J.}\ \bibnamefont
  {Kim}}, \bibinfo {author} {\bibfnamefont {S.~K.}\ \bibnamefont {Kim}},
  \bibinfo {author} {\bibfnamefont {Y.}~\bibnamefont {Hirata}}, \bibinfo
  {author} {\bibfnamefont {S.-H.}\ \bibnamefont {Oh}}, \bibinfo {author}
  {\bibfnamefont {T.}~\bibnamefont {Tono}}, \bibinfo {author} {\bibfnamefont
  {D.-H.}\ \bibnamefont {Kim}}, \bibinfo {author} {\bibfnamefont
  {T.}~\bibnamefont {Okuno}}, \bibinfo {author} {\bibfnamefont {W.~S.}\
  \bibnamefont {Ham}}, \bibinfo {author} {\bibfnamefont {S.}~\bibnamefont
  {Kim}}, \bibinfo {author} {\bibfnamefont {G.}~\bibnamefont {Go}}, \bibinfo
  {author} {\bibfnamefont {Y.}~\bibnamefont {Tserkovnyak}}, \bibinfo {author}
  {\bibfnamefont {A.}~\bibnamefont {Tsukamoto}}, \bibinfo {author}
  {\bibfnamefont {T.}~\bibnamefont {Moriyama}}, \bibinfo {author}
  {\bibfnamefont {K.-J.}\ \bibnamefont {Lee}},\ and\ \bibinfo {author}
  {\bibfnamefont {T.}~\bibnamefont {Ono}},\ }\bibfield  {title} {\bibinfo
  {title} {Fast domain wall motion in the vicinity of the angular momentum
  compensation temperature of ferrimagnets},\ }\href@noop {} {\bibfield
  {journal} {\bibinfo  {journal} {Nature Materials}\ }\textbf {\bibinfo
  {volume} {16}},\ \bibinfo {pages} {1187} (\bibinfo {year}
  {2017})}\BibitemShut {NoStop}%
\bibitem [{\citenamefont {Yang}\ \emph {et~al.}(2021)\citenamefont {Yang},
  \citenamefont {Naaman}, \citenamefont {Paltiel},\ and\ \citenamefont
  {Parkin}}]{Yang:2021wy}%
  \BibitemOpen
  \bibfield  {author} {\bibinfo {author} {\bibfnamefont {S.-H.}\ \bibnamefont
  {Yang}}, \bibinfo {author} {\bibfnamefont {R.}~\bibnamefont {Naaman}},
  \bibinfo {author} {\bibfnamefont {Y.}~\bibnamefont {Paltiel}},\ and\ \bibinfo
  {author} {\bibfnamefont {S.~S.~P.}\ \bibnamefont {Parkin}},\ }\bibfield
  {title} {\bibinfo {title} {Chiral spintronics},\ }\href@noop {} {\bibfield
  {journal} {\bibinfo  {journal} {Nature Reviews Physics}\ }\textbf {\bibinfo
  {volume} {3}},\ \bibinfo {pages} {328} (\bibinfo {year} {2021})}\BibitemShut
  {NoStop}%
\bibitem [{\citenamefont {Guan}\ \emph {et~al.}(2021)\citenamefont {Guan},
  \citenamefont {Zhou}, \citenamefont {Li}, \citenamefont {Ma}, \citenamefont
  {Yang},\ and\ \citenamefont {Parkin}}]{Guan:2021vo}%
  \BibitemOpen
  \bibfield  {author} {\bibinfo {author} {\bibfnamefont {Y.}~\bibnamefont
  {Guan}}, \bibinfo {author} {\bibfnamefont {X.}~\bibnamefont {Zhou}}, \bibinfo
  {author} {\bibfnamefont {F.}~\bibnamefont {Li}}, \bibinfo {author}
  {\bibfnamefont {T.}~\bibnamefont {Ma}}, \bibinfo {author} {\bibfnamefont
  {S.-H.}\ \bibnamefont {Yang}},\ and\ \bibinfo {author} {\bibfnamefont
  {S.~S.~P.}\ \bibnamefont {Parkin}},\ }\bibfield  {title} {\bibinfo {title}
  {Ionitronic manipulation of current-induced domain wall motion in synthetic
  antiferromagnets},\ }\href@noop {} {\bibfield  {journal} {\bibinfo  {journal}
  {Nature Communications}\ }\textbf {\bibinfo {volume} {12}},\ \bibinfo {pages}
  {5002} (\bibinfo {year} {2021})}\BibitemShut {NoStop}%
\bibitem [{\citenamefont {Bl{\"a}sing}\ \emph {et~al.}(2018)\citenamefont
  {Bl{\"a}sing}, \citenamefont {Ma}, \citenamefont {Yang}, \citenamefont
  {Garg}, \citenamefont {Dejene}, \citenamefont {N'Diaye}, \citenamefont
  {Chen}, \citenamefont {Liu},\ and\ \citenamefont {Parkin}}]{Blasing:2018vb}%
  \BibitemOpen
  \bibfield  {author} {\bibinfo {author} {\bibfnamefont {R.}~\bibnamefont
  {Bl{\"a}sing}}, \bibinfo {author} {\bibfnamefont {T.}~\bibnamefont {Ma}},
  \bibinfo {author} {\bibfnamefont {S.-H.}\ \bibnamefont {Yang}}, \bibinfo
  {author} {\bibfnamefont {C.}~\bibnamefont {Garg}}, \bibinfo {author}
  {\bibfnamefont {F.~K.}\ \bibnamefont {Dejene}}, \bibinfo {author}
  {\bibfnamefont {A.~T.}\ \bibnamefont {N'Diaye}}, \bibinfo {author}
  {\bibfnamefont {G.}~\bibnamefont {Chen}}, \bibinfo {author} {\bibfnamefont
  {K.}~\bibnamefont {Liu}},\ and\ \bibinfo {author} {\bibfnamefont {S.~S.~P.}\
  \bibnamefont {Parkin}},\ }\bibfield  {title} {\bibinfo {title} {Exchange
  coupling torque in ferrimagnetic co/gd bilayer maximized near angular
  momentum compensation temperature},\ }\href@noop {} {\bibfield  {journal}
  {\bibinfo  {journal} {Nature Communications}\ }\textbf {\bibinfo {volume}
  {9}},\ \bibinfo {pages} {4984} (\bibinfo {year} {2018})}\BibitemShut
  {NoStop}%
\bibitem [{\citenamefont {Yoshimura}\ \emph {et~al.}(2016)\citenamefont
  {Yoshimura}, \citenamefont {Kim}, \citenamefont {Taniguchi}, \citenamefont
  {Tono}, \citenamefont {Ueda}, \citenamefont {Hiramatsu}, \citenamefont
  {Moriyama}, \citenamefont {Yamada}, \citenamefont {Nakatani},\ and\
  \citenamefont {Ono}}]{Yoshimura:2016uk}%
  \BibitemOpen
  \bibfield  {author} {\bibinfo {author} {\bibfnamefont {Y.}~\bibnamefont
  {Yoshimura}}, \bibinfo {author} {\bibfnamefont {K.-J.}\ \bibnamefont {Kim}},
  \bibinfo {author} {\bibfnamefont {T.}~\bibnamefont {Taniguchi}}, \bibinfo
  {author} {\bibfnamefont {T.}~\bibnamefont {Tono}}, \bibinfo {author}
  {\bibfnamefont {K.}~\bibnamefont {Ueda}}, \bibinfo {author} {\bibfnamefont
  {R.}~\bibnamefont {Hiramatsu}}, \bibinfo {author} {\bibfnamefont
  {T.}~\bibnamefont {Moriyama}}, \bibinfo {author} {\bibfnamefont
  {K.}~\bibnamefont {Yamada}}, \bibinfo {author} {\bibfnamefont
  {Y.}~\bibnamefont {Nakatani}},\ and\ \bibinfo {author} {\bibfnamefont
  {T.}~\bibnamefont {Ono}},\ }\bibfield  {title} {\bibinfo {title}
  {Soliton-like magnetic domain wall motion induced by the interfacial
  dzyaloshinskii--moriya interaction},\ }\href@noop {} {\bibfield  {journal}
  {\bibinfo  {journal} {Nature Physics}\ }\textbf {\bibinfo {volume} {12}},\
  \bibinfo {pages} {157} (\bibinfo {year} {2016})}\BibitemShut {NoStop}%
\bibitem [{\citenamefont {Jiang}\ \emph {et~al.}(2015)\citenamefont {Jiang},
  \citenamefont {Upadhyaya}, \citenamefont {Zhang}, \citenamefont {Yu},
  \citenamefont {Jungfleisch}, \citenamefont {Fradin}, \citenamefont {Pearson},
  \citenamefont {Tserkovnyak}, \citenamefont {Wang}, \citenamefont {Heinonen},
  \citenamefont {te~Velthuis},\ and\ \citenamefont {Hoffmann}}]{Jiang283}%
  \BibitemOpen
  \bibfield  {author} {\bibinfo {author} {\bibfnamefont {W.}~\bibnamefont
  {Jiang}}, \bibinfo {author} {\bibfnamefont {P.}~\bibnamefont {Upadhyaya}},
  \bibinfo {author} {\bibfnamefont {W.}~\bibnamefont {Zhang}}, \bibinfo
  {author} {\bibfnamefont {G.}~\bibnamefont {Yu}}, \bibinfo {author}
  {\bibfnamefont {M.~B.}\ \bibnamefont {Jungfleisch}}, \bibinfo {author}
  {\bibfnamefont {F.~Y.}\ \bibnamefont {Fradin}}, \bibinfo {author}
  {\bibfnamefont {J.~E.}\ \bibnamefont {Pearson}}, \bibinfo {author}
  {\bibfnamefont {Y.}~\bibnamefont {Tserkovnyak}}, \bibinfo {author}
  {\bibfnamefont {K.~L.}\ \bibnamefont {Wang}}, \bibinfo {author}
  {\bibfnamefont {O.}~\bibnamefont {Heinonen}}, \bibinfo {author}
  {\bibfnamefont {S.~G.~E.}\ \bibnamefont {te~Velthuis}},\ and\ \bibinfo
  {author} {\bibfnamefont {A.}~\bibnamefont {Hoffmann}},\ }\bibfield  {title}
  {\bibinfo {title} {Blowing magnetic skyrmion bubbles},\ }\href@noop {}
  {\bibfield  {journal} {\bibinfo  {journal} {Science}\ }\textbf {\bibinfo
  {volume} {349}},\ \bibinfo {pages} {283} (\bibinfo {year}
  {2015})}\BibitemShut {NoStop}%
\bibitem [{\citenamefont {Guo}\ \emph {et~al.}(2022)\citenamefont {Guo},
  \citenamefont {Zhang},\ and\ \citenamefont {Cheng}}]{hantao2022prb}%
  \BibitemOpen
  \bibfield  {author} {\bibinfo {author} {\bibfnamefont {M.}~\bibnamefont
  {Guo}}, \bibinfo {author} {\bibfnamefont {H.}~\bibnamefont {Zhang}},\ and\
  \bibinfo {author} {\bibfnamefont {R.}~\bibnamefont {Cheng}},\ }\bibfield
  {title} {\bibinfo {title} {Manipulating ferrimagnets by fields and
  currents},\ }\href@noop {} {\bibfield  {journal} {\bibinfo  {journal} {Phys.
  Rev. B}\ }\textbf {\bibinfo {volume} {105}},\ \bibinfo {pages} {064410}
  (\bibinfo {year} {2022})}\BibitemShut {NoStop}%
\bibitem [{\citenamefont {Jin}\ \emph {et~al.}(2021)\citenamefont {Jin},
  \citenamefont {Hong}, \citenamefont {Kim}, \citenamefont {Lee},\ and\
  \citenamefont {Kim}}]{sekwon2021prb}%
  \BibitemOpen
  \bibfield  {author} {\bibinfo {author} {\bibfnamefont {M.}~\bibnamefont
  {Jin}}, \bibinfo {author} {\bibfnamefont {I.-S.}\ \bibnamefont {Hong}},
  \bibinfo {author} {\bibfnamefont {D.-H.}\ \bibnamefont {Kim}}, \bibinfo
  {author} {\bibfnamefont {K.-J.}\ \bibnamefont {Lee}},\ and\ \bibinfo {author}
  {\bibfnamefont {S.~K.}\ \bibnamefont {Kim}},\ }\bibfield  {title} {\bibinfo
  {title} {Domain-wall motion driven by a rotating field in a ferrimagnet},\
  }\href@noop {} {\bibfield  {journal} {\bibinfo  {journal} {Phys. Rev. B}\
  }\textbf {\bibinfo {volume} {104}},\ \bibinfo {pages} {184431} (\bibinfo
  {year} {2021})}\BibitemShut {NoStop}%
\bibitem [{\citenamefont {Donges}\ \emph {et~al.}(2020)\citenamefont {Donges},
  \citenamefont {Grimm}, \citenamefont {Jakobs}, \citenamefont {Selzer},
  \citenamefont {Ritzmann}, \citenamefont {Atxitia},\ and\ \citenamefont
  {Nowak}}]{donges2022prr}%
  \BibitemOpen
  \bibfield  {author} {\bibinfo {author} {\bibfnamefont {A.}~\bibnamefont
  {Donges}}, \bibinfo {author} {\bibfnamefont {N.}~\bibnamefont {Grimm}},
  \bibinfo {author} {\bibfnamefont {F.}~\bibnamefont {Jakobs}}, \bibinfo
  {author} {\bibfnamefont {S.}~\bibnamefont {Selzer}}, \bibinfo {author}
  {\bibfnamefont {U.}~\bibnamefont {Ritzmann}}, \bibinfo {author}
  {\bibfnamefont {U.}~\bibnamefont {Atxitia}},\ and\ \bibinfo {author}
  {\bibfnamefont {U.}~\bibnamefont {Nowak}},\ }\bibfield  {title} {\bibinfo
  {title} {Unveiling domain wall dynamics of ferrimagnets in thermal magnon
  currents: Competition of angular momentum transfer and entropic torque},\
  }\href@noop {} {\bibfield  {journal} {\bibinfo  {journal} {Phys. Rev.
  Research}\ }\textbf {\bibinfo {volume} {2}},\ \bibinfo {pages} {013293}
  (\bibinfo {year} {2020})}\BibitemShut {NoStop}%
\bibitem [{\citenamefont {\ifmmode~\check{Z}\else \v{Z}\fi{}elezn\'y}\ \emph
  {et~al.}(2014)\citenamefont {\ifmmode~\check{Z}\else \v{Z}\fi{}elezn\'y},
  \citenamefont {Gao}, \citenamefont {V\'yborn\'y}, \citenamefont {Zemen},
  \citenamefont {Ma\ifmmode~\check{s}\else \v{s}\fi{}ek}, \citenamefont
  {Manchon}, \citenamefont {Wunderlich}, \citenamefont {Sinova},\ and\
  \citenamefont {Jungwirth}}]{Sinova2014prl}%
  \BibitemOpen
  \bibfield  {author} {\bibinfo {author} {\bibfnamefont {J.}~\bibnamefont
  {\ifmmode~\check{Z}\else \v{Z}\fi{}elezn\'y}}, \bibinfo {author}
  {\bibfnamefont {H.}~\bibnamefont {Gao}}, \bibinfo {author} {\bibfnamefont
  {K.}~\bibnamefont {V\'yborn\'y}}, \bibinfo {author} {\bibfnamefont
  {J.}~\bibnamefont {Zemen}}, \bibinfo {author} {\bibfnamefont
  {J.}~\bibnamefont {Ma\ifmmode~\check{s}\else \v{s}\fi{}ek}}, \bibinfo
  {author} {\bibfnamefont {A.}~\bibnamefont {Manchon}}, \bibinfo {author}
  {\bibfnamefont {J.}~\bibnamefont {Wunderlich}}, \bibinfo {author}
  {\bibfnamefont {J.}~\bibnamefont {Sinova}},\ and\ \bibinfo {author}
  {\bibfnamefont {T.}~\bibnamefont {Jungwirth}},\ }\bibfield  {title} {\bibinfo
  {title} {Relativistic n\'eel-order fields induced by electrical current in
  antiferromagnets},\ }\href@noop {} {\bibfield  {journal} {\bibinfo  {journal}
  {Phys. Rev. Lett.}\ }\textbf {\bibinfo {volume} {113}},\ \bibinfo {pages}
  {157201} (\bibinfo {year} {2014})}\BibitemShut {NoStop}%
\bibitem [{\citenamefont {Gomonay}\ \emph {et~al.}(2016)\citenamefont
  {Gomonay}, \citenamefont {Jungwirth},\ and\ \citenamefont
  {Sinova}}]{PhysRevLett.117.017202}%
  \BibitemOpen
  \bibfield  {author} {\bibinfo {author} {\bibfnamefont {O.}~\bibnamefont
  {Gomonay}}, \bibinfo {author} {\bibfnamefont {T.}~\bibnamefont {Jungwirth}},\
  and\ \bibinfo {author} {\bibfnamefont {J.}~\bibnamefont {Sinova}},\
  }\bibfield  {title} {\bibinfo {title} {High antiferromagnetic domain wall
  velocity induced by n\'eel spin-orbit torques},\ }\href@noop {} {\bibfield
  {journal} {\bibinfo  {journal} {Phys. Rev. Lett.}\ }\textbf {\bibinfo
  {volume} {117}},\ \bibinfo {pages} {017202} (\bibinfo {year}
  {2016})}\BibitemShut {NoStop}%
\bibitem [{\citenamefont {Kumar}\ \emph {et~al.}(2022)\citenamefont {Kumar},
  \citenamefont {Jin}, \citenamefont {Sbiaa}, \citenamefont {Kl{\"a}ui},
  \citenamefont {Bedanta}, \citenamefont {Fukami}, \citenamefont {Ravelosona},
  \citenamefont {Yang}, \citenamefont {Liu},\ and\ \citenamefont
  {Piramanayagam}}]{Kumar:2022vy}%
  \BibitemOpen
  \bibfield  {author} {\bibinfo {author} {\bibfnamefont {D.}~\bibnamefont
  {Kumar}}, \bibinfo {author} {\bibfnamefont {T.}~\bibnamefont {Jin}}, \bibinfo
  {author} {\bibfnamefont {R.}~\bibnamefont {Sbiaa}}, \bibinfo {author}
  {\bibfnamefont {M.}~\bibnamefont {Kl{\"a}ui}}, \bibinfo {author}
  {\bibfnamefont {S.}~\bibnamefont {Bedanta}}, \bibinfo {author} {\bibfnamefont
  {S.}~\bibnamefont {Fukami}}, \bibinfo {author} {\bibfnamefont
  {D.}~\bibnamefont {Ravelosona}}, \bibinfo {author} {\bibfnamefont {S.-H.}\
  \bibnamefont {Yang}}, \bibinfo {author} {\bibfnamefont {X.}~\bibnamefont
  {Liu}},\ and\ \bibinfo {author} {\bibfnamefont {S.~N.}\ \bibnamefont
  {Piramanayagam}},\ }\bibfield  {title} {\bibinfo {title} {Domain wall memory:
  Physics, materials, and devices},\ }\bibfield  {booktitle} {\emph {\bibinfo
  {booktitle} {Domain Wall Memory: Physics, Materials, and Devices}},\
  }\href@noop {} {\bibfield  {journal} {\bibinfo  {journal} {Physics Reports}\
  }\textbf {\bibinfo {volume} {958}},\ \bibinfo {pages} {1} (\bibinfo {year}
  {2022})}\BibitemShut {NoStop}%
\bibitem [{\citenamefont {Luo}\ and\ \citenamefont
  {You}(2021)}]{doi:10.1063/5.0042917}%
  \BibitemOpen
  \bibfield  {author} {\bibinfo {author} {\bibfnamefont {S.}~\bibnamefont
  {Luo}}\ and\ \bibinfo {author} {\bibfnamefont {L.}~\bibnamefont {You}},\
  }\bibfield  {title} {\bibinfo {title} {Skyrmion devices for memory and logic
  applications},\ }\href@noop {} {\bibfield  {journal} {\bibinfo  {journal}
  {APL Materials}\ }\textbf {\bibinfo {volume} {9}},\ \bibinfo {pages} {050901}
  (\bibinfo {year} {2021})}\BibitemShut {NoStop}%
\bibitem [{\citenamefont {Grollier}\ \emph {et~al.}(2020)\citenamefont
  {Grollier}, \citenamefont {Querlioz}, \citenamefont {Camsari}, \citenamefont
  {Everschor-Sitte}, \citenamefont {Fukami},\ and\ \citenamefont
  {Stiles}}]{Grollier:2020ur}%
  \BibitemOpen
  \bibfield  {author} {\bibinfo {author} {\bibfnamefont {J.}~\bibnamefont
  {Grollier}}, \bibinfo {author} {\bibfnamefont {D.}~\bibnamefont {Querlioz}},
  \bibinfo {author} {\bibfnamefont {K.~Y.}\ \bibnamefont {Camsari}}, \bibinfo
  {author} {\bibfnamefont {K.}~\bibnamefont {Everschor-Sitte}}, \bibinfo
  {author} {\bibfnamefont {S.}~\bibnamefont {Fukami}},\ and\ \bibinfo {author}
  {\bibfnamefont {M.~D.}\ \bibnamefont {Stiles}},\ }\bibfield  {title}
  {\bibinfo {title} {Neuromorphic spintronics},\ }\href@noop {} {\bibfield
  {journal} {\bibinfo  {journal} {Nature Electronics}\ }\textbf {\bibinfo
  {volume} {3}},\ \bibinfo {pages} {360} (\bibinfo {year} {2020})}\BibitemShut
  {NoStop}%
\bibitem [{\citenamefont {Tserkovnyak}\ and\ \citenamefont
  {Xiao}(2018)}]{PhysRevLett.121.127701}%
  \BibitemOpen
  \bibfield  {author} {\bibinfo {author} {\bibfnamefont {Y.}~\bibnamefont
  {Tserkovnyak}}\ and\ \bibinfo {author} {\bibfnamefont {J.}~\bibnamefont
  {Xiao}},\ }\bibfield  {title} {\bibinfo {title} {Energy storage via
  topological spin textures},\ }\href@noop {} {\bibfield  {journal} {\bibinfo
  {journal} {Phys. Rev. Lett.}\ }\textbf {\bibinfo {volume} {121}},\ \bibinfo
  {pages} {127701} (\bibinfo {year} {2018})}\BibitemShut {NoStop}%
\bibitem [{\citenamefont {Jones}\ \emph {et~al.}(2020)\citenamefont {Jones},
  \citenamefont {Zou}, \citenamefont {Zhang},\ and\ \citenamefont
  {Tserkovnyak}}]{Daltonenergy}%
  \BibitemOpen
  \bibfield  {author} {\bibinfo {author} {\bibfnamefont {D.}~\bibnamefont
  {Jones}}, \bibinfo {author} {\bibfnamefont {J.}~\bibnamefont {Zou}}, \bibinfo
  {author} {\bibfnamefont {S.}~\bibnamefont {Zhang}},\ and\ \bibinfo {author}
  {\bibfnamefont {Y.}~\bibnamefont {Tserkovnyak}},\ }\bibfield  {title}
  {\bibinfo {title} {Energy storage in magnetic textures driven by vorticity
  flow},\ }\href@noop {} {\bibfield  {journal} {\bibinfo  {journal} {Phys. Rev.
  B}\ }\textbf {\bibinfo {volume} {102}},\ \bibinfo {pages} {140411} (\bibinfo
  {year} {2020})}\BibitemShut {NoStop}%
\bibitem [{\citenamefont {Zang}\ \emph {et~al.}(2018)\citenamefont {Zang},
  \citenamefont {Cros},\ and\ \citenamefont {Hoffmann}}]{TopologyinMagnetism}%
  \BibitemOpen
  \bibinfo {editor} {\bibfnamefont {J.}~\bibnamefont {Zang}}, \bibinfo {editor}
  {\bibfnamefont {V.}~\bibnamefont {Cros}},\ and\ \bibinfo {editor}
  {\bibfnamefont {A.}~\bibnamefont {Hoffmann}},\ eds.,\ \href@noop {} {\emph
  {\bibinfo {title} {Topology in Magnetism}}}\ (\bibinfo  {publisher} {Springer
  International Publishing},\ \bibinfo {year} {2018})\BibitemShut {NoStop}%
\bibitem [{\citenamefont {Liu}\ \emph {et~al.}(2020)\citenamefont {Liu},
  \citenamefont {Hou}, \citenamefont {Han},\ and\ \citenamefont
  {Zang}}]{Zangprl}%
  \BibitemOpen
  \bibfield  {author} {\bibinfo {author} {\bibfnamefont {Y.}~\bibnamefont
  {Liu}}, \bibinfo {author} {\bibfnamefont {W.}~\bibnamefont {Hou}}, \bibinfo
  {author} {\bibfnamefont {X.}~\bibnamefont {Han}},\ and\ \bibinfo {author}
  {\bibfnamefont {J.}~\bibnamefont {Zang}},\ }\bibfield  {title} {\bibinfo
  {title} {Three-dimensional dynamics of a magnetic hopfion driven by spin
  transfer torque},\ }\href@noop {} {\bibfield  {journal} {\bibinfo  {journal}
  {Phys. Rev. Lett.}\ }\textbf {\bibinfo {volume} {124}},\ \bibinfo {pages}
  {127204} (\bibinfo {year} {2020})}\BibitemShut {NoStop}%
\bibitem [{\citenamefont {Zou}\ \emph {et~al.}(2020)\citenamefont {Zou},
  \citenamefont {Zhang},\ and\ \citenamefont {Tserkovnyak}}]{jiprl2020}%
  \BibitemOpen
  \bibfield  {author} {\bibinfo {author} {\bibfnamefont {J.}~\bibnamefont
  {Zou}}, \bibinfo {author} {\bibfnamefont {S.}~\bibnamefont {Zhang}},\ and\
  \bibinfo {author} {\bibfnamefont {Y.}~\bibnamefont {Tserkovnyak}},\
  }\bibfield  {title} {\bibinfo {title} {Topological transport of deconfined
  hedgehogs in magnets},\ }\href@noop {} {\bibfield  {journal} {\bibinfo
  {journal} {Phys. Rev. Lett.}\ }\textbf {\bibinfo {volume} {125}},\ \bibinfo
  {pages} {267201} (\bibinfo {year} {2020})}\BibitemShut {NoStop}%
\bibitem [{\citenamefont {Zou}\ \emph {et~al.}(2019)\citenamefont {Zou},
  \citenamefont {Kim},\ and\ \citenamefont {Tserkovnyak}}]{jivortex}%
  \BibitemOpen
  \bibfield  {author} {\bibinfo {author} {\bibfnamefont {J.}~\bibnamefont
  {Zou}}, \bibinfo {author} {\bibfnamefont {S.~K.}\ \bibnamefont {Kim}},\ and\
  \bibinfo {author} {\bibfnamefont {Y.}~\bibnamefont {Tserkovnyak}},\
  }\bibfield  {title} {\bibinfo {title} {Topological transport of vorticity in
  heisenberg magnets},\ }\href@noop {} {\bibfield  {journal} {\bibinfo
  {journal} {Phys. Rev. B}\ }\textbf {\bibinfo {volume} {99}},\ \bibinfo
  {pages} {180402} (\bibinfo {year} {2019})}\BibitemShut {NoStop}%
\bibitem [{\citenamefont {Tserkovnyak}\ and\ \citenamefont
  {Zou}(2019)}]{quantumvortex}%
  \BibitemOpen
  \bibfield  {author} {\bibinfo {author} {\bibfnamefont {Y.}~\bibnamefont
  {Tserkovnyak}}\ and\ \bibinfo {author} {\bibfnamefont {J.}~\bibnamefont
  {Zou}},\ }\bibfield  {title} {\bibinfo {title} {Quantum hydrodynamics of
  vorticity},\ }\href@noop {} {\bibfield  {journal} {\bibinfo  {journal} {Phys.
  Rev. Research}\ }\textbf {\bibinfo {volume} {1}},\ \bibinfo {pages} {033071}
  (\bibinfo {year} {2019})}\BibitemShut {NoStop}%
\bibitem [{\citenamefont {Tserkovnyak}\ \emph {et~al.}(2020)\citenamefont
  {Tserkovnyak}, \citenamefont {Zou}, \citenamefont {Kim},\ and\ \citenamefont
  {Takei}}]{Yqwinding}%
  \BibitemOpen
  \bibfield  {author} {\bibinfo {author} {\bibfnamefont {Y.}~\bibnamefont
  {Tserkovnyak}}, \bibinfo {author} {\bibfnamefont {J.}~\bibnamefont {Zou}},
  \bibinfo {author} {\bibfnamefont {S.~K.}\ \bibnamefont {Kim}},\ and\ \bibinfo
  {author} {\bibfnamefont {S.}~\bibnamefont {Takei}},\ }\bibfield  {title}
  {\bibinfo {title} {Quantum hydrodynamics of spin winding},\ }\href@noop {}
  {\bibfield  {journal} {\bibinfo  {journal} {Phys. Rev. B}\ }\textbf {\bibinfo
  {volume} {102}},\ \bibinfo {pages} {224433} (\bibinfo {year}
  {2020})}\BibitemShut {NoStop}%
\bibitem [{\citenamefont {Parkin}\ \emph {et~al.}(2008)\citenamefont {Parkin},
  \citenamefont {Hayashi},\ and\ \citenamefont {Thomas}}]{Parkin190}%
  \BibitemOpen
  \bibfield  {author} {\bibinfo {author} {\bibfnamefont {S.~S.~P.}\
  \bibnamefont {Parkin}}, \bibinfo {author} {\bibfnamefont {M.}~\bibnamefont
  {Hayashi}},\ and\ \bibinfo {author} {\bibfnamefont {L.}~\bibnamefont
  {Thomas}},\ }\bibfield  {title} {\bibinfo {title} {Magnetic domain-wall
  racetrack memory},\ }\href@noop {} {\bibfield  {journal} {\bibinfo  {journal}
  {Science}\ }\textbf {\bibinfo {volume} {320}},\ \bibinfo {pages} {190}
  (\bibinfo {year} {2008})}\BibitemShut {NoStop}%
\bibitem [{\citenamefont {Shor}(1994)}]{shor_1994}%
  \BibitemOpen
  \bibfield  {author} {\bibinfo {author} {\bibfnamefont {P.}~\bibnamefont
  {Shor}},\ }\bibfield  {title} {\bibinfo {title} {Algorithms for quantum
  computation: discrete logarithms and factoring},\ }in\ \href@noop {} {\emph
  {\bibinfo {booktitle} {Proceedings 35th Annual Symposium on Foundations of
  Computer Science}}}\ (\bibinfo {year} {1994})\ pp.\ \bibinfo {pages}
  {124--134}\BibitemShut {NoStop}%
\bibitem [{\citenamefont {Blinov}\ \emph {et~al.}(2004)\citenamefont {Blinov},
  \citenamefont {Moehring}, \citenamefont {Duan},\ and\ \citenamefont
  {Monroe}}]{Blinov2004nature}%
  \BibitemOpen
  \bibfield  {author} {\bibinfo {author} {\bibfnamefont {B.~B.}\ \bibnamefont
  {Blinov}}, \bibinfo {author} {\bibfnamefont {D.~L.}\ \bibnamefont
  {Moehring}}, \bibinfo {author} {\bibfnamefont {L.~M.}\ \bibnamefont {Duan}},\
  and\ \bibinfo {author} {\bibfnamefont {C.}~\bibnamefont {Monroe}},\
  }\bibfield  {title} {\bibinfo {title} {Observation of entanglement between a
  single trapped atom and a single photon},\ }\href@noop {} {\bibfield
  {journal} {\bibinfo  {journal} {Nature}\ }\textbf {\bibinfo {volume} {428}},\
  \bibinfo {pages} {153} (\bibinfo {year} {2004})}\BibitemShut {NoStop}%
\bibitem [{\citenamefont {Volz}\ \emph {et~al.}(2006)\citenamefont {Volz},
  \citenamefont {Weber}, \citenamefont {Schlenk}, \citenamefont {Rosenfeld},
  \citenamefont {Vrana}, \citenamefont {Saucke}, \citenamefont {Kurtsiefer},\
  and\ \citenamefont {Weinfurter}}]{volz2006prl}%
  \BibitemOpen
  \bibfield  {author} {\bibinfo {author} {\bibfnamefont {J.}~\bibnamefont
  {Volz}}, \bibinfo {author} {\bibfnamefont {M.}~\bibnamefont {Weber}},
  \bibinfo {author} {\bibfnamefont {D.}~\bibnamefont {Schlenk}}, \bibinfo
  {author} {\bibfnamefont {W.}~\bibnamefont {Rosenfeld}}, \bibinfo {author}
  {\bibfnamefont {J.}~\bibnamefont {Vrana}}, \bibinfo {author} {\bibfnamefont
  {K.}~\bibnamefont {Saucke}}, \bibinfo {author} {\bibfnamefont
  {C.}~\bibnamefont {Kurtsiefer}},\ and\ \bibinfo {author} {\bibfnamefont
  {H.}~\bibnamefont {Weinfurter}},\ }\bibfield  {title} {\bibinfo {title}
  {Observation of entanglement of a single photon with a trapped atom},\
  }\href@noop {} {\bibfield  {journal} {\bibinfo  {journal} {Phys. Rev. Lett.}\
  }\textbf {\bibinfo {volume} {96}},\ \bibinfo {pages} {030404} (\bibinfo
  {year} {2006})}\BibitemShut {NoStop}%
\bibitem [{\citenamefont {Blatt}\ and\ \citenamefont
  {Wineland}(2008)}]{Blatt2008nature}%
  \BibitemOpen
  \bibfield  {author} {\bibinfo {author} {\bibfnamefont {R.}~\bibnamefont
  {Blatt}}\ and\ \bibinfo {author} {\bibfnamefont {D.}~\bibnamefont
  {Wineland}},\ }\bibfield  {title} {\bibinfo {title} {Entangled states of
  trapped atomic ions},\ }\href@noop {} {\bibfield  {journal} {\bibinfo
  {journal} {Nature}\ }\textbf {\bibinfo {volume} {453}},\ \bibinfo {pages}
  {1008} (\bibinfo {year} {2008})}\BibitemShut {NoStop}%
\bibitem [{\citenamefont {Koch}\ \emph {et~al.}(2007)\citenamefont {Koch},
  \citenamefont {Yu}, \citenamefont {Gambetta}, \citenamefont {Houck},
  \citenamefont {Schuster}, \citenamefont {Majer}, \citenamefont {Blais},
  \citenamefont {Devoret}, \citenamefont {Girvin},\ and\ \citenamefont
  {Schoelkopf}}]{Koch_pra_2007}%
  \BibitemOpen
  \bibfield  {author} {\bibinfo {author} {\bibfnamefont {J.}~\bibnamefont
  {Koch}}, \bibinfo {author} {\bibfnamefont {T.~M.}\ \bibnamefont {Yu}},
  \bibinfo {author} {\bibfnamefont {J.}~\bibnamefont {Gambetta}}, \bibinfo
  {author} {\bibfnamefont {A.~A.}\ \bibnamefont {Houck}}, \bibinfo {author}
  {\bibfnamefont {D.~I.}\ \bibnamefont {Schuster}}, \bibinfo {author}
  {\bibfnamefont {J.}~\bibnamefont {Majer}}, \bibinfo {author} {\bibfnamefont
  {A.}~\bibnamefont {Blais}}, \bibinfo {author} {\bibfnamefont {M.~H.}\
  \bibnamefont {Devoret}}, \bibinfo {author} {\bibfnamefont {S.~M.}\
  \bibnamefont {Girvin}},\ and\ \bibinfo {author} {\bibfnamefont {R.~J.}\
  \bibnamefont {Schoelkopf}},\ }\bibfield  {title} {\bibinfo {title}
  {Charge-insensitive qubit design derived from the cooper pair box},\
  }\href@noop {} {\bibfield  {journal} {\bibinfo  {journal} {Phys. Rev. A}\
  }\textbf {\bibinfo {volume} {76}},\ \bibinfo {pages} {042319} (\bibinfo
  {year} {2007})}\BibitemShut {NoStop}%
\bibitem [{\citenamefont {Barends}\ \emph {et~al.}(2013)\citenamefont
  {Barends}, \citenamefont {Kelly}, \citenamefont {Megrant}, \citenamefont
  {Sank}, \citenamefont {Jeffrey}, \citenamefont {Chen}, \citenamefont {Yin},
  \citenamefont {Chiaro}, \citenamefont {Mutus}, \citenamefont {Neill},
  \citenamefont {O'Malley}, \citenamefont {Roushan}, \citenamefont {Wenner},
  \citenamefont {White}, \citenamefont {Cleland},\ and\ \citenamefont
  {Martinis}}]{Barends_prl_2013}%
  \BibitemOpen
  \bibfield  {author} {\bibinfo {author} {\bibfnamefont {R.}~\bibnamefont
  {Barends}}, \bibinfo {author} {\bibfnamefont {J.}~\bibnamefont {Kelly}},
  \bibinfo {author} {\bibfnamefont {A.}~\bibnamefont {Megrant}}, \bibinfo
  {author} {\bibfnamefont {D.}~\bibnamefont {Sank}}, \bibinfo {author}
  {\bibfnamefont {E.}~\bibnamefont {Jeffrey}}, \bibinfo {author} {\bibfnamefont
  {Y.}~\bibnamefont {Chen}}, \bibinfo {author} {\bibfnamefont {Y.}~\bibnamefont
  {Yin}}, \bibinfo {author} {\bibfnamefont {B.}~\bibnamefont {Chiaro}},
  \bibinfo {author} {\bibfnamefont {J.}~\bibnamefont {Mutus}}, \bibinfo
  {author} {\bibfnamefont {C.}~\bibnamefont {Neill}}, \bibinfo {author}
  {\bibfnamefont {P.}~\bibnamefont {O'Malley}}, \bibinfo {author}
  {\bibfnamefont {P.}~\bibnamefont {Roushan}}, \bibinfo {author} {\bibfnamefont
  {J.}~\bibnamefont {Wenner}}, \bibinfo {author} {\bibfnamefont {T.~C.}\
  \bibnamefont {White}}, \bibinfo {author} {\bibfnamefont {A.~N.}\ \bibnamefont
  {Cleland}},\ and\ \bibinfo {author} {\bibfnamefont {J.~M.}\ \bibnamefont
  {Martinis}},\ }\bibfield  {title} {\bibinfo {title} {Coherent josephson qubit
  suitable for scalable quantum integrated circuits},\ }\href@noop {}
  {\bibfield  {journal} {\bibinfo  {journal} {Phys. Rev. Lett.}\ }\textbf
  {\bibinfo {volume} {111}},\ \bibinfo {pages} {080502} (\bibinfo {year}
  {2013})}\BibitemShut {NoStop}%
\bibitem [{\citenamefont {Loss}\ and\ \citenamefont
  {DiVincenzo}(1998)}]{PhysRevA.57.120}%
  \BibitemOpen
  \bibfield  {author} {\bibinfo {author} {\bibfnamefont {D.}~\bibnamefont
  {Loss}}\ and\ \bibinfo {author} {\bibfnamefont {D.~P.}\ \bibnamefont
  {DiVincenzo}},\ }\bibfield  {title} {\bibinfo {title} {Quantum computation
  with quantum dots},\ }\href@noop {} {\bibfield  {journal} {\bibinfo
  {journal} {Phys. Rev. A}\ }\textbf {\bibinfo {volume} {57}},\ \bibinfo
  {pages} {120} (\bibinfo {year} {1998})}\BibitemShut {NoStop}%
\bibitem [{\citenamefont {Basso~Basset}\ \emph {et~al.}(2019)\citenamefont
  {Basso~Basset}, \citenamefont {Rota}, \citenamefont {Schimpf}, \citenamefont
  {Tedeschi}, \citenamefont {Zeuner}, \citenamefont {Covre~da Silva},
  \citenamefont {Reindl}, \citenamefont {Zwiller}, \citenamefont {J\"ons},
  \citenamefont {Rastelli},\ and\ \citenamefont {Trotta}}]{Basso2019prl}%
  \BibitemOpen
  \bibfield  {author} {\bibinfo {author} {\bibfnamefont {F.}~\bibnamefont
  {Basso~Basset}}, \bibinfo {author} {\bibfnamefont {M.~B.}\ \bibnamefont
  {Rota}}, \bibinfo {author} {\bibfnamefont {C.}~\bibnamefont {Schimpf}},
  \bibinfo {author} {\bibfnamefont {D.}~\bibnamefont {Tedeschi}}, \bibinfo
  {author} {\bibfnamefont {K.~D.}\ \bibnamefont {Zeuner}}, \bibinfo {author}
  {\bibfnamefont {S.~F.}\ \bibnamefont {Covre~da Silva}}, \bibinfo {author}
  {\bibfnamefont {M.}~\bibnamefont {Reindl}}, \bibinfo {author} {\bibfnamefont
  {V.}~\bibnamefont {Zwiller}}, \bibinfo {author} {\bibfnamefont {K.~D.}\
  \bibnamefont {J\"ons}}, \bibinfo {author} {\bibfnamefont {A.}~\bibnamefont
  {Rastelli}},\ and\ \bibinfo {author} {\bibfnamefont {R.}~\bibnamefont
  {Trotta}},\ }\bibfield  {title} {\bibinfo {title} {Entanglement swapping with
  photons generated on demand by a quantum dot},\ }\href@noop {} {\bibfield
  {journal} {\bibinfo  {journal} {Phys. Rev. Lett.}\ }\textbf {\bibinfo
  {volume} {123}},\ \bibinfo {pages} {160501} (\bibinfo {year}
  {2019})}\BibitemShut {NoStop}%
\bibitem [{\citenamefont {Qiao}\ \emph {et~al.}(2020)\citenamefont {Qiao},
  \citenamefont {Kandel}, \citenamefont {Manikandan}, \citenamefont {Jordan},
  \citenamefont {Fallahi}, \citenamefont {Gardner}, \citenamefont {Manfra},\
  and\ \citenamefont {Nichol}}]{Qiao:2020ncom}%
  \BibitemOpen
  \bibfield  {author} {\bibinfo {author} {\bibfnamefont {H.}~\bibnamefont
  {Qiao}}, \bibinfo {author} {\bibfnamefont {Y.~P.}\ \bibnamefont {Kandel}},
  \bibinfo {author} {\bibfnamefont {S.~K.}\ \bibnamefont {Manikandan}},
  \bibinfo {author} {\bibfnamefont {A.~N.}\ \bibnamefont {Jordan}}, \bibinfo
  {author} {\bibfnamefont {S.}~\bibnamefont {Fallahi}}, \bibinfo {author}
  {\bibfnamefont {G.~C.}\ \bibnamefont {Gardner}}, \bibinfo {author}
  {\bibfnamefont {M.~J.}\ \bibnamefont {Manfra}},\ and\ \bibinfo {author}
  {\bibfnamefont {J.~M.}\ \bibnamefont {Nichol}},\ }\bibfield  {title}
  {\bibinfo {title} {Conditional teleportation of quantum-dot spin states},\
  }\href@noop {} {\bibfield  {journal} {\bibinfo  {journal} {Nature
  Communications}\ }\textbf {\bibinfo {volume} {11}},\ \bibinfo {pages} {3022}
  (\bibinfo {year} {2020})}\BibitemShut {NoStop}%
\bibitem [{\citenamefont {Hendrickx}\ \emph {et~al.}(2021)\citenamefont
  {Hendrickx}, \citenamefont {Lawrie}, \citenamefont {Russ}, \citenamefont {van
  Riggelen}, \citenamefont {de~Snoo}, \citenamefont {Schouten}, \citenamefont
  {Sammak}, \citenamefont {Scappucci},\ and\ \citenamefont
  {Veldhorst}}]{Hendrickx:2021tv}%
  \BibitemOpen
  \bibfield  {author} {\bibinfo {author} {\bibfnamefont {N.~W.}\ \bibnamefont
  {Hendrickx}}, \bibinfo {author} {\bibfnamefont {W.~I.~L.}\ \bibnamefont
  {Lawrie}}, \bibinfo {author} {\bibfnamefont {M.}~\bibnamefont {Russ}},
  \bibinfo {author} {\bibfnamefont {F.}~\bibnamefont {van Riggelen}}, \bibinfo
  {author} {\bibfnamefont {S.~L.}\ \bibnamefont {de~Snoo}}, \bibinfo {author}
  {\bibfnamefont {R.~N.}\ \bibnamefont {Schouten}}, \bibinfo {author}
  {\bibfnamefont {A.}~\bibnamefont {Sammak}}, \bibinfo {author} {\bibfnamefont
  {G.}~\bibnamefont {Scappucci}},\ and\ \bibinfo {author} {\bibfnamefont
  {M.}~\bibnamefont {Veldhorst}},\ }\bibfield  {title} {\bibinfo {title} {A
  four-qubit germanium quantum processor},\ }\href@noop {} {\bibfield
  {journal} {\bibinfo  {journal} {Nature}\ }\textbf {\bibinfo {volume} {591}},\
  \bibinfo {pages} {580} (\bibinfo {year} {2021})}\BibitemShut {NoStop}%
\bibitem [{\citenamefont {Philips}\ \emph {et~al.}(2022)\citenamefont
  {Philips}, \citenamefont {{Mahale}}, \citenamefont {Amitonov}, \citenamefont
  {de~Snoo}, \citenamefont {Russ}, \citenamefont {Kalhor}, \citenamefont
  {Volk}, \citenamefont {Lawrie}, \citenamefont {Brousse}, \citenamefont
  {Tryputen}, \citenamefont {Wuetz}, \citenamefont {Sammak}, \citenamefont
  {Veldhorst}, \citenamefont {Scappucci},\ and\ \citenamefont
  {Vandersypen}}]{Philips:2022ur}%
  \BibitemOpen
  \bibfield  {author} {\bibinfo {author} {\bibfnamefont {S.~G.~J.}\
  \bibnamefont {Philips}}, \bibinfo {author} {\bibfnamefont {M.~T.}\
  \bibnamefont {{Mahale}}, \bibfnamefont {Pratibhadzik}}, \bibinfo {author}
  {\bibfnamefont {S.~V.}\ \bibnamefont {Amitonov}}, \bibinfo {author}
  {\bibfnamefont {S.~L.}\ \bibnamefont {de~Snoo}}, \bibinfo {author}
  {\bibfnamefont {M.}~\bibnamefont {Russ}}, \bibinfo {author} {\bibfnamefont
  {N.}~\bibnamefont {Kalhor}}, \bibinfo {author} {\bibfnamefont
  {C.}~\bibnamefont {Volk}}, \bibinfo {author} {\bibfnamefont {W.~I.~L.}\
  \bibnamefont {Lawrie}}, \bibinfo {author} {\bibfnamefont {D.}~\bibnamefont
  {Brousse}}, \bibinfo {author} {\bibfnamefont {L.}~\bibnamefont {Tryputen}},
  \bibinfo {author} {\bibfnamefont {B.~P.}\ \bibnamefont {Wuetz}}, \bibinfo
  {author} {\bibfnamefont {A.}~\bibnamefont {Sammak}}, \bibinfo {author}
  {\bibfnamefont {M.}~\bibnamefont {Veldhorst}}, \bibinfo {author}
  {\bibfnamefont {G.}~\bibnamefont {Scappucci}},\ and\ \bibinfo {author}
  {\bibfnamefont {L.~M.~K.}\ \bibnamefont {Vandersypen}},\ }\bibfield  {title}
  {\bibinfo {title} {Universal control of a six-qubit quantum processor in
  silicon},\ }\href@noop {} {\bibfield  {journal} {\bibinfo  {journal}
  {Nature}\ }\textbf {\bibinfo {volume} {609}},\ \bibinfo {pages} {919}
  (\bibinfo {year} {2022})}\BibitemShut {NoStop}%
\bibitem [{\citenamefont {Mills}\ \emph {et~al.}(2022)\citenamefont {Mills},
  \citenamefont {Guinn}, \citenamefont {Gullans}, \citenamefont {Sigillito},
  \citenamefont {Feldman}, \citenamefont {Nielsen},\ and\ \citenamefont
  {Petta}}]{petta2022sa}%
  \BibitemOpen
  \bibfield  {author} {\bibinfo {author} {\bibfnamefont {A.~R.}\ \bibnamefont
  {Mills}}, \bibinfo {author} {\bibfnamefont {C.~R.}\ \bibnamefont {Guinn}},
  \bibinfo {author} {\bibfnamefont {M.~J.}\ \bibnamefont {Gullans}}, \bibinfo
  {author} {\bibfnamefont {A.~J.}\ \bibnamefont {Sigillito}}, \bibinfo {author}
  {\bibfnamefont {M.~M.}\ \bibnamefont {Feldman}}, \bibinfo {author}
  {\bibfnamefont {E.}~\bibnamefont {Nielsen}},\ and\ \bibinfo {author}
  {\bibfnamefont {J.~R.}\ \bibnamefont {Petta}},\ }\bibfield  {title} {\bibinfo
  {title} {Two-qubit silicon quantum processor with operation fidelity
  exceeding 99
  {Science Advances}\ }\textbf {\bibinfo {volume} {8}},\ \bibinfo {pages}
  {eabn5130} (\bibinfo {year} {2022})}\BibitemShut {NoStop}%
\bibitem [{\citenamefont {Psaroudaki}\ and\ \citenamefont
  {Panagopoulos}(2021)}]{christina_prl_2021}%
  \BibitemOpen
  \bibfield  {author} {\bibinfo {author} {\bibfnamefont {C.}~\bibnamefont
  {Psaroudaki}}\ and\ \bibinfo {author} {\bibfnamefont {C.}~\bibnamefont
  {Panagopoulos}},\ }\bibfield  {title} {\bibinfo {title} {Skyrmion qubits: A
  new class of quantum logic elements based on nanoscale magnetization},\
  }\href@noop {} {\bibfield  {journal} {\bibinfo  {journal} {Phys. Rev. Lett.}\
  }\textbf {\bibinfo {volume} {127}},\ \bibinfo {pages} {067201} (\bibinfo
  {year} {2021})}\BibitemShut {NoStop}%
\bibitem [{\citenamefont {Xia}\ \emph {et~al.}(2023)\citenamefont {Xia},
  \citenamefont {Zhang}, \citenamefont {Liu}, \citenamefont {Zhou},\ and\
  \citenamefont {Ezawa}}]{xia2023universal}%
  \BibitemOpen
  \bibfield  {author} {\bibinfo {author} {\bibfnamefont {J.}~\bibnamefont
  {Xia}}, \bibinfo {author} {\bibfnamefont {X.}~\bibnamefont {Zhang}}, \bibinfo
  {author} {\bibfnamefont {X.}~\bibnamefont {Liu}}, \bibinfo {author}
  {\bibfnamefont {Y.}~\bibnamefont {Zhou}},\ and\ \bibinfo {author}
  {\bibfnamefont {M.}~\bibnamefont {Ezawa}},\ }\bibfield  {title} {\bibinfo
  {title} {Universal quantum computation based on nanoscale skyrmion helicity
  qubits in frustrated magnets},\ }\href@noop {} {\bibfield  {journal}
  {\bibinfo  {journal} {Physical Review Letters}\ }\textbf {\bibinfo {volume}
  {130}},\ \bibinfo {pages} {106701} (\bibinfo {year} {2023})}\BibitemShut
  {NoStop}%
\bibitem [{\citenamefont {Xia}\ \emph {et~al.}(2022)\citenamefont {Xia},
  \citenamefont {Zhang}, \citenamefont {Liu}, \citenamefont {Zhou},\ and\
  \citenamefont {Ezawa}}]{xia2022qubits}%
  \BibitemOpen
  \bibfield  {author} {\bibinfo {author} {\bibfnamefont {J.}~\bibnamefont
  {Xia}}, \bibinfo {author} {\bibfnamefont {X.}~\bibnamefont {Zhang}}, \bibinfo
  {author} {\bibfnamefont {X.}~\bibnamefont {Liu}}, \bibinfo {author}
  {\bibfnamefont {Y.}~\bibnamefont {Zhou}},\ and\ \bibinfo {author}
  {\bibfnamefont {M.}~\bibnamefont {Ezawa}},\ }\bibfield  {title} {\bibinfo
  {title} {Qubits based on merons in magnetic nanodisks},\ }\href@noop {}
  {\bibfield  {journal} {\bibinfo  {journal} {Communications Materials}\
  }\textbf {\bibinfo {volume} {3}},\ \bibinfo {pages} {88} (\bibinfo {year}
  {2022})}\BibitemShut {NoStop}%
\bibitem [{\citenamefont {Takei}\ \emph {et~al.}(2017)\citenamefont {Takei},
  \citenamefont {Tserkovnyak},\ and\ \citenamefont
  {Mohseni}}]{PhysRevB.95.144402}%
  \BibitemOpen
  \bibfield  {author} {\bibinfo {author} {\bibfnamefont {S.}~\bibnamefont
  {Takei}}, \bibinfo {author} {\bibfnamefont {Y.}~\bibnamefont {Tserkovnyak}},\
  and\ \bibinfo {author} {\bibfnamefont {M.}~\bibnamefont {Mohseni}},\
  }\bibfield  {title} {\bibinfo {title} {Spin superfluid josephson quantum
  devices},\ }\href@noop {} {\bibfield  {journal} {\bibinfo  {journal} {Phys.
  Rev. B}\ }\textbf {\bibinfo {volume} {95}},\ \bibinfo {pages} {144402}
  (\bibinfo {year} {2017})}\BibitemShut {NoStop}%
\bibitem [{\citenamefont {Takei}\ and\ \citenamefont
  {Mohseni}(2018)}]{PhysRevB.97.064401}%
  \BibitemOpen
  \bibfield  {author} {\bibinfo {author} {\bibfnamefont {S.}~\bibnamefont
  {Takei}}\ and\ \bibinfo {author} {\bibfnamefont {M.}~\bibnamefont
  {Mohseni}},\ }\bibfield  {title} {\bibinfo {title} {Quantum control of
  topological defects in magnetic systems},\ }\href@noop {} {\bibfield
  {journal} {\bibinfo  {journal} {Phys. Rev. B}\ }\textbf {\bibinfo {volume}
  {97}},\ \bibinfo {pages} {064401} (\bibinfo {year} {2018})}\BibitemShut
  {NoStop}%
\bibitem [{\citenamefont {Ahari}\ and\ \citenamefont
  {Tserkovnyak}(2023)}]{ahari2023andreev}%
  \BibitemOpen
  \bibfield  {author} {\bibinfo {author} {\bibfnamefont {M.~T.}\ \bibnamefont
  {Ahari}}\ and\ \bibinfo {author} {\bibfnamefont {Y.}~\bibnamefont
  {Tserkovnyak}},\ }\bibfield  {title} {\bibinfo {title} {Andreev spin qubit: A
  nonadiabatic geometric gate},\ }\href@noop {} {\bibfield  {journal} {\bibinfo
   {journal} {arXiv preprint arXiv:2303.04344}\ } (\bibinfo {year}
  {2023})}\BibitemShut {NoStop}%
\bibitem [{\citenamefont {Nadj-Perge}\ \emph {et~al.}(2010)\citenamefont
  {Nadj-Perge}, \citenamefont {Frolov}, \citenamefont {Bakkers},\ and\
  \citenamefont {Kouwenhoven}}]{Nadj-Perge:2010tb}%
  \BibitemOpen
  \bibfield  {author} {\bibinfo {author} {\bibfnamefont {S.}~\bibnamefont
  {Nadj-Perge}}, \bibinfo {author} {\bibfnamefont {S.~M.}\ \bibnamefont
  {Frolov}}, \bibinfo {author} {\bibfnamefont {E.~P. A.~M.}\ \bibnamefont
  {Bakkers}},\ and\ \bibinfo {author} {\bibfnamefont {L.~P.}\ \bibnamefont
  {Kouwenhoven}},\ }\bibfield  {title} {\bibinfo {title} {Spin--orbit qubit in
  a semiconductor nanowire},\ }\href@noop {} {\bibfield  {journal} {\bibinfo
  {journal} {Nature}\ }\textbf {\bibinfo {volume} {468}},\ \bibinfo {pages}
  {1084} (\bibinfo {year} {2010})}\BibitemShut {NoStop}%
\bibitem [{\citenamefont {Froning}\ \emph
  {et~al.}(2021{\natexlab{a}})\citenamefont {Froning}, \citenamefont
  {Camenzind}, \citenamefont {van~der Molen}, \citenamefont {Li}, \citenamefont
  {Bakkers}, \citenamefont {Zumb{\"u}hl},\ and\ \citenamefont
  {Braakman}}]{Froning:2021wk}%
  \BibitemOpen
  \bibfield  {author} {\bibinfo {author} {\bibfnamefont {F.~N.~M.}\
  \bibnamefont {Froning}}, \bibinfo {author} {\bibfnamefont {L.~C.}\
  \bibnamefont {Camenzind}}, \bibinfo {author} {\bibfnamefont {O.~A.~H.}\
  \bibnamefont {van~der Molen}}, \bibinfo {author} {\bibfnamefont
  {A.}~\bibnamefont {Li}}, \bibinfo {author} {\bibfnamefont {E.~P. A.~M.}\
  \bibnamefont {Bakkers}}, \bibinfo {author} {\bibfnamefont {D.~M.}\
  \bibnamefont {Zumb{\"u}hl}},\ and\ \bibinfo {author} {\bibfnamefont {F.~R.}\
  \bibnamefont {Braakman}},\ }\bibfield  {title} {\bibinfo {title} {Ultrafast
  hole spin qubit with gate-tunable spin--orbit switch functionality},\
  }\href@noop {} {\bibfield  {journal} {\bibinfo  {journal} {Nature
  Nanotechnology}\ }\textbf {\bibinfo {volume} {16}},\ \bibinfo {pages} {308}
  (\bibinfo {year} {2021}{\natexlab{a}})}\BibitemShut {NoStop}%
\bibitem [{\citenamefont {Froning}\ \emph
  {et~al.}(2021{\natexlab{b}})\citenamefont {Froning}, \citenamefont
  {Ran\ifmmode \check{c}\else \v{c}\fi{}i\ifmmode~\acute{c}\else \'{c}\fi{}},
  \citenamefont {Het\'enyi}, \citenamefont {Bosco}, \citenamefont {Rehmann},
  \citenamefont {Li}, \citenamefont {Bakkers}, \citenamefont {Zwanenburg},
  \citenamefont {Loss}, \citenamefont {Zumb\"uhl},\ and\ \citenamefont
  {Braakman}}]{froning2021prr}%
  \BibitemOpen
  \bibfield  {author} {\bibinfo {author} {\bibfnamefont {F.~N.~M.}\
  \bibnamefont {Froning}}, \bibinfo {author} {\bibfnamefont {M.~J.}\
  \bibnamefont {Ran\ifmmode \check{c}\else \v{c}\fi{}i\ifmmode~\acute{c}\else
  \'{c}\fi{}}}, \bibinfo {author} {\bibfnamefont {B.}~\bibnamefont
  {Het\'enyi}}, \bibinfo {author} {\bibfnamefont {S.}~\bibnamefont {Bosco}},
  \bibinfo {author} {\bibfnamefont {M.~K.}\ \bibnamefont {Rehmann}}, \bibinfo
  {author} {\bibfnamefont {A.}~\bibnamefont {Li}}, \bibinfo {author}
  {\bibfnamefont {E.~P. A.~M.}\ \bibnamefont {Bakkers}}, \bibinfo {author}
  {\bibfnamefont {F.~A.}\ \bibnamefont {Zwanenburg}}, \bibinfo {author}
  {\bibfnamefont {D.}~\bibnamefont {Loss}}, \bibinfo {author} {\bibfnamefont
  {D.~M.}\ \bibnamefont {Zumb\"uhl}},\ and\ \bibinfo {author} {\bibfnamefont
  {F.~R.}\ \bibnamefont {Braakman}},\ }\bibfield  {title} {\bibinfo {title}
  {Strong spin-orbit interaction and $g$-factor renormalization of hole spins
  in ge/si nanowire quantum dots},\ }\href@noop {} {\bibfield  {journal}
  {\bibinfo  {journal} {Phys. Rev. Research}\ }\textbf {\bibinfo {volume}
  {3}},\ \bibinfo {pages} {013081} (\bibinfo {year}
  {2021}{\natexlab{b}})}\BibitemShut {NoStop}%
\bibitem [{\citenamefont {Gu}\ \emph {et~al.}(2022)\citenamefont {Gu},
  \citenamefont {Guan}, \citenamefont {Hazra}, \citenamefont {Deniz},
  \citenamefont {Migliorini}, \citenamefont {Zhang},\ and\ \citenamefont
  {Parkin}}]{gu2022three}%
  \BibitemOpen
  \bibfield  {author} {\bibinfo {author} {\bibfnamefont {K.}~\bibnamefont
  {Gu}}, \bibinfo {author} {\bibfnamefont {Y.}~\bibnamefont {Guan}}, \bibinfo
  {author} {\bibfnamefont {B.~K.}\ \bibnamefont {Hazra}}, \bibinfo {author}
  {\bibfnamefont {H.}~\bibnamefont {Deniz}}, \bibinfo {author} {\bibfnamefont
  {A.}~\bibnamefont {Migliorini}}, \bibinfo {author} {\bibfnamefont
  {W.}~\bibnamefont {Zhang}},\ and\ \bibinfo {author} {\bibfnamefont {S.~S.}\
  \bibnamefont {Parkin}},\ }\bibfield  {title} {\bibinfo {title}
  {Three-dimensional racetrack memory devices designed from freestanding
  magnetic heterostructures},\ }\href@noop {} {\bibfield  {journal} {\bibinfo
  {journal} {Nature Nanotechnology}\ }\textbf {\bibinfo {volume} {17}},\
  \bibinfo {pages} {1065} (\bibinfo {year} {2022})}\BibitemShut {NoStop}%
\bibitem [{\citenamefont {Dasgupta}\ and\ \citenamefont
  {Zou}(2021)}]{ji_2021_neel}%
  \BibitemOpen
  \bibfield  {author} {\bibinfo {author} {\bibfnamefont {S.}~\bibnamefont
  {Dasgupta}}\ and\ \bibinfo {author} {\bibfnamefont {J.}~\bibnamefont {Zou}},\
  }\bibfield  {title} {\bibinfo {title} {Zeeman term for the n\'eel vector in a
  two sublattice antiferromagnet},\ }\href@noop {} {\bibfield  {journal}
  {\bibinfo  {journal} {Phys. Rev. B}\ }\textbf {\bibinfo {volume} {104}},\
  \bibinfo {pages} {064415} (\bibinfo {year} {2021})}\BibitemShut {NoStop}%
\bibitem [{\citenamefont {Loss}\ \emph {et~al.}(1992)\citenamefont {Loss},
  \citenamefont {DiVincenzo},\ and\ \citenamefont
  {Grinstein}}]{daniel_1992_prl}%
  \BibitemOpen
  \bibfield  {author} {\bibinfo {author} {\bibfnamefont {D.}~\bibnamefont
  {Loss}}, \bibinfo {author} {\bibfnamefont {D.~P.}\ \bibnamefont
  {DiVincenzo}},\ and\ \bibinfo {author} {\bibfnamefont {G.}~\bibnamefont
  {Grinstein}},\ }\bibfield  {title} {\bibinfo {title} {Suppression of
  tunneling by interference in half-integer-spin particles},\ }\href@noop {}
  {\bibfield  {journal} {\bibinfo  {journal} {Phys. Rev. Lett.}\ }\textbf
  {\bibinfo {volume} {69}},\ \bibinfo {pages} {3232} (\bibinfo {year}
  {1992})}\BibitemShut {NoStop}%
\bibitem [{\citenamefont {Chiolero}\ and\ \citenamefont
  {Loss}(1997)}]{daniel_1997_prb}%
  \BibitemOpen
  \bibfield  {author} {\bibinfo {author} {\bibfnamefont {A.}~\bibnamefont
  {Chiolero}}\ and\ \bibinfo {author} {\bibfnamefont {D.}~\bibnamefont
  {Loss}},\ }\bibfield  {title} {\bibinfo {title} {Macroscopic quantum
  coherence in ferrimagnets},\ }\href@noop {} {\bibfield  {journal} {\bibinfo
  {journal} {Phys. Rev. B}\ }\textbf {\bibinfo {volume} {56}},\ \bibinfo
  {pages} {738} (\bibinfo {year} {1997})}\BibitemShut {NoStop}%
\bibitem [{\citenamefont {Altland}\ and\ \citenamefont
  {Simons}(2010)}]{atland}%
  \BibitemOpen
  \bibfield  {author} {\bibinfo {author} {\bibfnamefont {A.}~\bibnamefont
  {Altland}}\ and\ \bibinfo {author} {\bibfnamefont {B.~D.}\ \bibnamefont
  {Simons}},\ }\href@noop {} {\emph {\bibinfo {title} {Condensed matter field
  theory}}},\ \bibinfo {edition} {2nd}\ ed.\ (\bibinfo  {publisher} {Cambridge
  University Press},\ \bibinfo {year} {2010})\BibitemShut {NoStop}%
\bibitem [{dws()}]{dwsize}%
  \BibitemOpen
  \href@noop {} {}\bibinfo {note} {This number is experimentally feasible. The
  domain wall size can be made to be a few nanometers, the racetrack can be
  atomically thin, and the width of the track can be around 10 nm with the
  state-of-the-art technology.}\BibitemShut {Stop}%
\bibitem [{anh()}]{anharmo}%
  \BibitemOpen
  \href@noop {} {}\bibinfo {note} {We use $b_y=0.15$ in our estimation in which
  case the anharmonicity $\hbar\omega_0/t_g\approx 12$. Our computational space
  is thus well-isolated from higher excited states.}\BibitemShut {Stop}%
\bibitem [{\citenamefont {Golovach}\ \emph {et~al.}(2006)\citenamefont
  {Golovach}, \citenamefont {Borhani},\ and\ \citenamefont
  {Loss}}]{golovach2006prb}%
  \BibitemOpen
  \bibfield  {author} {\bibinfo {author} {\bibfnamefont {V.~N.}\ \bibnamefont
  {Golovach}}, \bibinfo {author} {\bibfnamefont {M.}~\bibnamefont {Borhani}},\
  and\ \bibinfo {author} {\bibfnamefont {D.}~\bibnamefont {Loss}},\ }\bibfield
  {title} {\bibinfo {title} {Electric-dipole-induced spin resonance in quantum
  dots},\ }\href@noop {} {\bibfield  {journal} {\bibinfo  {journal} {Phys. Rev.
  B}\ }\textbf {\bibinfo {volume} {74}},\ \bibinfo {pages} {165319} (\bibinfo
  {year} {2006})}\BibitemShut {NoStop}%
\bibitem [{\citenamefont {Bosco}\ \emph {et~al.}(2022)\citenamefont {Bosco},
  \citenamefont {Scarlino}, \citenamefont {Klinovaja},\ and\ \citenamefont
  {Loss}}]{stefano2022prl}%
  \BibitemOpen
  \bibfield  {author} {\bibinfo {author} {\bibfnamefont {S.}~\bibnamefont
  {Bosco}}, \bibinfo {author} {\bibfnamefont {P.}~\bibnamefont {Scarlino}},
  \bibinfo {author} {\bibfnamefont {J.}~\bibnamefont {Klinovaja}},\ and\
  \bibinfo {author} {\bibfnamefont {D.}~\bibnamefont {Loss}},\ }\bibfield
  {title} {\bibinfo {title} {Fully tunable longitudinal spin-photon
  interactions in si and ge quantum dots},\ }\href@noop {} {\bibfield
  {journal} {\bibinfo  {journal} {Phys. Rev. Lett.}\ }\textbf {\bibinfo
  {volume} {129}},\ \bibinfo {pages} {066801} (\bibinfo {year}
  {2022})}\BibitemShut {NoStop}%
\bibitem [{\citenamefont {Bosco}\ \emph
  {et~al.}(2021{\natexlab{a}})\citenamefont {Bosco}, \citenamefont {Benito},
  \citenamefont {Adelsberger},\ and\ \citenamefont {Loss}}]{stefano2021prb}%
  \BibitemOpen
  \bibfield  {author} {\bibinfo {author} {\bibfnamefont {S.}~\bibnamefont
  {Bosco}}, \bibinfo {author} {\bibfnamefont {M.}~\bibnamefont {Benito}},
  \bibinfo {author} {\bibfnamefont {C.}~\bibnamefont {Adelsberger}},\ and\
  \bibinfo {author} {\bibfnamefont {D.}~\bibnamefont {Loss}},\ }\bibfield
  {title} {\bibinfo {title} {Squeezed hole spin qubits in ge quantum dots with
  ultrafast gates at low power},\ }\href@noop {} {\bibfield  {journal}
  {\bibinfo  {journal} {Phys. Rev. B}\ }\textbf {\bibinfo {volume} {104}},\
  \bibinfo {pages} {115425} (\bibinfo {year} {2021}{\natexlab{a}})}\BibitemShut
  {NoStop}%
\bibitem [{\citenamefont {Kolodrubetz}\ \emph {et~al.}(2017)\citenamefont
  {Kolodrubetz}, \citenamefont {Sels}, \citenamefont {Mehta},\ and\
  \citenamefont {Polkovnikov}}]{anatoli_2017_phyreport}%
  \BibitemOpen
  \bibfield  {author} {\bibinfo {author} {\bibfnamefont {M.}~\bibnamefont
  {Kolodrubetz}}, \bibinfo {author} {\bibfnamefont {D.}~\bibnamefont {Sels}},
  \bibinfo {author} {\bibfnamefont {P.}~\bibnamefont {Mehta}},\ and\ \bibinfo
  {author} {\bibfnamefont {A.}~\bibnamefont {Polkovnikov}},\ }\bibfield
  {title} {\bibinfo {title} {Geometry and non-adiabatic response in quantum and
  classical systems},\ }\href@noop {} {\bibfield  {journal} {\bibinfo
  {journal} {Physics Reports}\ }\textbf {\bibinfo {volume} {697}},\ \bibinfo
  {pages} {1} (\bibinfo {year} {2017})},\ \bibinfo {note} {geometry and
  non-adiabatic response in quantum and classical systems}\BibitemShut
  {NoStop}%
\bibitem [{\citenamefont {Fasth}\ \emph {et~al.}(2007)\citenamefont {Fasth},
  \citenamefont {Fuhrer}, \citenamefont {Samuelson}, \citenamefont {Golovach},\
  and\ \citenamefont {Loss}}]{fasth2007prl}%
  \BibitemOpen
  \bibfield  {author} {\bibinfo {author} {\bibfnamefont {C.}~\bibnamefont
  {Fasth}}, \bibinfo {author} {\bibfnamefont {A.}~\bibnamefont {Fuhrer}},
  \bibinfo {author} {\bibfnamefont {L.}~\bibnamefont {Samuelson}}, \bibinfo
  {author} {\bibfnamefont {V.~N.}\ \bibnamefont {Golovach}},\ and\ \bibinfo
  {author} {\bibfnamefont {D.}~\bibnamefont {Loss}},\ }\bibfield  {title}
  {\bibinfo {title} {Direct measurement of the spin-orbit interaction in a
  two-electron inas nanowire quantum dot},\ }\href@noop {} {\bibfield
  {journal} {\bibinfo  {journal} {Phys. Rev. Lett.}\ }\textbf {\bibinfo
  {volume} {98}},\ \bibinfo {pages} {266801} (\bibinfo {year}
  {2007})}\BibitemShut {NoStop}%
\bibitem [{\citenamefont {Camenzind}\ \emph {et~al.}(2022)\citenamefont
  {Camenzind}, \citenamefont {Geyer}, \citenamefont {Fuhrer}, \citenamefont
  {Warburton}, \citenamefont {Zumb{\"u}hl},\ and\ \citenamefont
  {Kuhlmann}}]{Camenzind:2022tr}%
  \BibitemOpen
  \bibfield  {author} {\bibinfo {author} {\bibfnamefont {L.~C.}\ \bibnamefont
  {Camenzind}}, \bibinfo {author} {\bibfnamefont {S.}~\bibnamefont {Geyer}},
  \bibinfo {author} {\bibfnamefont {A.}~\bibnamefont {Fuhrer}}, \bibinfo
  {author} {\bibfnamefont {R.~J.}\ \bibnamefont {Warburton}}, \bibinfo {author}
  {\bibfnamefont {D.~M.}\ \bibnamefont {Zumb{\"u}hl}},\ and\ \bibinfo {author}
  {\bibfnamefont {A.~V.}\ \bibnamefont {Kuhlmann}},\ }\bibfield  {title}
  {\bibinfo {title} {A hole spin qubit in a fin field-effect transistor above 4
  kelvin},\ }\href@noop {} {\bibfield  {journal} {\bibinfo  {journal} {Nature
  Electronics}\ }\textbf {\bibinfo {volume} {5}},\ \bibinfo {pages} {178}
  (\bibinfo {year} {2022})}\BibitemShut {NoStop}%
\bibitem [{\citenamefont {Li}\ \emph {et~al.}(2015)\citenamefont {Li},
  \citenamefont {Hudson}, \citenamefont {Dzurak},\ and\ \citenamefont
  {Hamilton}}]{Li:2015us}%
  \BibitemOpen
  \bibfield  {author} {\bibinfo {author} {\bibfnamefont {R.}~\bibnamefont
  {Li}}, \bibinfo {author} {\bibfnamefont {F.~E.}\ \bibnamefont {Hudson}},
  \bibinfo {author} {\bibfnamefont {A.~S.}\ \bibnamefont {Dzurak}},\ and\
  \bibinfo {author} {\bibfnamefont {A.~R.}\ \bibnamefont {Hamilton}},\
  }\bibfield  {title} {\bibinfo {title} {Pauli spin blockade of heavy holes in
  a silicon double quantum dot},\ }\bibfield  {booktitle} {\emph {\bibinfo
  {booktitle} {Nano Letters}},\ }\href@noop {} {\bibfield  {journal} {\bibinfo
  {journal} {Nano Letters}\ }\textbf {\bibinfo {volume} {15}},\ \bibinfo
  {pages} {7314} (\bibinfo {year} {2015})}\BibitemShut {NoStop}%
\bibitem [{\citenamefont {Bosco}\ \emph
  {et~al.}(2021{\natexlab{b}})\citenamefont {Bosco}, \citenamefont
  {Het\'enyi},\ and\ \citenamefont {Loss}}]{stefano2021prx}%
  \BibitemOpen
  \bibfield  {author} {\bibinfo {author} {\bibfnamefont {S.}~\bibnamefont
  {Bosco}}, \bibinfo {author} {\bibfnamefont {B.}~\bibnamefont {Het\'enyi}},\
  and\ \bibinfo {author} {\bibfnamefont {D.}~\bibnamefont {Loss}},\ }\bibfield
  {title} {\bibinfo {title} {Hole spin qubits in $\mathrm{Si}$ finfets with
  fully tunable spin-orbit coupling and sweet spots for charge noise},\
  }\href@noop {} {\bibfield  {journal} {\bibinfo  {journal} {PRX Quantum}\
  }\textbf {\bibinfo {volume} {2}},\ \bibinfo {pages} {010348} (\bibinfo {year}
  {2021}{\natexlab{b}})}\BibitemShut {NoStop}%
\bibitem [{\citenamefont {Bosco}\ and\ \citenamefont
  {Loss}(2021)}]{stefano2021prl}%
  \BibitemOpen
  \bibfield  {author} {\bibinfo {author} {\bibfnamefont {S.}~\bibnamefont
  {Bosco}}\ and\ \bibinfo {author} {\bibfnamefont {D.}~\bibnamefont {Loss}},\
  }\bibfield  {title} {\bibinfo {title} {Fully tunable hyperfine interactions
  of hole spin qubits in si and ge quantum dots},\ }\href@noop {} {\bibfield
  {journal} {\bibinfo  {journal} {Phys. Rev. Lett.}\ }\textbf {\bibinfo
  {volume} {127}},\ \bibinfo {pages} {190501} (\bibinfo {year}
  {2021})}\BibitemShut {NoStop}%
\bibitem [{\citenamefont {Dmytruk}\ \emph {et~al.}(2018)\citenamefont
  {Dmytruk}, \citenamefont {Chevallier}, \citenamefont {Loss},\ and\
  \citenamefont {Klinovaja}}]{dmytruk2018prb}%
  \BibitemOpen
  \bibfield  {author} {\bibinfo {author} {\bibfnamefont {O.}~\bibnamefont
  {Dmytruk}}, \bibinfo {author} {\bibfnamefont {D.}~\bibnamefont {Chevallier}},
  \bibinfo {author} {\bibfnamefont {D.}~\bibnamefont {Loss}},\ and\ \bibinfo
  {author} {\bibfnamefont {J.}~\bibnamefont {Klinovaja}},\ }\bibfield  {title}
  {\bibinfo {title} {Renormalization of the quantum dot $g$-factor in
  superconducting rashba nanowires},\ }\href@noop {} {\bibfield  {journal}
  {\bibinfo  {journal} {Phys. Rev. B}\ }\textbf {\bibinfo {volume} {98}},\
  \bibinfo {pages} {165403} (\bibinfo {year} {2018})}\BibitemShut {NoStop}%
\bibitem [{\citenamefont {Krantz}\ \emph {et~al.}(2019)\citenamefont {Krantz},
  \citenamefont {Kjaergaard}, \citenamefont {Yan}, \citenamefont {Orlando},
  \citenamefont {Gustavsson},\ and\ \citenamefont {Oliver}}]{Krantz_2019_apr}%
  \BibitemOpen
  \bibfield  {author} {\bibinfo {author} {\bibfnamefont {P.}~\bibnamefont
  {Krantz}}, \bibinfo {author} {\bibfnamefont {M.}~\bibnamefont {Kjaergaard}},
  \bibinfo {author} {\bibfnamefont {F.}~\bibnamefont {Yan}}, \bibinfo {author}
  {\bibfnamefont {T.~P.}\ \bibnamefont {Orlando}}, \bibinfo {author}
  {\bibfnamefont {S.}~\bibnamefont {Gustavsson}},\ and\ \bibinfo {author}
  {\bibfnamefont {W.~D.}\ \bibnamefont {Oliver}},\ }\bibfield  {title}
  {\bibinfo {title} {A quantum engineer's guide to superconducting qubits},\
  }\href@noop {} {\bibfield  {journal} {\bibinfo  {journal} {Applied Physics
  Reviews}\ }\textbf {\bibinfo {volume} {6}},\ \bibinfo {pages} {021318}
  (\bibinfo {year} {2019})}\BibitemShut {NoStop}%
\bibitem [{\citenamefont {Nielsen}\ and\ \citenamefont
  {Chuang}(2011)}]{nielsen}%
  \BibitemOpen
  \bibfield  {author} {\bibinfo {author} {\bibfnamefont {M.~A.}\ \bibnamefont
  {Nielsen}}\ and\ \bibinfo {author} {\bibfnamefont {I.~L.}\ \bibnamefont
  {Chuang}},\ }\href@noop {} {\emph {\bibinfo {title} {Quantum Computation and
  Quantum Information}}},\ \bibinfo {edition} {10th}\ ed.\ (\bibinfo
  {publisher} {Cambridge University Press},\ \bibinfo {year} {January 31,
  2011})\BibitemShut {NoStop}%
\bibitem [{\citenamefont {Parkin}\ \emph {et~al.}(1990)\citenamefont {Parkin},
  \citenamefont {More},\ and\ \citenamefont {Roche}}]{parkin_1990_prl}%
  \BibitemOpen
  \bibfield  {author} {\bibinfo {author} {\bibfnamefont {S.~S.~P.}\
  \bibnamefont {Parkin}}, \bibinfo {author} {\bibfnamefont {N.}~\bibnamefont
  {More}},\ and\ \bibinfo {author} {\bibfnamefont {K.~P.}\ \bibnamefont
  {Roche}},\ }\bibfield  {title} {\bibinfo {title} {Oscillations in exchange
  coupling and magnetoresistance in metallic superlattice structures: Co/ru,
  co/cr, and fe/cr},\ }\href@noop {} {\bibfield  {journal} {\bibinfo  {journal}
  {Physical Review Letters}\ }\textbf {\bibinfo {volume} {64}},\ \bibinfo
  {pages} {2304} (\bibinfo {year} {1990})}\BibitemShut {NoStop}%
\bibitem [{\citenamefont {Parkin}\ \emph {et~al.}(1991)\citenamefont {Parkin},
  \citenamefont {Bhadra},\ and\ \citenamefont {Roche}}]{parkin_1991_prl}%
  \BibitemOpen
  \bibfield  {author} {\bibinfo {author} {\bibfnamefont {S.~S.~P.}\
  \bibnamefont {Parkin}}, \bibinfo {author} {\bibfnamefont {R.}~\bibnamefont
  {Bhadra}},\ and\ \bibinfo {author} {\bibfnamefont {K.~P.}\ \bibnamefont
  {Roche}},\ }\bibfield  {title} {\bibinfo {title} {Oscillatory magnetic
  exchange coupling through thin copper layers},\ }\href@noop {} {\bibfield
  {journal} {\bibinfo  {journal} {Physical Review Letters}\ }\textbf {\bibinfo
  {volume} {66}},\ \bibinfo {pages} {2152} (\bibinfo {year}
  {1991})}\BibitemShut {NoStop}%
\bibitem [{\citenamefont {Okada}\ \emph {et~al.}(2017)\citenamefont {Okada},
  \citenamefont {He}, \citenamefont {Gu}, \citenamefont {Kanai}, \citenamefont
  {Soumyanarayanan}, \citenamefont {Lim}, \citenamefont {Tran}, \citenamefont
  {Mori}, \citenamefont {Maekawa}, \citenamefont {Matsukura}, \citenamefont
  {Ohno},\ and\ \citenamefont {Panagopoulos}}]{okada2017}%
  \BibitemOpen
  \bibfield  {author} {\bibinfo {author} {\bibfnamefont {A.}~\bibnamefont
  {Okada}}, \bibinfo {author} {\bibfnamefont {S.}~\bibnamefont {He}}, \bibinfo
  {author} {\bibfnamefont {B.}~\bibnamefont {Gu}}, \bibinfo {author}
  {\bibfnamefont {S.}~\bibnamefont {Kanai}}, \bibinfo {author} {\bibfnamefont
  {A.}~\bibnamefont {Soumyanarayanan}}, \bibinfo {author} {\bibfnamefont
  {S.~T.}\ \bibnamefont {Lim}}, \bibinfo {author} {\bibfnamefont
  {M.}~\bibnamefont {Tran}}, \bibinfo {author} {\bibfnamefont {M.}~\bibnamefont
  {Mori}}, \bibinfo {author} {\bibfnamefont {S.}~\bibnamefont {Maekawa}},
  \bibinfo {author} {\bibfnamefont {F.}~\bibnamefont {Matsukura}}, \bibinfo
  {author} {\bibfnamefont {H.}~\bibnamefont {Ohno}},\ and\ \bibinfo {author}
  {\bibfnamefont {C.}~\bibnamefont {Panagopoulos}},\ }\bibfield  {title}
  {\bibinfo {title} {Magnetization dynamics and its scattering mechanism in
  thin cofeb films with interfacial anisotropy},\ }\href@noop {} {\bibfield
  {journal} {\bibinfo  {journal} {Proceedings of the National Academy of
  Sciences}\ }\textbf {\bibinfo {volume} {114}},\ \bibinfo {pages} {3815}
  (\bibinfo {year} {2017})}\BibitemShut {NoStop}%
\bibitem [{\citenamefont {Maier-Flaig}\ \emph {et~al.}(2017)\citenamefont
  {Maier-Flaig}, \citenamefont {Klingler}, \citenamefont {Dubs}, \citenamefont
  {Surzhenko}, \citenamefont {Gross}, \citenamefont {Weiler}, \citenamefont
  {Huebl},\ and\ \citenamefont {Goennenwein}}]{marier2017}%
  \BibitemOpen
  \bibfield  {author} {\bibinfo {author} {\bibfnamefont {H.}~\bibnamefont
  {Maier-Flaig}}, \bibinfo {author} {\bibfnamefont {S.}~\bibnamefont
  {Klingler}}, \bibinfo {author} {\bibfnamefont {C.}~\bibnamefont {Dubs}},
  \bibinfo {author} {\bibfnamefont {O.}~\bibnamefont {Surzhenko}}, \bibinfo
  {author} {\bibfnamefont {R.}~\bibnamefont {Gross}}, \bibinfo {author}
  {\bibfnamefont {M.}~\bibnamefont {Weiler}}, \bibinfo {author} {\bibfnamefont
  {H.}~\bibnamefont {Huebl}},\ and\ \bibinfo {author} {\bibfnamefont
  {S.~T.~B.}\ \bibnamefont {Goennenwein}},\ }\bibfield  {title} {\bibinfo
  {title} {Temperature-dependent magnetic damping of yttrium iron garnet
  spheres},\ }\href@noop {} {\bibfield  {journal} {\bibinfo  {journal} {Phys.
  Rev. B}\ }\textbf {\bibinfo {volume} {95}},\ \bibinfo {pages} {214423}
  (\bibinfo {year} {2017})}\BibitemShut {NoStop}%
\bibitem [{\citenamefont {Trifunovic}\ \emph {et~al.}(2013)\citenamefont
  {Trifunovic}, \citenamefont {Pedrocchi},\ and\ \citenamefont
  {Loss}}]{Daniel2013prx}%
  \BibitemOpen
  \bibfield  {author} {\bibinfo {author} {\bibfnamefont {L.}~\bibnamefont
  {Trifunovic}}, \bibinfo {author} {\bibfnamefont {F.~L.}\ \bibnamefont
  {Pedrocchi}},\ and\ \bibinfo {author} {\bibfnamefont {D.}~\bibnamefont
  {Loss}},\ }\bibfield  {title} {\bibinfo {title} {Long-distance entanglement
  of spin qubits via ferromagnet},\ }\href@noop {} {\bibfield  {journal}
  {\bibinfo  {journal} {Phys. Rev. X}\ }\textbf {\bibinfo {volume} {3}},\
  \bibinfo {pages} {041023} (\bibinfo {year} {2013})}\BibitemShut {NoStop}%
\bibitem [{\citenamefont {Fukami}\ \emph {et~al.}(2021)\citenamefont {Fukami},
  \citenamefont {Candido}, \citenamefont {Awschalom},\ and\ \citenamefont
  {Flatt\'e}}]{Fukami2021prx}%
  \BibitemOpen
  \bibfield  {author} {\bibinfo {author} {\bibfnamefont {M.}~\bibnamefont
  {Fukami}}, \bibinfo {author} {\bibfnamefont {D.~R.}\ \bibnamefont {Candido}},
  \bibinfo {author} {\bibfnamefont {D.~D.}\ \bibnamefont {Awschalom}},\ and\
  \bibinfo {author} {\bibfnamefont {M.~E.}\ \bibnamefont {Flatt\'e}},\
  }\bibfield  {title} {\bibinfo {title} {Opportunities for long-range
  magnon-mediated entanglement of spin qubits via on- and off-resonant
  coupling},\ }\href@noop {} {\bibfield  {journal} {\bibinfo  {journal} {PRX
  Quantum}\ }\textbf {\bibinfo {volume} {2}},\ \bibinfo {pages} {040314}
  (\bibinfo {year} {2021})}\BibitemShut {NoStop}%
\bibitem [{\citenamefont {Candido}\ \emph {et~al.}(2020)\citenamefont
  {Candido}, \citenamefont {Fuchs}, \citenamefont {Johnston-Halperin},\ and\
  \citenamefont {Flatt{\'{e}}}}]{Candido_2020}%
  \BibitemOpen
  \bibfield  {author} {\bibinfo {author} {\bibfnamefont {D.~R.}\ \bibnamefont
  {Candido}}, \bibinfo {author} {\bibfnamefont {G.~D.}\ \bibnamefont {Fuchs}},
  \bibinfo {author} {\bibfnamefont {E.}~\bibnamefont {Johnston-Halperin}},\
  and\ \bibinfo {author} {\bibfnamefont {M.~E.}\ \bibnamefont {Flatt{\'{e}}}},\
  }\bibfield  {title} {\bibinfo {title} {Predicted strong coupling of
  solid-state spins via a single magnon mode},\ }\href@noop {} {\bibfield
  {journal} {\bibinfo  {journal} {Materials for Quantum Technology}\ }\textbf
  {\bibinfo {volume} {1}},\ \bibinfo {pages} {011001} (\bibinfo {year}
  {2020})}\BibitemShut {NoStop}%
\bibitem [{\citenamefont {M\"uhlherr}\ \emph {et~al.}(2019)\citenamefont
  {M\"uhlherr}, \citenamefont {Shkolnikov},\ and\ \citenamefont
  {Burkard}}]{Burkard2019prb}%
  \BibitemOpen
  \bibfield  {author} {\bibinfo {author} {\bibfnamefont {C.}~\bibnamefont
  {M\"uhlherr}}, \bibinfo {author} {\bibfnamefont {V.~O.}\ \bibnamefont
  {Shkolnikov}},\ and\ \bibinfo {author} {\bibfnamefont {G.}~\bibnamefont
  {Burkard}},\ }\bibfield  {title} {\bibinfo {title} {Magnetic resonance in
  defect spins mediated by spin waves},\ }\href@noop {} {\bibfield  {journal}
  {\bibinfo  {journal} {Phys. Rev. B}\ }\textbf {\bibinfo {volume} {99}},\
  \bibinfo {pages} {195413} (\bibinfo {year} {2019})}\BibitemShut {NoStop}%
\bibitem [{\citenamefont {Zou}\ \emph {et~al.}(2022)\citenamefont {Zou},
  \citenamefont {Zhang},\ and\ \citenamefont {Tserkovnyak}}]{zou2022prb}%
  \BibitemOpen
  \bibfield  {author} {\bibinfo {author} {\bibfnamefont {J.}~\bibnamefont
  {Zou}}, \bibinfo {author} {\bibfnamefont {S.}~\bibnamefont {Zhang}},\ and\
  \bibinfo {author} {\bibfnamefont {Y.}~\bibnamefont {Tserkovnyak}},\
  }\bibfield  {title} {\bibinfo {title} {Bell-state generation for spin qubits
  via dissipative coupling},\ }\href@noop {} {\bibfield  {journal} {\bibinfo
  {journal} {Physical Review B}\ }\textbf {\bibinfo {volume} {106}},\ \bibinfo
  {pages} {L180406} (\bibinfo {year} {2022})}\BibitemShut {NoStop}%
\bibitem [{\citenamefont {Song}\ \emph {et~al.}(2021)\citenamefont {Song},
  \citenamefont {Sun}, \citenamefont {Anderson}, \citenamefont {Wang},
  \citenamefont {Qian}, \citenamefont {Taniguchi}, \citenamefont {Watanabe},
  \citenamefont {McGuire}, \citenamefont {St{\"o}hr}, \citenamefont {Xiao},
  \citenamefont {Cao}, \citenamefont {Wrachtrup},\ and\ \citenamefont
  {Xu}}]{Song_science_2021}%
  \BibitemOpen
  \bibfield  {author} {\bibinfo {author} {\bibfnamefont {T.}~\bibnamefont
  {Song}}, \bibinfo {author} {\bibfnamefont {Q.-C.}\ \bibnamefont {Sun}},
  \bibinfo {author} {\bibfnamefont {E.}~\bibnamefont {Anderson}}, \bibinfo
  {author} {\bibfnamefont {C.}~\bibnamefont {Wang}}, \bibinfo {author}
  {\bibfnamefont {J.}~\bibnamefont {Qian}}, \bibinfo {author} {\bibfnamefont
  {T.}~\bibnamefont {Taniguchi}}, \bibinfo {author} {\bibfnamefont
  {K.}~\bibnamefont {Watanabe}}, \bibinfo {author} {\bibfnamefont {M.~A.}\
  \bibnamefont {McGuire}}, \bibinfo {author} {\bibfnamefont {R.}~\bibnamefont
  {St{\"o}hr}}, \bibinfo {author} {\bibfnamefont {D.}~\bibnamefont {Xiao}},
  \bibinfo {author} {\bibfnamefont {T.}~\bibnamefont {Cao}}, \bibinfo {author}
  {\bibfnamefont {J.}~\bibnamefont {Wrachtrup}},\ and\ \bibinfo {author}
  {\bibfnamefont {X.}~\bibnamefont {Xu}},\ }\bibfield  {title} {\bibinfo
  {title} {Direct visualization of magnetic domains and moire magnetism in
  twisted 2d magnets},\ }\href@noop {} {\bibfield  {journal} {\bibinfo
  {journal} {Science}\ }\textbf {\bibinfo {volume} {374}},\ \bibinfo {pages}
  {1140} (\bibinfo {year} {2021})}\BibitemShut {NoStop}%
\bibitem [{\citenamefont {Finco}\ \emph {et~al.}(2021)\citenamefont {Finco},
  \citenamefont {Haykal}, \citenamefont {Tanos}, \citenamefont {Fabre},
  \citenamefont {Chouaieb}, \citenamefont {Akhtar}, \citenamefont
  {Robert-Philip}, \citenamefont {Legrand}, \citenamefont {Ajejas},
  \citenamefont {Bouzehouane}, \citenamefont {Reyren}, \citenamefont
  {Devolder}, \citenamefont {Adam}, \citenamefont {Kim}, \citenamefont {Cros},\
  and\ \citenamefont {Jacques}}]{Finco_Natcom_2021}%
  \BibitemOpen
  \bibfield  {author} {\bibinfo {author} {\bibfnamefont {A.}~\bibnamefont
  {Finco}}, \bibinfo {author} {\bibfnamefont {A.}~\bibnamefont {Haykal}},
  \bibinfo {author} {\bibfnamefont {R.}~\bibnamefont {Tanos}}, \bibinfo
  {author} {\bibfnamefont {F.}~\bibnamefont {Fabre}}, \bibinfo {author}
  {\bibfnamefont {S.}~\bibnamefont {Chouaieb}}, \bibinfo {author}
  {\bibfnamefont {W.}~\bibnamefont {Akhtar}}, \bibinfo {author} {\bibfnamefont
  {I.}~\bibnamefont {Robert-Philip}}, \bibinfo {author} {\bibfnamefont
  {W.}~\bibnamefont {Legrand}}, \bibinfo {author} {\bibfnamefont
  {F.}~\bibnamefont {Ajejas}}, \bibinfo {author} {\bibfnamefont
  {K.}~\bibnamefont {Bouzehouane}}, \bibinfo {author} {\bibfnamefont
  {N.}~\bibnamefont {Reyren}}, \bibinfo {author} {\bibfnamefont
  {T.}~\bibnamefont {Devolder}}, \bibinfo {author} {\bibfnamefont {J.-P.}\
  \bibnamefont {Adam}}, \bibinfo {author} {\bibfnamefont {J.-V.}\ \bibnamefont
  {Kim}}, \bibinfo {author} {\bibfnamefont {V.}~\bibnamefont {Cros}},\ and\
  \bibinfo {author} {\bibfnamefont {V.}~\bibnamefont {Jacques}},\ }\bibfield
  {title} {\bibinfo {title} {Imaging non-collinear antiferromagnetic textures
  via single spin relaxometry},\ }\href@noop {} {\bibfield  {journal} {\bibinfo
   {journal} {Nature Communications}\ }\textbf {\bibinfo {volume} {12}},\
  \bibinfo {pages} {767} (\bibinfo {year} {2021})}\BibitemShut {NoStop}%
\bibitem [{\citenamefont {Elzerman}\ \emph {et~al.}(2004)\citenamefont
  {Elzerman}, \citenamefont {Hanson}, \citenamefont {Willems~van Beveren},
  \citenamefont {Witkamp}, \citenamefont {Vandersypen},\ and\ \citenamefont
  {Kouwenhoven}}]{Elzerman:2004wt}%
  \BibitemOpen
  \bibfield  {author} {\bibinfo {author} {\bibfnamefont {J.~M.}\ \bibnamefont
  {Elzerman}}, \bibinfo {author} {\bibfnamefont {R.}~\bibnamefont {Hanson}},
  \bibinfo {author} {\bibfnamefont {L.~H.}\ \bibnamefont {Willems~van
  Beveren}}, \bibinfo {author} {\bibfnamefont {B.}~\bibnamefont {Witkamp}},
  \bibinfo {author} {\bibfnamefont {L.~M.~K.}\ \bibnamefont {Vandersypen}},\
  and\ \bibinfo {author} {\bibfnamefont {L.~P.}\ \bibnamefont {Kouwenhoven}},\
  }\bibfield  {title} {\bibinfo {title} {Single-shot read-out of an individual
  electron spin in a quantum dot},\ }\href@noop {} {\bibfield  {journal}
  {\bibinfo  {journal} {Nature}\ }\textbf {\bibinfo {volume} {430}},\ \bibinfo
  {pages} {431} (\bibinfo {year} {2004})}\BibitemShut {NoStop}%
\bibitem [{\citenamefont {Lai}\ \emph {et~al.}(2011)\citenamefont {Lai},
  \citenamefont {Lim}, \citenamefont {Yang}, \citenamefont {Zwanenburg},
  \citenamefont {Coish}, \citenamefont {Qassemi}, \citenamefont {Morello},\
  and\ \citenamefont {Dzurak}}]{Lai:2011vv}%
  \BibitemOpen
  \bibfield  {author} {\bibinfo {author} {\bibfnamefont {N.~S.}\ \bibnamefont
  {Lai}}, \bibinfo {author} {\bibfnamefont {W.~H.}\ \bibnamefont {Lim}},
  \bibinfo {author} {\bibfnamefont {C.~H.}\ \bibnamefont {Yang}}, \bibinfo
  {author} {\bibfnamefont {F.~A.}\ \bibnamefont {Zwanenburg}}, \bibinfo
  {author} {\bibfnamefont {W.~A.}\ \bibnamefont {Coish}}, \bibinfo {author}
  {\bibfnamefont {F.}~\bibnamefont {Qassemi}}, \bibinfo {author} {\bibfnamefont
  {A.}~\bibnamefont {Morello}},\ and\ \bibinfo {author} {\bibfnamefont {A.~S.}\
  \bibnamefont {Dzurak}},\ }\bibfield  {title} {\bibinfo {title} {Pauli spin
  blockade in a highly tunable silicon double quantum dot},\ }\href@noop {}
  {\bibfield  {journal} {\bibinfo  {journal} {Scientific Reports}\ }\textbf
  {\bibinfo {volume} {1}},\ \bibinfo {pages} {110} (\bibinfo {year}
  {2011})}\BibitemShut {NoStop}%
\bibitem [{\citenamefont {Pla}\ \emph {et~al.}(2013)\citenamefont {Pla},
  \citenamefont {Tan}, \citenamefont {Dehollain}, \citenamefont {Lim},
  \citenamefont {Morton}, \citenamefont {Zwanenburg}, \citenamefont {Jamieson},
  \citenamefont {Dzurak},\ and\ \citenamefont {Morello}}]{Pla:2013ta}%
  \BibitemOpen
  \bibfield  {author} {\bibinfo {author} {\bibfnamefont {J.~J.}\ \bibnamefont
  {Pla}}, \bibinfo {author} {\bibfnamefont {K.~Y.}\ \bibnamefont {Tan}},
  \bibinfo {author} {\bibfnamefont {J.~P.}\ \bibnamefont {Dehollain}}, \bibinfo
  {author} {\bibfnamefont {W.~H.}\ \bibnamefont {Lim}}, \bibinfo {author}
  {\bibfnamefont {J.~J.~L.}\ \bibnamefont {Morton}}, \bibinfo {author}
  {\bibfnamefont {F.~A.}\ \bibnamefont {Zwanenburg}}, \bibinfo {author}
  {\bibfnamefont {D.~N.}\ \bibnamefont {Jamieson}}, \bibinfo {author}
  {\bibfnamefont {A.~S.}\ \bibnamefont {Dzurak}},\ and\ \bibinfo {author}
  {\bibfnamefont {A.}~\bibnamefont {Morello}},\ }\bibfield  {title} {\bibinfo
  {title} {High-fidelity readout and control of a nuclear spin qubit in
  silicon},\ }\href@noop {} {\bibfield  {journal} {\bibinfo  {journal}
  {Nature}\ }\textbf {\bibinfo {volume} {496}},\ \bibinfo {pages} {334}
  (\bibinfo {year} {2013})}\BibitemShut {NoStop}%
\bibitem [{\citenamefont {Harvey-Collard}\ \emph {et~al.}(2018)\citenamefont
  {Harvey-Collard}, \citenamefont {D'Anjou}, \citenamefont {Rudolph},
  \citenamefont {Jacobson}, \citenamefont {Dominguez}, \citenamefont
  {Ten~Eyck}, \citenamefont {Wendt}, \citenamefont {Pluym}, \citenamefont
  {Lilly}, \citenamefont {Coish}, \citenamefont {Pioro-Ladri\`ere},\ and\
  \citenamefont {Carroll}}]{PhysRevX.8.021046}%
  \BibitemOpen
  \bibfield  {author} {\bibinfo {author} {\bibfnamefont {P.}~\bibnamefont
  {Harvey-Collard}}, \bibinfo {author} {\bibfnamefont {B.}~\bibnamefont
  {D'Anjou}}, \bibinfo {author} {\bibfnamefont {M.}~\bibnamefont {Rudolph}},
  \bibinfo {author} {\bibfnamefont {N.~T.}\ \bibnamefont {Jacobson}}, \bibinfo
  {author} {\bibfnamefont {J.}~\bibnamefont {Dominguez}}, \bibinfo {author}
  {\bibfnamefont {G.~A.}\ \bibnamefont {Ten~Eyck}}, \bibinfo {author}
  {\bibfnamefont {J.~R.}\ \bibnamefont {Wendt}}, \bibinfo {author}
  {\bibfnamefont {T.}~\bibnamefont {Pluym}}, \bibinfo {author} {\bibfnamefont
  {M.~P.}\ \bibnamefont {Lilly}}, \bibinfo {author} {\bibfnamefont {W.~A.}\
  \bibnamefont {Coish}}, \bibinfo {author} {\bibfnamefont {M.}~\bibnamefont
  {Pioro-Ladri\`ere}},\ and\ \bibinfo {author} {\bibfnamefont {M.~S.}\
  \bibnamefont {Carroll}},\ }\bibfield  {title} {\bibinfo {title}
  {High-fidelity single-shot readout for a spin qubit via an enhanced latching
  mechanism},\ }\href@noop {} {\bibfield  {journal} {\bibinfo  {journal} {Phys.
  Rev. X}\ }\textbf {\bibinfo {volume} {8}},\ \bibinfo {pages} {021046}
  (\bibinfo {year} {2018})}\BibitemShut {NoStop}%
\bibitem [{\citenamefont {Morello}\ \emph {et~al.}(2010)\citenamefont
  {Morello}, \citenamefont {Pla}, \citenamefont {Zwanenburg}, \citenamefont
  {Chan}, \citenamefont {Tan}, \citenamefont {Huebl}, \citenamefont
  {M{\"o}tt{\"o}nen}, \citenamefont {Nugroho}, \citenamefont {Yang},
  \citenamefont {van Donkelaar}, \citenamefont {Alves}, \citenamefont
  {Jamieson}, \citenamefont {Escott}, \citenamefont {Hollenberg}, \citenamefont
  {Clark},\ and\ \citenamefont {Dzurak}}]{Morello:2010vm}%
  \BibitemOpen
  \bibfield  {author} {\bibinfo {author} {\bibfnamefont {A.}~\bibnamefont
  {Morello}}, \bibinfo {author} {\bibfnamefont {J.~J.}\ \bibnamefont {Pla}},
  \bibinfo {author} {\bibfnamefont {F.~A.}\ \bibnamefont {Zwanenburg}},
  \bibinfo {author} {\bibfnamefont {K.~W.}\ \bibnamefont {Chan}}, \bibinfo
  {author} {\bibfnamefont {K.~Y.}\ \bibnamefont {Tan}}, \bibinfo {author}
  {\bibfnamefont {H.}~\bibnamefont {Huebl}}, \bibinfo {author} {\bibfnamefont
  {M.}~\bibnamefont {M{\"o}tt{\"o}nen}}, \bibinfo {author} {\bibfnamefont
  {C.~D.}\ \bibnamefont {Nugroho}}, \bibinfo {author} {\bibfnamefont
  {C.}~\bibnamefont {Yang}}, \bibinfo {author} {\bibfnamefont {J.~A.}\
  \bibnamefont {van Donkelaar}}, \bibinfo {author} {\bibfnamefont {A.~D.~C.}\
  \bibnamefont {Alves}}, \bibinfo {author} {\bibfnamefont {D.~N.}\ \bibnamefont
  {Jamieson}}, \bibinfo {author} {\bibfnamefont {C.~C.}\ \bibnamefont
  {Escott}}, \bibinfo {author} {\bibfnamefont {L.~C.~L.}\ \bibnamefont
  {Hollenberg}}, \bibinfo {author} {\bibfnamefont {R.~G.}\ \bibnamefont
  {Clark}},\ and\ \bibinfo {author} {\bibfnamefont {A.~S.}\ \bibnamefont
  {Dzurak}},\ }\bibfield  {title} {\bibinfo {title} {Single-shot readout of an
  electron spin in silicon},\ }\href@noop {} {\bibfield  {journal} {\bibinfo
  {journal} {Nature}\ }\textbf {\bibinfo {volume} {467}},\ \bibinfo {pages}
  {687} (\bibinfo {year} {2010})}\BibitemShut {NoStop}%
\bibitem [{\citenamefont {Neumann}\ \emph {et~al.}(2010)\citenamefont
  {Neumann}, \citenamefont {Beck}, \citenamefont {Steiner}, \citenamefont
  {Rempp}, \citenamefont {Fedder}, \citenamefont {Hemmer}, \citenamefont
  {Wrachtrup},\ and\ \citenamefont {Jelezko}}]{pnsci2010}%
  \BibitemOpen
  \bibfield  {author} {\bibinfo {author} {\bibfnamefont {P.}~\bibnamefont
  {Neumann}}, \bibinfo {author} {\bibfnamefont {J.}~\bibnamefont {Beck}},
  \bibinfo {author} {\bibfnamefont {M.}~\bibnamefont {Steiner}}, \bibinfo
  {author} {\bibfnamefont {F.}~\bibnamefont {Rempp}}, \bibinfo {author}
  {\bibfnamefont {H.}~\bibnamefont {Fedder}}, \bibinfo {author} {\bibfnamefont
  {P.~R.}\ \bibnamefont {Hemmer}}, \bibinfo {author} {\bibfnamefont
  {J.}~\bibnamefont {Wrachtrup}},\ and\ \bibinfo {author} {\bibfnamefont
  {F.}~\bibnamefont {Jelezko}},\ }\bibfield  {title} {\bibinfo {title}
  {Single-shot readout of a single nuclear spin},\ }\href@noop {} {\bibfield
  {journal} {\bibinfo  {journal} {Science}\ }\textbf {\bibinfo {volume}
  {329}},\ \bibinfo {pages} {542} (\bibinfo {year} {2010})}\BibitemShut
  {NoStop}%
\bibitem [{\citenamefont {Fradkin}(2013)}]{qftcmt}%
  \BibitemOpen
  \bibfield  {author} {\bibinfo {author} {\bibfnamefont {E.}~\bibnamefont
  {Fradkin}},\ }\href@noop {} {\emph {\bibinfo {title} {Field Theories of
  Condensed Matter Physics}}},\ \bibinfo {edition} {2nd}\ ed.\ (\bibinfo
  {publisher} {Cambridge University Press},\ \bibinfo {year}
  {2013})\BibitemShut {NoStop}%
\bibitem [{\citenamefont {Kim}\ and\ \citenamefont
  {Tchernyshyov}(2022)}]{kim2022}%
  \BibitemOpen
  \bibfield  {author} {\bibinfo {author} {\bibfnamefont {S.~K.}\ \bibnamefont
  {Kim}}\ and\ \bibinfo {author} {\bibfnamefont {O.}~\bibnamefont
  {Tchernyshyov}},\ }\href@noop {} {\bibinfo {title} {Mechanics of a
  ferromagnetic domain wall}} (\bibinfo {year} {2022})\BibitemShut {NoStop}%
\bibitem [{\citenamefont {Braun}\ and\ \citenamefont
  {Loss}(1996)}]{daniel2016prb}%
  \BibitemOpen
  \bibfield  {author} {\bibinfo {author} {\bibfnamefont {H.-B.}\ \bibnamefont
  {Braun}}\ and\ \bibinfo {author} {\bibfnamefont {D.}~\bibnamefont {Loss}},\
  }\bibfield  {title} {\bibinfo {title} {Berry's phase and quantum dynamics of
  ferromagnetic solitons},\ }\href@noop {} {\bibfield  {journal} {\bibinfo
  {journal} {Phys. Rev. B}\ }\textbf {\bibinfo {volume} {53}},\ \bibinfo
  {pages} {3237} (\bibinfo {year} {1996})}\BibitemShut {NoStop}%
\end{thebibliography}
%

\end{document}